\documentclass[preprint,12pt,final]{elsarticle}

\usepackage{graphics}
\usepackage{graphicx}
\usepackage{epsfig}
\usepackage{amsmath,amsfonts}
\usepackage{amssymb}

\newcounter{bla}

\journal{Computer Physics Communications}

\usepackage{epstopdf}
\usepackage{hyperref}
\usepackage{xcolor}
\usepackage[normalem]{ulem}
\usepackage{parskip}

\newcommand{\Core}{\texttt{Core}}
\newcommand{\Cpp}{\texttt{C++}}
\newcommand{\dd}{\mathrm{d}}
\newcommand{\Mathematica}{\texttt{Mathe\-matica}}
\newcommand{\MathLink}{\texttt{MathLink}}
\newcommand{\MSbar}{\overline{\mathrm{MS}}}
\newcommand{\Python}{\texttt{Python}}
\newcommand{\REvolver}{\texttt{REvolver}}

\begin{document}

\begin{frontmatter}

\title{\textbf{REvolver}\\ Automated running and matching of couplings and masses in QCD}

\author[a,b]{Andr\'e H. Hoang}
\ead{andre.hoang@univie.ac.at}
\author[a]{Christopher Lepenik}
\ead{christopher.lepenik@univie.ac.at}
\author[c,d]{Vicent Mateu}
\ead{vmateu@usal.es}

\address[a]{University of Vienna, Faculty of Physics, Boltzmanngasse 5, A-1090 Wien, Austria}
\address[b]{Erwin Schr\"odinger International Institute for Mathematical Physics,\\
University of Vienna, Boltzmanngasse 9, A-1090 Wien, Austria}
\address[c]{Departamento de F\'isica Fundamental e IUFFyM,\\Universidad de Salamanca, E-37008 Salamanca, Spain}
\address[d]{Instituto de F\'isica Te\'orica UAM-CSIC, E-28049 Madrid, Spain}

\begin{abstract}
In this article we present \REvolver{}, a \Cpp{} library for renormalization group evolution and automatic flavor matching of the QCD coupling and quark masses, as well as precise conversion between various quark mass renormalization schemes. The library systematically accounts for the renormalization group evolution of low-scale short-distance masses which depend linearly on the renormalization scale and sums logarithmic terms of high and low scales that are missed by the common logarithmic renormalization scale evolution. The library can also be accessed through \Mathematica{} and \Python{} interfaces and provides renormalization group evolution for complex renormalization scales as well.

\end{abstract}


\end{frontmatter}
\vspace{-540pt}
\hfill{\footnotesize UWThPh-2020-18, IFT-UAM/CSIC-21-5}

\clearpage
{\bf PROGRAM SUMMARY}

\begin{small}
{\em Program Title:} \REvolver{}\\
{\em Licensing provisions:} GPLv3 or later\\
{\em Programming language:} \Cpp{}, \Python{}, \texttt{Wolfram Language}\\
{\em Program obtainable from:} \url{https://gitlab.com/REvolver-hep/REvolver}\\
{\em Operating system:} Linux, MacOS, partially Windows\\
{\em Required RAM:} insignificant for a limited number of instances of the \Core{} class\\
{\em Number of processors used:} one\\
{\em Running time:} fractions of seconds for single commands\\
{\em Supplementary material:} this article, demo programs, doxygen documentation

{\em Nature of problem:}\\
The strong coupling and the quark masses are fundamental parameters of QCD that are scheme and renormalization-scale dependent. The choice of scheme depends on the active number of flavors and the range of scales, and is dictated by the requirements to minimize the size of corrections and to sum large logarithmic corrections to all orders. For the strong coupling and the quark masses at high scales, the $\overline{\mbox{MS}}$ scheme with logarithmic scale dependence is used. For quark masses at low scales, short-distance mass schemes with linear scale-dependence are used. The \REvolver{} library provides conversions for the strong coupling and the most common quark mass schemes, with renormalization scale evolution implemented such that all types of large logarithmic terms are summed to all orders, accounting for flavor threshold effects and state-or-the-art correction terms. The pole mass, which is not a short-distance mass and contains a sizable renormalon ambiguity, is treated as a derived quantity.

{\em Solution method:}\\
Renormalization group equations are solved for complex-valued scales to machine precision based on fast-converging iterative algorithms and analytic all-order expressions. Matching relations for the strong coupling at flavor thresholds are computed in a way that gives equal results for upward and downward evolution. \Core{} objects allow to define an arbitrary number of physical scenarios for strong coupling values and quark mass spectra, where options for precision and matching scales can be set freely, and values for quark masses in all common schemes including the pole mass can be extracted. All \REvolver{} routines are implemented entirely in \Cpp{} and can be accessed through \Mathematica{} and \Python{} interfaces.

\end{small}

\clearpage
\tableofcontents
\clearpage

\section{Introduction}
\label{sec:intro}

Quark masses are fundamental parameters of quantum chromodynamics\break (QCD) and their precise determination in adequate schemes and at appropriate renormalization scales is of high interest for theoretical as well as experimental studies of many processes. These can be governed by energy scales ranging from a few GeV (e.g.\ for hadronic states) up to several hundred GeV and even TeV scales (e.g. for particle collisions that take place at the Large Hadron Collider). One needs to employ the renormalization group evolution equations to reliably relate the values of quark masses defined at such widely different energy scales. An interesting situation arises if the dynamical scale governing the quark mass dependence of an observable is much smaller than the quark mass itself. In this case, the common running ${\overline{\rm MS}}$ mass scheme, which obeys a renormalization group equation with logarithmic scale dependence, cannot be employed for high-precision applications, because it is only meaningful for scales of the order or larger than the mass. Rather, so-called low-scale short-distance masses must be used, which obey renormalization group equations with linear scale dependence. The numerical impact of the renormalization group evolution is particularly important for the top quark mass where, due to its large value, significant scale hierarchies can arise.

Here we present \REvolver{}, a \Cpp{} library with routines that provide renor\-mal\-iza\-tion-group resummed conversions between quark mass schemes defined at different renormalization scales, including scales much lower than the mass, where low-scale short-distance masses are employed, as well as above the mass value, where the ${\overline{\rm MS}}$ mass is used. The routines are based on the creation of so-called \Core{} objects, each of which representing a certain physical scenario for the heavy quark masses (charm, bottom and top quarks, as well as hypothetical heavier flavors), the number of massless quarks and the strong coupling $\alpha_s$. In a single session, an (in principle) arbitrary number of \Core{} objects can be created and managed. Each \Core{} object can then provide values for the quark masses in the most popular low-scale short-distance schemes as well as for the ${\overline{\rm MS}}$ mass and the strong coupling at any (real or complex-valued) renormalization scale and in any flavor number scheme, consistently accounting for flavor threshold effects and the resummation of large logarithms of all kinds. The basis of the quark mass evolution equations for scales below the respective mass is the renormalization group equation of the natural MSR mass (here simply called the MSR mass), which was provided in Refs.~\cite{Hoang:2017suc} together with a full treatment of flavor matching corrections when the evolution crosses the thresholds related to lighter massive quarks~\cite{Hoang:2017btd}. Furthermore, for each \Core{} object, options can be set to specify the perturbative precision in the flavor matching and the renormalization group evolution. Using all available theoretical input in the literature,
see Sec.~\ref{sec:theoryinput} for a detailed listing, it is possible to relate the ${\overline{\rm MS}}$, MSR and most other low-scale short-distance quark masses with a theoretical precision of $10$ to $20$\,MeV (neglecting any parametric uncertainties). Quark mass values in the pole scheme cannot be defined at the same level of precision due to the pole mass renormalon ambiguity which decreases the accuracy by an order of magnitude~\cite{Beneke:2016cbu,Hoang:2017btd}. \REvolver{} offers the possibility to set up \Core{} objects using pole masses as an input or to extract pole mass values from a \Core{} object, but it treats the pole mass as a derived quantity where the user has to specify the way in which the pole mass value is defined. Furthermore, \REvolver{} provides various options to account for the asymptotic higher order corrections of the pole mass and the pole mass renormalon ambiguity. All \REvolver{} routines are implemented entirely in \Cpp{} and can be accessed through \Mathematica{} and \Python{} interfaces.

There is an existing \Cpp{} library accompanied by a \Mathematica{} package called \texttt{CRunDec} and \texttt{RunDec}~\cite{Chetyrkin:2000yt,Herren:2017osy}, respectively, which already provide many functionalities included in the \REvolver{} library. We have cross checked in detail that any theoretical (perturbative) input implemented in \texttt{CRunDec} and \texttt{RunDec} agrees with the corresponding one employed for \REvolver{}. We have furthermore checked that the numerical output of the routines provided in \texttt{CRunDec}/\texttt{RunDec} is in agreement with the equivalent routines of \REvolver{}.
The \REvolver{} library, however, exceeds \texttt{CRunDec}/\texttt{RunDec}
\begin{itemize}
\item[(i)] by providing the \Core{} concept that allows to automatically create, extend and manage an arbitrary number of scenarios for strong coupling values, mass spectra and theory settings, and to extract quark masses and the QCD coupling in all flavor number schemes and at all scales,
\item[(ii)] by accounting for the renormalization group resummation of large logarithms and lighter massive quark flavor thresholds when dealing with quark masses at renormalization scales smaller than the quark mass, as well as for low-scale short-distance masses,
\item[(iii)] by giving access to machine-precision numerical routines that provide quasi-exact solutions of the renormalization group equations for the running masses and the strong coupling at complex scales, and
\item[(iv)] by providing routines to determine the asymptotic series for the pole mass to an arbitrary order that allow to extract different pole mass definitions and to quantify the pole mass renormalon ambiguity with various methods.
\end{itemize}
The emphasis of all \REvolver{} functionalities is to provide integrated and easy-to-use routines, while maintaining the possibility to deviate from default settings and specify all available options, useful for high-precision phenomenological and conceptual QCD studies aiming for uncertainties at the level of $10$ to $20$\,MeV for short-distance masses.

This article is organized as follows:
In Sec.~\ref{sec:terminology} essential terminology used for the description of the \REvolver{} package is explained.
In Sec.~\ref{sec:MSRevol} we succinctly review the MSR mass and the R-evolution concepts~\cite{Hoang:2017suc,Hoang:2008yj} which are essential for the resummation of the logarithms mentioned above in bullet point (ii).
Section~\ref{sec:setup} provides general information concerning the \REvolver{} installation and setting up the \Cpp{}, \Mathematica{} and \Python{} interfaces. The philosophy of the \Core{} concept is explained in Sec.~\ref{sec:structure}, and
Sec.~\ref{sec:methalgo} provides a structured introduction to all available \REvolver{} routines. In Sec.~\ref{sec:examples} a sizable number of pedagogical examples for applications of \REvolver{} routines are provided, partly using the routine's default settings, partly using alternative optional parameter setting, to demonstrate the versatility of \REvolver{} for important phenomenological applications in the literature. It is recommended that the user consults the examples shown in this section, which are also collected in a \Mathematica{} notebook, a Jupyter Notebook using the \Python{} interface, and a \Cpp{} source file provided with the \REvolver{} package.
In Sec.~\ref{sec:theoryinput} important references are provided which were used as the source for the higher order corrections implemented for the strong coupling and the mass schemes supported by \REvolver{}. Here, also a detailed citation recommendation for these higher order corrections is provided.
Section~\ref{sec:conclusions} contains a summary. Finally, some details concerning the algorithms used for \Core{} creation and the quasi-exact solution of renormalization group equations are given in Appendix~\ref{sec:algo}. In addition, a number of essential formulae for implementation-dependent quantities are provided which cannot be found in the literature in the form used in \REvolver{}.

\section{Terminology}
\label{sec:terminology}

This article employs a particular terminology when referring to renor\-mal\-iza\-tion-scale dependent mass schemes and the pole mass:
\begin{itemize}

\item \textit{Running quark mass in the $n_f$-flavor scheme $m^{(n_f)}_q(\mu)$}: Refers to the $\MSbar$ mass if the flavor
number $n_f$ includes this massive quark, and the MSR mass otherwise. For example, the running top quark mass
at the scale $\mu$ in the 6-flavor scheme $m_t^{(6)}(\mu)$ refers to the $\MSbar$ mass $\overline m_t^{(6)}(\mu)$, and the running top quark mass at the scale $\mu$ in the 4-flavor scheme $m_t^{(4)}(\mu)$ refers to the MSR mass $m_t^{{\rm MSR},(4)}(\mu)$.

\item \textit{Standard running mass $\overline m_q$}: Refers to the $\MSbar$ mass in the flavor number scheme where all lighter
quarks along with this quark are treated dynamically, evaluated at the scale of this mass. For
example, the standard running top mass is the $6$-flavor $\MSbar$ quark mass evaluated at the
scale of this top mass: $\overline m_t\equiv\overline m_t^{(6)}\!(\overline m_t^{(6)})$.
The mass dependence of all flavor-threshold corrections is expressed in terms of the standard running mass.

\item \textit{Asymptotic pole mass}: Refers to the pole mass value obtained from the running mass defined by summing the perturbative series to the order of the minimal correction.

\item \textit{Order-dependent pole mass}: Refers to the pole mass value obtained from the running mass by truncating the perturbative series at a specified order.

\end{itemize}

\section{The MSR mass and R-evolution}
\label{sec:MSRevol}

The natural MSR mass of a massive quark defined in Ref.~\cite{Hoang:2017suc} (and called just the MSR mass here) plays a central role in \texttt{REvolver} and is a renormalization scale and flavor-number-dependent low-scale short-distance mass. It is derived from the $\MSbar$ mass and treated as the natural extension of the $\MSbar$ mass for renormalization scales below the mass of the quark. This combination of the scale-dependent $\MSbar$ and MSR masses extends the well known concept of flavor-number dependent renormalization group evolution and flavor threshold matching for scales above the quark mass (where the $\MSbar$ mass scheme is appropriate) to lower scales. In contrast to usual logarithmic renormalization scale evolution (as known from the $\MSbar$ masses or the strong coupling), the MSR mass renormalization group evolution is linear. This is consequence of the linear dynamical scaling that arises when the off-shell massive quark quantum fluctuations are integrated out in the nonrelativistic limit.
Together with the flavor number dependent strong $\MSbar$ coupling, the $\MSbar$ and MSR masses form the basis of the core concept of REvolver and allow to also resum large logarithms involving low-scale short-distance mass schemes other than the MSR mass. This functionality is used by default in the REvolver routines (but can also be switched off by the user on demand). The MSR mass~\cite{Hoang:2017suc,Hoang:2017btd,Hoang:2008yj} has already been used in a number of applications, but
we still find it warranted to briefly review its main concepts in this section.
For simplicity, we consider the case in which all $n_\ell$ quarks lighter than $q$ are massless. The reader is referred to
Ref.~\cite{Hoang:2017btd} for the case with massive lighter quarks.

To define the MSR mass, one starts with the relation between the standard running mass and pole mass,
\begin{equation}\label{eq:msbarpoleseries}
m_q^{\rm pole} - \overline m_q = \overline m_q\,\sum_{n=1}^\infty\,a_n^{\overline{\rm MS}} (n_\ell,n_h)
\Biggl[\frac{\alpha_s^{(n_\ell+n_h)}(\overline m_q)}{4\pi}\Biggr]^{\!n} \,,
\end{equation}
where $n_f = n_\ell + n_h$ is the number of active flavors, with $n_\ell$ being the number of massless quarks and $n_h=1$ referring to the quark $q$.
Since the MSR mass is employed for renormalization scales below $m_q$, one integrates out the virtual heavy quark loops by
setting $n_h=0$. This allows to define a renormalization scale $R$ smaller than $m_q$ and the corrections to the pole mass having a linear dependence on $R$ to implement a consistent nonrelativistic scaling behavior. The MSR mass is thus defined by furthermore setting $\overline m_q \to R$:
\begin{align}\label{eq:MSRn}
m_q^{\rm pole}-m_q^{{\rm MSR},(n_\ell)}(R)= R\sum_{n=1}^\infty a_n^{\MSbar}(n_\ell,0)\!\Biggl[\frac{\alpha_s^{(n_\ell)}(R)}{4\pi}\Biggr]^{\!n}\,.
\end{align}
In contrast to the $\MSbar$ mass, which has only logarithmic dependence on the scale $\mu$, the MSR mass has
an additional linear dependence on $R$. The $\MSbar$ and MSR masses can be related perturbatively and unambiguously through Eqs.~\eqref{eq:msbarpoleseries} and
\eqref{eq:MSRn} because the pole mass in both equalities is identical.\footnote{This is consistent since the two series on the RHS of Eqs.~\eqref{eq:msbarpoleseries} and
\eqref{eq:MSRn} have the same leading linear and mass-independent renormalon ambiguity.}
The resulting perturbative series for the difference of two MSR masses at different renormalization scales $R_1$ and $R_2$ is renormalon free, as long as it is expressed in powers of the strong coupling at the same renormalization scale. As a result, for disparate values of $R_1$ and $R_2$ large logarithms will appear. These logarithms can be consistently summed up with the renormalization group equation
\begin{equation}\label{eq:Revol}
-\!\frac{\rm d}{{\rm d} R}m_q^{{\rm MSR},(n_\ell)}(R)=\gamma_R^M[\alpha_s^{(n_\ell)}(R)]=\sum_{n=0}^\infty \gamma_n^R(n_\ell)\! \Biggl[\frac{\alpha_s^{(n_\ell)}(R)}{4\pi}\Biggr]^{\!n+1},
\end{equation}
In contrast to the logarithmic renormalization group equations for the $\MSbar$ mass and the strong coupling, it shows a {\it linear power scaling} and has therefore been dubbed as the R-evolution equation.
The anomalous dimension coefficients $\gamma_n^R(n_\ell)$ can be calculated from the relation [\,see Eq.~\eqref{eq:alphaRGE} for the definition of the QCD $\beta$-function coefficients $\beta_i$\,]
\begin{align}
\gamma_n^R(n_\ell) = a^{\MSbar}_{n+1}(n_\ell,0)-2\sum_{j=0}^{n-1} (n-j)\,\beta_j\, a^{\MSbar}_{n-j}(n_\ell,0)\,.
\end{align}
The coefficients of the R-evolution equation have the following explicit form:
\begin{align}\label{eqn:gammaRn}
\gamma_0^{R}(n_\ell) & = {\textstyle \frac{16}{3}}\,,\\
\gamma_1^{R}(n_\ell) & = 96.1039 - 9.55076\, n_\ell\,,\nonumber\\
\gamma_2^{R}(n_\ell) & = 1595.75 - 269.953\, n_\ell - 2.65945\, n_\ell^2\,,\nonumber\\
\gamma_3^{R}(n_\ell) & = (12319.\pm417.) - (9103.\pm10.)\, n_\ell + 610.264\, n_\ell^2 - 6.515\, n_\ell^3\,.\nonumber
\end{align}
The uncertainties in the ${\cal O}(\alpha_s^4)$ coefficient arise from the numerical uncertainties in the relation between the $\MSbar$ and the pole masses at this order.
In App.~\ref{sec:MSRmass} we present an efficient algorithm to exactly integrate Eq.~\eqref{eq:Revol}.

Adopting appropriate values for $R$ and $n_\ell$ the MSR mass $m_q^{{\rm MSR},(n_\ell)}(R)$ can be related in a renormalon-free manner to any other low-scale short-distance mass without the appearance of large logarithms and can thus be used to also resum potentially large R-evolution logarithms in the relation of other low-scale short-distance mass schemes. In the presence of massive quarks with masses lighter than $m_q$, the MSR mass has an $n_\ell$-dependent renormalization group evolution and flavor threshold corrections in close analogy to the renormalization group evolution of the strong coupling and the $\MSbar$ mass. This allows for the resummation of large logarithms involving the masses of the lighter massive quarks. For details we refer to Ref.~\cite{Hoang:2017btd}.

\section{Setup}
\label{sec:setup}

There are three ways to access the functionalities of the \REvolver{} library:
\begin{itemize}
\item via the \Cpp{} library directly, which might be most suitable for extensive automated tasks and to interface with other libraries and codes,

\item via the Wolfram \Mathematica{}~\cite{mathematica} interface (using~\texttt{WSTP} / \MathLink{}), which is suitable for interactive tasks and for using in parallel with other \Mathematica{} features,

\item via the \Python{}~\cite{Rossum:1995:PRM:869369} interface (generated using \texttt{SWIG}~\cite{swig}) for usage in scripts and interactive execution in Jupyter notebooks~\cite{soton403913}.
\end{itemize}

\subsection{Installation}

Note that slightly more detailed instructions for installing the code, including Windows-specific commands, are given in the \texttt{README.md} file provided with the source code. Here we only describe the installation procedure for Linux and MacOS, and only for \REvolver{} itself (not for \texttt{CMake} and other auxiliaries).

For the compilation of \REvolver{} a \texttt{C++11} compatible compiler is needed. The recommended (and tested) choices are \texttt{gcc} on Linux, Apple \texttt{Clang} on MacOS and \texttt{MinGW} on Windows. It is expected that \REvolver{} compiles on other platforms and with different compilers as well, although this has not been tested and we do not provide any specific instructions.

We provide a \texttt{CMake} script with various options controlling which interfaces and demonstration codes are built. To use the script, at least version~3.1 of \texttt{CMake} is required.

If \REvolver{} is to be used via \Mathematica{}, Wolfram \Mathematica{} is required in a
version which supports \texttt{WSTP} or \MathLink{}.

For the \Python{} interface, at least version 3 of \Python{} has to be installed, including the development packages. We note that the \Python{} interface is currently supported only on Linux and MacOS.

After downloading the code to the local hard drive, open a command line interface, navigate to the directory \texttt{code/} and run the commands
\begin{verbatim}
  $ mkdir build
  $ cd build/
\end{verbatim}

In the next step, the \texttt{CMake} script will be executed. Depending on which interfaces are to be prepared, various flags can be set:
\begin{itemize}
\item \texttt{wolfr} determines if the \Mathematica{} interface is prepared (default: \texttt{OFF}),

\item \texttt{py} determines if the \Python{} interface is prepared (default: \texttt{OFF}),

\item \verb|cpp_demo| determines if the \Cpp{} demo executable is built (default: \texttt{ON}).
\end{itemize}
The static \REvolver{} \Cpp{} library is always compiled. Note that to compile the \Cpp{} demo executable the library \texttt{Quadpack++} \cite{quadpackpp} is used. However, no additional steps are required by the user since the library is provided with \REvolver{}. Note that the \verb|Quadpack++| library is only compiled if the flag \verb|cpp_demo| is set
to \verb|ON|.

To execute the \texttt{CMake} script with default flags and compile the code, one has to run the terminal commands
\begin{verbatim}
  $ cmake ..
  $ make install
\end{verbatim}
or in general
\begin{verbatim}
  $ cmake [(-D <flag>={ON|OFF})...] ..
  $ make install
\end{verbatim}
where \texttt{<flag>} is a placeholder for one of the flags listed above. For example, to prepare the \Mathematica{} interface, but not the compilation of the \Cpp{} demo code one would use
\begin{verbatim}
  $ cmake -D wolfr=ON -D cpp_demo=OFF ..
  $ make install
\end{verbatim}

The directory \texttt{code/build/} can be safely removed after the compiling is done.
The resulting libraries and executables can be found at the following locations:
\begin{itemize}
\item the static \Cpp{} library file:\\
\texttt{code/lib/libREvolver.a}

\item the \texttt{MathLink / WSTP} executable, ready to be loaded in a \Mathematica{} notebook:\\
\texttt{code/bin/REvolver}

\item the \Python{} module file and dynamic library, ready to be imported in a \Python{} script:\\
\texttt{code/pyREvolver/lib/pyREvolver.py}\\
\verb|code/pyREvolver/lib/_pyREvolver.so|

\item the \Cpp{} demo executable:\\
\texttt{code/bin/examples}
\end{itemize}

\subsection{General Usage}
\label{sec:usage}

\subsubsection*{\Cpp{} Interface}

To use the \REvolver{} \Cpp{} static library, the respective header file has to be included which is done via
\begin{verbatim}
  #include REvolver.h
\end{verbatim}
and the library has to be properly linked when compiling the code. After including the header file, the implemented classes and routines are available in the namespace \texttt{revo} and accessible with the scope resolution prefix \texttt{revo::} unless the instruction \texttt{using namespace revo;} has been invoked, such that the resolution prefix is not necessary.

For a demonstration, we refer to the source file \texttt{code/examples/examples.cpp} and the related executable \texttt{code/bin/examples} (if \verb|cpp_demo=ON| was set).

\subsubsection*{\Mathematica{} Interface}

To load the \texttt{WSTP} executable in a \Mathematica{} notebook, execute
\begin{verbatim}
  Install["<path to executable>/REvolver"]
\end{verbatim}
with \verb|<path to executable>| referring to the directory path where the executable is located. For future convenience it might be useful to execute
\begin{verbatim}
  CopyFile["<path to executable>/REvolver",
           $UserBaseDirectory <> "/Applications/REvolver"]
\end{verbatim}
which copies the executable to the user base directory of \Mathematica{}, making it possible to load the executable with the command
\begin{verbatim}
  Install["REvolver"]
\end{verbatim}
in the future.

For a demonstration of how to load \REvolver{} in a \Mathematica{} notebook and a general overview of the available functions, see the demonstration notebook \texttt{code/examples/examples.nb} provided with the package.

\subsubsection*{\Python{} Interface}

The module can be loaded in a \Python{} script or Jupyter notebook with the usual syntax
\begin{verbatim}
  import pyREvolver
\end{verbatim}
assuming that \texttt{pyREvolver.py} and the shared library file (\texttt{*.so}) are located in the same folder as the script or notebook, or have been added to the module search path with the following command
\begin{verbatim}
  import sys
  sys.path.append('<path to pyREvolver.py>')
\end{verbatim}

For a demonstration of how to load and use the module, see the Jupyter notebook \texttt{code/examples/examples.ipynb} provided with the package.

\section{\Core{} Structure}
\label{sec:structure}

As described in the introduction, all functionalities of the library are centered around instances of the class \texttt{revo::}\Core{} (simply called ``\Core{} objects'' or ``\Core{}s'' in the following) each representing a certain physical scenario for the quark mass spectrum and the strong coupling and from which numerical values for quark masses and the strong coupling in specified schemes and at specified scales can be extracted. In principle, the number of \Core{} objects defined at the same time is only limited by the available memory, regardless of the interface used.

\begin{figure}
\center
\includegraphics[width=0.53\textwidth]{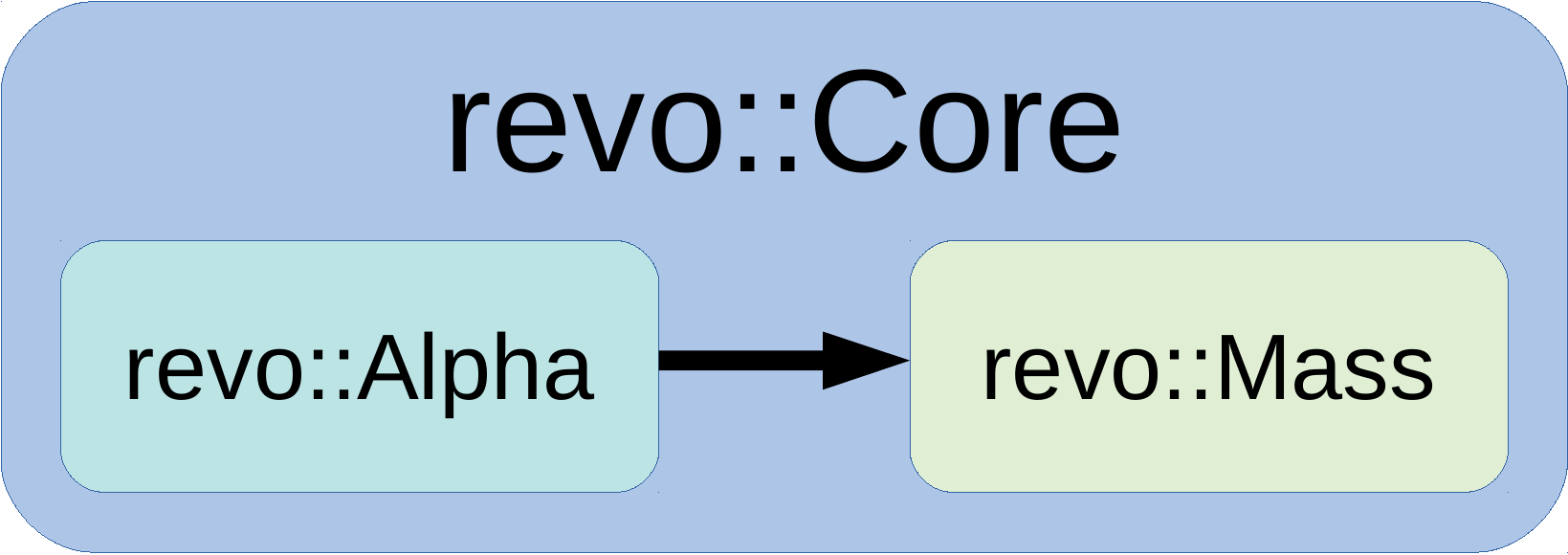}
\caption{\label{fig:corestructure}Schematic structure of the class \texttt{revo::}\Core{} in \Cpp{}: the class contains an instance of the class \texttt{revo::Alpha} as a member providing functionalities related to the strong coupling such as running and matching, and an instance of the class \texttt{revo::Mass} providing functionalities related to the quark masses such as running, matching and conversion. \texttt{revo::Mass} uses \texttt{revo::Alpha} for coupling evolution, and \texttt{revo} is the namespace where all relevant \REvolver{} classes are defined.}
\end{figure}

The schematic structure of the class \texttt{revo::Core}
in \Cpp{} is depicted in Fig.~\ref{fig:corestructure}: the class has objects of the classes \texttt{revo::Alpha} and \texttt{revo::Mass} as members. The class \texttt{revo::Alpha} has various member functions related to the strong coupling like running, matching and for obtaining the QCD scale $\Lambda_\mathrm{QCD}$. The class \texttt{revo::\allowbreak Mass} has member functions related to the evolution and matching of the running masses as well as the extraction of quark mass values in specified schemes. It uses an instance of the class \texttt{revo::\allowbreak Alpha} to obtain the necessary coupling values. The respective member objects of the \texttt{revo::}\Core{} class can be accessed through the member functions \texttt{revo::\allowbreak Core::\allowbreak alpha()} and
\texttt{revo::\allowbreak Core::\allowbreak masses()}, respectively. The class \texttt{revo::}\Core{} itself represents the frame to access these member functions and provides additional functionalities related to setting up a physical coupling and quark mass spectrum scenario, and extending an existing scenario by adding additional heavier massive quarks.

Although in principle possible, helper classes such as \texttt{revo::\allowbreak Mass} and \texttt{revo::\allowbreak Alpha} are not meant to be used outside \Core{} objects. All available functionalities can (and should) be accessed through \Core{} objects.

When using the \Mathematica{} interface, the specific structure of the classes are not relevant since the wrapper hides most details to fit into the Wolfram language syntax. To preserve the possibility to have multiple \Core{}s defined at the same time in \Mathematica{}, a unique name has to be specified for each \Core{} instance, which is referred to when extracting mass and coupling values or when extending scenarios.

For concrete usage and examples we refer to Secs.~\ref{sec:methalgo} and \ref{sec:examples}.

\section{Implemented Functions}
\label{sec:methalgo}

In the following descriptions and examples we will assume that the namespace \texttt{revo} has been introduced in the \Cpp{} code with the instruction
\begin{verbatim}
  using namespace revo;
\end{verbatim}
such that the scope resolution prefix \texttt{revo::} can be omitted,
and that in \Python{} the module was loaded with
\begin{verbatim}
  from pyREvolver import *
\end{verbatim}
to keep code snippets uncluttered.

The syntax and interface structure in \Python{} is the same as in \Cpp{}, with a few exceptions:
\begin{itemize}
\item If a constant of an \texttt{enum class} type has to be provided as an input, the scope resolution operator \texttt{::} has to be exchanged with \verb|_|, e.g.\ the enumerator \texttt{MSbar} of type \texttt{MScheme} has to be provided using \verb|MScheme_MSbar|
instead of \texttt{MScheme::MSbar}.

\item The \Cpp{} function \texttt{Mass::mPole} allows for two optional pointer-type inputs to provide the possibility of accessing several output values (see Sec.~\ref{sec:pole}). In \Python{}, instead, within the class \texttt{Mass} the additional member function \texttt{mPoleDetailed} is provided which returns a tuple of values. In \Mathematica{} the same functionality is provided by the function \texttt{MassPoleDetailed}.

\item \Cpp{} specific syntax cannot be used, e.g.\ initializing an \texttt{std::vector} with an initializer list.
\end{itemize}

We will treat the \Cpp{} and \Mathematica{} interfaces on an equal footing, always stating the \Cpp{} function prototypes and definitions first with the \Mathematica{} ones following. We will then briefly describe the inputs and outputs, and in most cases give short examples. If not stated otherwise, the related \Python{} syntax is the same as in \Cpp{}. Also, to focus on the essential functionalities first, we will present all commands without optional parameters at the beginning and describe additional options in a second step. Note that the given function prototypes do not always correspond exactly to the ones present in the source codes to make the descriptions more transparent, e.g.\ for template functions in \Cpp{} or type restrictions in \Mathematica{}.

For a more detailed and technical documentation of the full functionality and interface structure of the \Cpp{} library, please consider reading the online doxygen documentation (see \url{https://revolver-hep.gitlab.io/REvolver}).

A detailed documentation of the functions accessible via the \Mathematica{} interface is available through the \Mathematica{} internal documentation.

\subsection{Constructing a \Core{} and Accessing Scenario Parameters}

In the following we describe how to construct \Core{}s in the various interfaces and how to read out their scenario parameters. The scenario parameters of a \Core{} uniquely reflect its physical scenario. They include the total flavor number, the flavor number scheme, value as well as scale of the strong coupling specified at \Core{} creation, the running masses at reference scales, the flavor matching scales, and the parameters that specify the precision of the theoretical input and scheme choices. The latter include the perturbative orders of renormalization group equations and threshold matching relations, the lambda parameters setting variations in renormalization group equations, the variation related to the uncertainty of the perturbative \mbox{4-loop} pole-$\MSbar$ mass coefficient, and the coefficients of the QCD $\beta$-function. All scenario parameters, except for coupling and quark mass values, acquire default values if not specified at \Core{} creation.

\subsubsection{\Cpp{} / \Python{} only: \texttt{RunPar} and \texttt{RunParV}}
\label{sec:runpar}

In the \Cpp{} and \Python{} interfaces, the \texttt{RunPar} \texttt{struct}
\begin{verbatim}
  struct RunPar {
    int nf;
    double value;
    double scale;
  };
\end{verbatim}
is used to collect the parameters of the running coupling and masses. \texttt{RunPar} \texttt{struct}s contain the active number of flavors \texttt{nf} specifying the flavor number scheme, the parameter (coupling or mass) value \texttt{value} and the respective renormalization scale \texttt{scale}. All numbers referring to quantities with dimensions of energy handled by REvolver (e.g. masses, renormalization scales or $\Lambda_{\rm QCD}$) are understood in GeV units.

The related type \texttt{RunParV} is an alias for \texttt{std::vector<RunPar>}, i.e.\ a collection of \texttt{RunPar}s.

In the following \Cpp{} example we define the \texttt{RunPar} \texttt{struct}s \texttt{alphaPar} and \texttt{alphaPar2} specifying flavor number schemes, values and renormalization scales for the strong coupling, and the \texttt{RunParV}s \texttt{mPar} and \texttt{mPar2} containing three \texttt{RunPar}s each, specifying values for running masses of charm, bottom and top quarks. \texttt{alphaPar} sets a realistic value for the strong coupling \mbox{$\alpha_s^{(5)}(m_Z)=0.1181$}, while \texttt{alphaPar2} contains parameters to specify the strong coupling $\alpha_s^{(4)}(4.2\,\mathrm{GeV})$. \texttt{mPar} defines standard running masses with realistic values of charm, bottom and top quarks, namely $\overline m_c = 1.3$\,GeV, \mbox{$\overline m_b \!=\! 4.2$\,GeV\!}, and $\overline m_t \!=\! 163$\,GeV\!, while \texttt{mPar2} defines values for different flavor number schemes and scales, specifically $m^{(6)}_c\!(163.0\,\mathrm{GeV})=\overline m_c^{(6)}\!(163.0\,\mathrm{GeV})$, $m_b^{(4)}(4.2\,\mathrm{GeV})=m_b^{{\rm MSR},(4)}(4.2\,\mathrm{GeV})$, and $m_t^{(5)}(4.2\,\mathrm{GeV})=m_t^{{\rm MSR},(5)}(4.2\,\mathrm{GeV})$. In the \Python{} example we only define \texttt{alphaPar} and \texttt{mPar} for brevity.

\subsubsection*{\Cpp{} example}
\begin{verbatim}
  RunPar alphaPar = {5, 0.1181, 91.187};
  RunPar alphaPar2 = {4, 0.22491680889566054, 4.2};

  RunParV mPar;
  mPar.push_back({4, 1.3, 1.3});
  mPar.push_back({5, 4.2, 4.2});
  mPar.push_back({6, 163.0, 163.0});

  RunParV mPar2;
  mPar2.push_back({6, 0.6173718176865822, 163.0});
  mPar2.push_back({4, 4.20502733598667, 4.2});
  mPar2.push_back({5, 172.37293079716443, 4.2});
\end{verbatim}

\subsubsection*{\Python{} example}
\begin{verbatim}
  alphaPar = RunPar(5, 0.1181, 91.187)

  mPar = RunParV(3)
  mPar[0] = RunPar(4, 1.3, 1.3)
  mPar[1] = RunPar(5, 4.2, 4.2)
  mPar[2] = RunPar(6, 163.0, 163.0)

  mPar2 = RunParV(3)
  mPar2[0] = RunPar(6, 0.6173718176865822, 163.0)
  mPar2[1] = RunPar(4, 4.20502733598667, 4.2)
  mPar2[2] = RunPar(5, 172.37293079716443, 4.2)
\end{verbatim}

\subsubsection{Constructing \Core{} objects with masses}
\label{sec:core}
The prototypes for the functions constructing \Core{}s in \Cpp{} and \Mathematica{}, respectively, are
\begin{verbatim}
  Core::Core(int nTot, const RunPar& alphaPar,
             const RunParV& mPar);
\end{verbatim}

\begin{verbatim}
  CoreCreate[CoreName_String, nTot_Integer, alphaPar_List,
             mPar_List]
\end{verbatim}
with the mandatory input \texttt{nTot}, specifying the total number of quark flavors in the scenario, as well as the input parameters for the strong coupling and the quark masses. In \Cpp{}, the coupling and mass parameters are given by \texttt{RunPar} \texttt{struct}s and \texttt{std::vector}s \texttt{RunParV}, respectively, which are described in Sec.~\ref{sec:runpar}, while in \Mathematica{}, the parameter collections are given by lists and lists of lists, respectively. The argument \texttt{CoreName} in \Mathematica{} specifies the user-defined unique name of the created \Core{} instance. The given masses must be sorted in increasing order with respect to their standard running mass values starting with the lightest. The number of massless quarks in a \Core{} is equal to \texttt{nTot} minus the number of elements in \texttt{mPar}. To construct a \Core{} without massive quarks, see Sec.~\ref{sec:coreml}.

\subsubsection*{\Cpp{} example}
The instructions
\begin{verbatim}
  Core core1(6, alphaPar, mPar);
  Core core2(6, alphaPar2, mPar2);
\end{verbatim}
construct two \Core{} objects named \texttt{core1} and \texttt{core2}, respectively, with a total flavor number of $6$, and the parameters determining the strong coupling and masses contained in \texttt{alphaPar}, \texttt{alphaPar2}, \texttt{mPar} and \texttt{mPar2} as defined in the example of Sec.~\ref{sec:runpar}. These are the minimal set of parameters that have to be specified to create \Core{} objects.

\subsubsection*{\Mathematica{} example}

To construct the same \Core{}s in \Mathematica{} one can use
\begin{verbatim}
  alphaPar = {5, amZdef, mZdef};
  mPar = {{4, 1.3, 1.3}, {5, 4.2, 4.2}, {6, 163.0, 163.0}};
  CoreCreate["core1", 6, alphaPar, mPar]

  alphaPar2 = {4, 0.22491680889566054, 4.2};
  mPar2 = {{6, 0.6173718176865822, 163.0},
           {4, 4.20502733598667, 4.2},
           {5, 172.37293079716443, 4.2}};
  CoreCreate["core2", 6, alphaPar2, mPar2]
\end{verbatim}
using the predefined parameters \texttt{amZdef = 0.1181} for the strong coupling and \texttt{mZdef = 91.187} for the Z-boson mass.

\subsubsection*{Optional parameters}

The \Core{} constructor allows to set a number of optional parameters to control the flavor matching scales, the perturbative order of matching relations and renormalization group equations, to perform scale variation of the renormalization group equations, and to vary the 4-loop pole-$\MSbar$ mass coefficient within its error band. The values of these optional parameters are a defining property of the physical scenario represented by a \Core{} object and respected by all functionalities related to numerical values of the strong coupling and the running masses.

The full \Cpp{} constructor prototype is
\begin{verbatim}
  Core::Core(int nTot, const RunPar& alphaPar,
             const RunParV& massPar,
             const doubleV& fMatch = doubleV(),
             int runAlpha = kMaxRunAlpha,
             double lambdaAlpha = 1.0,
             int orderAlpha = kMaxOrderAlpha,
             int runMSbar = kMaxRunMSbar,
             double lambdaMSbar = 1.0,
             int orderMSbar = kMaxOrderMSbar,
             int runMSR = kMaxRunMSR,
             double lambdaMSR = 1.0,
             int orderMSR = kMaxOrderMSR,
             double msBarDeltaError = 0.0);
\end{verbatim}
with \texttt{doubleV} being an alias for \texttt{std::vector<double>} set by \texttt{REvolver}. If one of the optional parameters shown in the constructor above is explicitly specified, all parameters appearing prior in the argument list must be specified as well. The values \texttt{kMaxRunAlpha}, \texttt{kMaxOrderAlpha}, \texttt{kMaxRunMSR}, \texttt{kMaxOrderMSR}, \texttt{kMaxRunMSbar} and \texttt{kMaxOrderMSbar} are predefined constants representing the respective defaults. In the \Mathematica{} interface, the optional parameters of the same name can be set individually via the options parameter syntax, i.e.\ by adding \texttt{opt->val} after the last regular function input, as shown in the examples below.

The meaning of the optional parameters is as follows:
\begin{itemize}
\item \texttt{fMatch}: a vector / list containing elements \{\texttt{f1, f2, ...}\}, where \texttt{fn} specifies that the flavor matching scale $\mu_n$
for the n-th lightest massive quark threshold is \texttt{fn} times the standard running mass: $\mu_n= \texttt{fn}\times \overline{m}_n$.
Default: all \texttt{fn} are set to $1.0$. (Note that the mass dependence of the flavor threshold corrections is expressed in terms of the standard running masses as well. The specification to use a different mass scheme to parameterize the flavor threshold corrections is not supported.)

\item \texttt{runAlpha}: the loop order used for the running of the strong coupling. Default: highest available order which is $5$.

\item \texttt{lambdaAlpha}: a parameter probing the renormalization scale dependence of the QCD $\beta$-function. With respect to the perturbative series of the $\beta$-function truncated at the order set by \texttt{runAlpha} (used for the value $1.0$), a (\texttt{runAlpha}$+1$) order term is estimated from renormalization scale variation (when a value different from $1.0$ is specified): the estimate is obtained by expanding the original perturbative series $\beta[\alpha_s(\mu)]$ (truncated at order \texttt{runAlpha}) in terms of \mbox{$\alpha_{s\!}(\mathtt{lambdaAlpha}\!\times\!\!\mu)$}, truncating at order \texttt{runAlpha}. The result is expanded in $\alpha_s(\mu)$ truncating again at order (\texttt{runAlpha}$+1$). A variation around $1.0$ of ${\cal O}(\pm 10\%)$ leads to an adequate uncertainty estimation for the known lower orders of the $\beta$-function, so that variations exceeding this range should be avoided. Default: $1.0$.

\item \texttt{orderAlpha}: the loop order used for the strong coupling flavor threshold matching relations. The renormalization scale dependence of these matching relations is precise to loop order \texttt{orderAlpha} and independent of the value specified for \texttt{runAlpha}. Default: highest available order which is $4$.

\item \texttt{runMSbar}: loop order used for the $\MSbar$ mass running. Default: highest available order which is $5$.

\item \texttt{lambdaMSbar}: a parameter probing the renormalization scale dependence of the anomalous dimension $\gamma$ of the $\MSbar$ mass in analogy to the parameter \texttt{lambdaAlpha}. The variation is performed by expanding the original series for $\gamma[\alpha_s(\mu)]$ (truncated at order \texttt{runMSbar}) in terms of $\alpha_s(\mathtt{lambdaMSbar}\times\mu)$, truncating at order \texttt{runMSbar}. A variation around $1.0$ of ${\cal O}(\pm 10\%)$ leads to an adequate uncertainty estimation for the known lower orders of $\gamma$, so that variations exceeding this range should be avoided. Default: $1.0$.

\item \texttt{orderMSbar}: loop order used for the flavor threshold matching relations of the $\MSbar$ masses. The renormalization scale dependence of these matching relations is precise to loop order \texttt{orderMSbar} and independent of the values specified for \texttt{runAlpha} and \texttt{orderAlpha}. Default:~highest available order which is $4$.

\item \texttt{runMSR}: the loop order used for the MSR mass running. Default: highest available order which is $4$.

\item \texttt{lambdaMSR}: a parameter probing the renormalization scale dependence of the anomalous dimension $\gamma^R$ of the MSR mass in analogy to the parameters \texttt{lambdaAlpha} and \texttt{lambdaMSbar}. The variation is performed by expanding the original $\gamma^R[\alpha_s(R)]$ series (truncated at order \texttt{runMSR}) in terms of $\alpha_s(\mathtt{lambdaMSR}\times R)$, truncating at order \texttt{runMSR}. A variation around $1.0$ by factors of around $0.5$ and $2$ leads to an adequate uncertainty estimation for the known lower orders of the $\gamma^R$, so that variations exceeding this range should be avoided. Default: $1.0$.

\item \texttt{orderMSR}: the loop order used for the flavor threshold matching relations of the MSR masses associated to the massive quark itself and all lighter massive quarks. The renormalization scale dependence of these matching relations is precise to loop order \texttt{orderMSR} and independent of the value specified for \texttt{runMSR}, \texttt{runAlpha} and \texttt{orderAlpha}. Default:~highest available order which is $4$.

\item \texttt{msBarDeltaError}: controls the error of the 4-loop coefficient in the pole-$\MSbar$ mass relation. Should be varied between $-1$ and $1$ to scan the standard deviation as quoted in Ref.~\cite{Marquard:2016dcn}. Default: $0.0$.

\item \texttt{precisionGoal}: the parameter setting the relative precision of all convergent infinite sums and iterative algorithms. The input value is clipped to the range $[10^{-6}, 10^{-15}]$. Default: $10^{-15}$ which we refer to as {\it machine precision}. The default should be adequate for most applications, but a lower precision goal may be specified for improving speed.
\end{itemize}

The quark mass dependence of all flavor matching relations (for the strong coupling and the running masses) is expressed in terms of the corresponding standard running masses $\overline m_q$. Changing this to an arbitrary mass scheme is not supported in \REvolver{}. This concerns flavor threshold matching as well as perturbative reexpansions of the strong coupling in other flavor number schemes. The resulting numerical differences are, however, tiny and smaller than the corresponding perturbative uncertainties.

\subsubsection*{\Cpp{} example}

The instruction
\begin{verbatim}
  Core core3(6, alphaPar, mPar, {2.0, 1.0, 1.0});
\end{verbatim}
constructs a \Core{} object named \texttt{core3} with a total flavor number of $6$ and the parameters specifying the strong coupling and masses contained in \texttt{alphaPar} and \texttt{mPar}, respectively, as defined in the example of Sec.~\ref{sec:runpar}. The matching scale of the flavor threshold related to the lightest massive particle is \mbox{$2\times\overline m_c=2.0\times 1.3$\,GeV}.

\subsubsection*{\Mathematica{} example}
The command
\begin{verbatim}
  CoreCreate["core3", 6, alphaPar, mPar,
             fMatch->{2.0, 1.0, 1.0}]
\end{verbatim}
has the same effect as the analogous \Cpp{} example, using the lists \texttt{alphaPar} and \texttt{mPar} defined in the previous example of this section.

\subsubsection{Constructing \Core{} objects without massive quarks}
\label{sec:coreml}

The functions with the prototypes
\begin{verbatim}
  Core::Core(const RunPar& alphaPar);
\end{verbatim}

\begin{verbatim}
  CoreCreate[CoreName_String, alphaPar_List]
\end{verbatim}
are used to construct a \Core{} object with massless quarks only and without specifying any optional parameters. The parameters are analogous to the massive case described in Sec.~\ref{sec:core}.

\subsubsection*{Optional parameters}

The full \Cpp{} constructor prototype for a \Core{} with only massless quarks is
\begin{verbatim}
  Core::Core(const RunPar& alphaPar,
             int runAlpha = kMaxRunAlpha,
             double lambdaAlpha = 1.0,
             const doubleV& beta = doubleV());
\end{verbatim}
where the optional variables \texttt{runAlpha} and \texttt{lambdaAlpha} are analogous to the massive case described in Sec.~\ref{sec:core} and can be set in \Mathematica{} using option parameters. With the optional input \texttt{beta} (which is a constant reference to a \texttt{std::vector<double>} of arbitrary length) one can specify an arbitrary number of custom $\beta$-function coefficients. Their default values are the common $\MSbar$ QCD $\beta$-function coefficients up to $5$ loops with all higher order coefficients set to zero. These are defined based on the $\beta$-function form
\begin{equation}\label{eq:alphaRGE}
\mu\frac{{\rm d} \alpha_s(\mu)}{{\rm d} \mu}=
\frac{{\rm d} \alpha_s(\mu)}{{\rm d} \ln(\mu)}=
\beta_{\rm QCD}(\alpha_s(\mu)) =
-2\alpha_s(\mu)\!\sum_{n=0}^\infty\beta_n
\biggl[\frac{\alpha_s(\mu)}{4\pi}\biggr]^{\!n+1}\,,
\end{equation}
where the elements of the \texttt{C++} container \verb|beta| correspond to the ordered list of coefficients $\beta_0,\cdots, \beta_n$. Note that adding masses to \Core{} objects with custom $\beta$-function coefficients is not supported and that, depending on the choice of \texttt{runAlpha} not all coefficients specified by the user may be used.

In \Mathematica{} the functionality of custom QCD $\beta$-function coefficients can be used with
\begin{verbatim}
  CoreCreate[CoreName_String, alphaPar_List, beta_List]
\end{verbatim}
with \verb|beta| being the list of $\beta$-function coefficients and the additional option parameters already explained before.

\subsubsection{\Mathematica{} only: listing and deleting \Core{}s}

\begin{verbatim}
  CoreList[]
  CoreDelete[CoreName_String]
  CoreDelete[CoreNames_List]
  CoreDeleteAll[]
\end{verbatim}

These commands list the names of the \Core{}s currently defined, and delete specific or all \Core{}s, respectively. The argument of \texttt{CoreDelete} is
a string referring to a \Core{} name or a list containing several \Core{} names.

\subsubsection*{Example}

\begin{verbatim}
  In[]:= CoreList[]
  Out[]= {core1, core2, core3}

  In[]:= CoreDelete["core3"]
         CoreList[]
  Out[]= {core1, core2}
\end{verbatim}
where the \Core{} named \texttt{core3} has been deleted from memory. We explicitly show \texttt{In[]} and \texttt{Out[]} to separate in- from out-put and assumed that definitions from previous examples are still valid.

\subsubsection{Accessing \Core{} parameters}
\label{sec:accesscoreparameters}

A \Core{} represents a certain physical scenario for the strong coupling and the quark mass spectrum that also depends on the theoretical approximations and conventions implemented (with optional/default parameters) specified at the time the \Core{} was created. The scenario parameters of a \Core{} (coupling, quark masses, theoretical approximations and conventions) unambiguously specify a given scenario and can be accessed by dedicated routines. Note that the scenario parameters of a \Core{} for the strong coupling depend on the way how the strong coupling was specified when the \Core{} has been created. Therefore it is possible to create two physically equivalent \Core{}s with differing scenario parameters for the strong coupling.

While in \Cpp{} the scenario parameters of \Core{} objects are returned from separate functions, some \Mathematica{} commands print collections of them. The respective function prototypes in \Cpp{} are
\begin{verbatim}
  int            Core::nTot() const;
  const RunPar&  Alpha::defParams() const;
  const doubleV& Core::standardMasses() const;
  const doubleV& Core::fMatch() const;
  int            Core::getOrder(OrderPar para) const;
  double         Core::getLambda(LambdaPar para) const;
  double         Core::msBarDeltaError() const;
  const doubleV& Core::betaCoefs(int nf) const;
\end{verbatim}
returning the total number of flavors, the defining \texttt{RunPar} related to the coupling, an \texttt{std::vector} of the standard running masses in increasing order, an \texttt{std::vector} with the \texttt{fn} factors specifying the flavor matching scales, the perturbative orders used for coupling and mass evolutions, the lambda scaling parameters set for coupling and mass evolution, the variation parameter of the 4-loop coefficient in the pole-$\MSbar$ evolution, and the $\beta$-function coefficients in the \texttt{nf}-flavor number scheme.

The function inputs of the types defined as
\begin{verbatim}
  enum class OrderPar {
    runAlpha,
    orderAlpha,
    runMSbar,
    runMSR,
    orderMSbar,
    orderMSR
  };
\end{verbatim}
and
\begin{verbatim}
  enum class LambdaPar { lambdaAlpha, lambdaMSbar, lambdaMSR };
\end{verbatim}
i.e.\ the enumerators of type \texttt{OrderPar} and \texttt{LambdaPar}, respectively, govern to which evolution (coupling, $\MSbar$, MSR) or matching procedure the output of the functions \texttt{getOrder} and \texttt{getLambda} refers to.

In \Mathematica{}, the parameters discussed above can be extracted using the functions
\begin{verbatim}
  CoreParams[CoreName_String]
  CoreParamsDetail[CoreName_String]
  BetaCoefs[CoreName_String]
\end{verbatim}

\texttt{CoreParams} returns from the specified \Core{} a list containing the total number of flavors, the flavor number scheme, value and renormalization scale of the strong coupling specified at \Core{} creation, and the standard running masses in increasing order. \texttt{CoreParamsDetail} returns the coupling and mass values at all flavor matching scales (in the flavor schemes above as well as below the corresponding threshold), the flavor number scheme, value as well as scale of the strong coupling specified at \Core{} creation, and all optional parameters set by \texttt{CoreCreate} and described in the \Cpp{} description above. \texttt{BetaCoefs} returns the $\beta$-function coefficients in all relevant flavor number schemes with the normalization as given in Eq.~\eqref{eq:alphaRGE}.

In addition to just printing the parameters, \texttt{CoreParamsDetail} allows for an optional parameter to which the parameters are saved as a nested list. The corresponding function prototype is
\begin{verbatim}
  CoreParamsDetail[CoreName_String, output_Symbol]
\end{verbatim}
where \texttt{output} is the symbol in which the list is stored.

\subsubsection*{\Cpp{} example}

The instructions
\begin{verbatim}
  core1.nTot();
  core1.getOrder(OrderPar::runAlpha);
\end{verbatim}
return the total number of flavors and the perturbative order used in the running of the strong coupling, respectively, for the \Core{} named \texttt{core1}. They correspond to the values \texttt{(int)6} and \texttt{(int)5}, respectively, given the definition of \texttt{core1} from the \Cpp{} example of Sec.~\ref{sec:core}.

\subsubsection*{\Mathematica{} example}

In the following we show how the output of the functions \texttt{CoreParams} and \texttt{BetaCoefs} looks like for the \Core{} named \texttt{core1} given in the \Mathematica{} example of Sec.~\ref{sec:core}:

\begin{verbatim}
  In[]:= CoreParams["core1"]
  Out[]= {6, {5, 0.1181, 91.187}, 1.3, 4.2, 163.}
\end{verbatim}
showing the list containing the total number of flavors, the flavor number scheme, value and renormalization scale of the strong coupling given at \Core{} creation, and the standard running masses in increasing order; and
\begin{verbatim}
  In[]:= BetaCoefs["core1"]
  Out[]= nl = 3: {9.,64.,643.833,12090.4,130378.}
         nl = 4: {8.33333,51.3333,406.352,8035.19,58310.6}
         nl = 5: {7.66667,38.6667,180.907,4826.16,15470.6}
         nl = 6: {7.,26.,-32.5,2472.28,271.428}
\end{verbatim}
showing the $\beta$-function coefficients in the relevant flavor number schemes.

The output of the function \texttt{CoreParamsDetail} is more extensive than that of \texttt{CoreParams}, giving
\scriptsize
\begin{verbatim}
  In[]:= CoreParamsDetail["core1"]
  Out[]=
\end{verbatim}\vspace*{-0.2cm}
\begin{equation*}
\hspace{25pt}
\begin{pmatrix}
\verb|[core1, 6]| & \texttt{m1-threshold} & \texttt{m2-threshold} & \texttt{m3-threshold}\\
\texttt{aS-values} & \verb|{0.385234, 0.383676}| & \verb|{0.224917, 0.224684}| & \verb|{0.108577, 0.108555}|\\
\texttt{m1-values} & \verb|{1.30636, 1.3}| & \verb|{0.947058, 0.945337}| & \verb|{0.617575, 0.617372}|\\
\texttt{m2-values} & \verb|{4.64086, 4.62912}| & \verb|{4.20503, 4.2}| & \verb|{2.7438, 2.7429}|\\
\texttt{m3-values} & \verb|{172.819, 172.807}| & \verb|{172.383, 172.373}| & \verb|{163.032, 163.}|\\
\end{pmatrix}
\end{equation*}
\begin{verbatim}
         aS input: {5,0.1181,91.187}
         Matching f-factors: {1.,1.,1.}
         runAlpha: 5
         lambdaAlpha: 1.
         orderAlpha: 4
         runMSbar: 5
         lambdaMSbar: 1.
         orderMSbar: 4
         runMSR: 4
         lambdaMSR: 1.
         orderMSR: 4
         msBarDeltaError: 0.
         precisionGoal: 1.*10^-15
\end{verbatim}
\normalsize
for the \Core{} named \texttt{core1}. The set of printed values represents the full physical content of a \Core{}. Creating a new \Core{} from this output in an arbitrary way leads to a physically equivalent core. So \Core{}s are created in a self-consistent way. This is made possible because \REvolver's algorithm to solve the renormalization group evolution for the strong coupling provides (machine precision) exact solutions and because its algorithm for the flavor matching is self-consistent, see the routines described in Sec.~\ref{sec:extractstandard}. Therefore, all massive quarks consistently affect each others flavor number scheme dependent running values depending on the parameters specified at \Core{} creation.

For a demonstration, consider the output of \texttt{CoreParamsDetail} of the \Core{} named \texttt{core2}
\scriptsize
\begin{verbatim}
  In[]:= CoreParamsDetail["core2"]
  Out[]=
\end{verbatim}\vspace*{-0.2cm}
\begin{equation*}
\hspace{25pt}
\begin{pmatrix}
\verb|[core2, 6]| & \texttt{m1-threshold} & \texttt{m2-threshold} & \texttt{m3-threshold}\\
\texttt{aS-values} & \verb|{0.385234, 0.383676}| & \verb|{0.224917, 0.224684}| & \verb|{0.108577, 0.108555}|\\
\texttt{m1-values} & \verb|{1.30636, 1.3}| & \verb|{0.947058, 0.945337}| & \verb|{0.617575, 0.617372}|\\
\texttt{m2-values} & \verb|{4.64086, 4.62912}| & \verb|{4.20503, 4.2}| & \verb|{2.7438, 2.7429}|\\
\texttt{m3-values} & \verb|{172.819, 172.807}| & \verb|{172.383, 172.373}| & \verb|{163.032, 163.}|\\
\end{pmatrix}
\end{equation*}
\begin{verbatim}
         aS input: {4,0.224917,4.2}
         Matching f-factors: {1.,1.,1.}
         runAlpha: 5
         lambdaAlpha: 1.
         orderAlpha: 4
         runMSbar: 5
         lambdaMSbar: 1.
         orderMSbar: 4
         runMSR: 4
         lambdaMSR: 1.
         orderMSR: 4
         msBarDeltaError: 0.
         precisionGoal: 1.*10^-15
\end{verbatim}
\normalsize
which is (apart from the \Core{} name and the strong coupling specifications at \Core{} creation) exactly the same. In fact, the numbers used to create \texttt{core2} in the examples of Sec.~\ref{sec:core} have been taken from the output of \texttt{CoreParamDetails["core1"]}.

\subsection{Extraction of Masses and Couplings}
\label{sec:extract}

The following section presents the routines to extract the values for the strong coupling and the running quark masses in any flavor number scheme at any scale, as well as values in other quark mass schemes from a given \Core{}.

\subsubsection{Running masses and strong coupling}
\label{sec:extractstandard}

The functions with the prototypes
\begin{verbatim}
  double Mass::mMS(int nfIn, double scale) const;
  double Alpha::operator()(double scale) const;
\end{verbatim}

\begin{verbatim}
  MassMS[CoreName_String, nfIn_Integer, scale_Real]
  AlphaQCD[CoreName_String, scale_Real]
\end{verbatim}
are used to extract a running mass (i.e.\ the $\MSbar$ mass if the flavor number scheme includes this massive quark, the MSR mass otherwise) and strong coupling at a specific renormalization scale \texttt{scale} from a \Core{} where the optional parameters valid in the creation of the \Core{} are respected. The functions furthermore use an {\it automatic matching convention}, which means that the flavor number scheme of the output is deduced automatically from \texttt{scale}, with flavor threshold matching at $\mu_n=\mathtt{fn}\times\overline m_n$, see Sec.~\ref{sec:core}.
\texttt{nfIn} specifies for which quark the running mass is returned and refers to the number of dynamical flavors of the associated standard running mass.

In \Cpp{}, as indicated by the scope resolution prefixes of \texttt{Mass::mMS} and \texttt{Alpha::\allowbreak operator()}, these functions are not by themselves members of the class \texttt{Core}, but members of the classes \texttt{Mass} and \texttt{Alpha} respectively. Consequently, in practice they are accessed through the \Core{} member functions \texttt{Core::alpha} and \texttt{Core::masses} (see Sec.~\ref{sec:structure} the following examples).

Note that matching at flavor thresholds is {\it always} done from below to above the thresholds, meaning that matching between $n_\ell$ and $n_\ell + 1$ flavor schemes always uses the perturbative expansion in $\alpha_s^{(n_\ell)}$. If this is not possible directly, e.g.\ when $\alpha_s^{(n_\ell+1)}$ is given and $\alpha_s^{(n_\ell)}$ still needs to be determined, the solution is computed iteratively. This convention is strictly applied everywhere, and in particular for the determination of the \Core{} parameters described in Sec.~\ref{sec:accesscoreparameters}. For the computation of the renormalization group evolution equations (at any specified order) we use algorithms which are exact, i.e.\ they provide results with {\it machine precision}. This, in combination with the flavor matching convention, has the advantage that \Core{}s are created in a self-consistent way. This means that a \Core{} that is created from the coupling and the masses at any renormalization scales extracted from an existing core \Core{} will lead to a \Core{} that is physically equivalent within machine precision if all optional parameters are set to equivalent values.

Both functions above are also available in a version permitting complex-valued input for \texttt{scale}, resulting in a complex-valued output. The function prototypes in this case are
\begin{verbatim}
  std::complex<double>
    Mass::mMS(int nfIn, std::complex<double> scale);
  std::complex<double>
    Alpha::operator()(std::complex<double> scale);
\end{verbatim}

\begin{verbatim}
  MassMS[CoreName_String, nfIn_Integer, scale_Complex]
  AlphaQCD[CoreName_String, scale_Complex]
\end{verbatim}
and the flavor number scheme of the output is determined from the automatic matching conventions based on the absolute value of \texttt{scale}.

Note that for applications where negative real values of \texttt{scale} are expected, the user of the \texttt{C++} or \texttt{Python} interfaces should declare \texttt{scale} as complex-valued explicitly from the start so that the intrinsic \texttt{C++} function evaluations can be performed. Otherwise, in such a case \texttt{NaN} will be returned. Using the \Mathematica{} interface, the complex-valued declaration is automatically employed if \texttt{scale} is negative real (or complex). In the way \texttt{C++} treats the branch cuts of complex-valued functions involved in the calculations, this corresponds to adding an infinitesimally small positive imaginary part to \texttt{scale}.

In case the input parameter \texttt{scale} is not provided when executing \texttt{Mass::mMS} and \texttt{MassMS}, the routines return the respective standard running mass.

\subsubsection*{\Cpp{} example}

The instruction
\begin{verbatim}
  core1.masses().mMS(6, 20.0);
\end{verbatim}
returns the mass value of the heaviest of the six quarks defined in the \Core{} named \texttt{core1}, referred to by the flavor number $6$ of the corresponding standard running mass flavor scheme, at the scale $20.0$\,GeV with automatic flavor matching (performed at the standard running mass of the heaviest quark in \texttt{core1}). Referring to the corresponding quark mass as $m_t$, the returned value \texttt{(double)171.046}
corresponds to $m_t^{\mathrm{MSR},(5)}(20.0\,\mathrm{GeV})$.

The instruction
\begin{verbatim}
  core1.alpha()(10.0);
\end{verbatim}
returns the value of the strong coupling defined in the \Core{} named \texttt{core1} at the scale $10.0$\,GeV. Due to automatic matching the returned value \texttt{(double)\allowbreak 0.178468} refers to $\alpha_s^{(5)}(10.0\,\rm{GeV})$. Note that the \Core{} member functions \texttt{Core::alpha} and \texttt{Core::masses} have been used to access member functions of the classes \texttt{Alpha} and \texttt{Mass}, see Sec.~\ref{sec:structure}.

\subsubsection*{\Mathematica{} example}
The \Mathematica{} command
\begin{verbatim}
  In[]:= MassMS["core1", 5, 30.0]
  Out[]= 3.199542552851507
\end{verbatim}
returns the mass value of the next-to-heaviest of the six quarks defined in the \Core{} named \texttt{core1}, referred to by the flavor number $5$, at the scale $30.0$\,GeV. Referring to the corresponding quark mass as $m_b$, due to automatic matching, the shown output corresponds to the $\MSbar$ mass $\overline{m}_b^{(5)}(30.0\,\mathrm{GeV})$. The command

\begin{verbatim}
  In[]:= AlphaQCD["core1", -10.0 + 0.1 I]
  Out[]= 0.11619771241600231 - 0.08307724827828394 I
\end{verbatim}
returns the value of the strong coupling defined in the \Core{} named \texttt{core1} at the complex scale $\mu=(-10.0+0.1\, i)$\,GeV. Due to the automatic matching convention, the flavor number scheme is automatically chosen to be $5$ since $|\mu| = 10.0005$\,GeV exceeds the standard running bottom mass in \texttt{core1}. Consequently the output refers to $\alpha_s^{(5)}((-10.0 + 0.1\, i)\,\mathrm{GeV})$.

\subsubsection*{Optional parameters}

Automatic matching at the flavor thresholds can be overruled using the additional input \texttt{nfOut}, which specifies the flavor number scheme. The function prototypes are
\begin{verbatim}
  double Mass::mMS(int nfIn, double scale,
                   int nfOut = kDefault) const;
  double Alpha::operator()(double scale,
                           int nfOut = kDefault) const;
\end{verbatim}

\begin{verbatim}
  MassMS[CoreName_String, nfIn_Integer, scale_Real,
         nfOut_Integer:kDefault]
  AlphaQCD[CoreName_String, scale_Real, nfOut_Integer:kDefault]
\end{verbatim}
and analogously for complex-valued input. In \Cpp{} as well as in \Mathematica{}, the value \texttt{kDefault} is a predefined constant, internally set to $-1$, specifying that the respective default values will be used.

\subsubsection{Other short-distance masses}
\label{sec:sdm}

From a given \Core{}, the extraction of values of quark masses in a number of other short-distance quark mass schemes is supported. This includes the renormalization group invariant (\texttt{RGI}) scheme \cite{Floratos:1978jb} and
the following low-scale short-distance mass schemes: 1S (\texttt{1S})~\cite{Hoang:1998ng,Hoang:1998hm,Hoang:1999ye}, kinetic (\texttt{Kin})~\cite{Czarnecki:1997sz}, potential subtracted (\texttt{PS})~\cite{Beneke:1998rk,Beneke:2005hg}, and renormalon subtracted\footnote{We implement only the ``unprimed'' version of the RS mass, which has a finite ${\cal O}(\alpha_s)$ term in its relation to the pole mass.} (\texttt{RS})~\cite{Pineda:2001zq}. Note that the optional parameters setting the loop orders of the conversion formulae used for the extraction of these masses are independent of the loop order parameters specified during \Core{} creation (\texttt{runAlpha}, \texttt{orderAlpha}, \texttt{runMSbar}, \texttt{orderMSbar}, \texttt{runMSR}, \texttt{orderMSR}), see Sec.~\ref{sec:core}.

The functions with the prototypes
\begin{verbatim}
  double Mass::mRGI(int nfIn) const;
  double Mass::m1S(int nfIn) const;
  double Mass::mKin(int nfIn, double scaleKin) const;
  double Mass::mPS(int nfIn, double muF) const;
  double Mass::mRS(int nfIn, double scaleRS) const;
\end{verbatim}

\begin{verbatim}
  MassRGI[CoreName_String, nfIn_Integer]
  Mass1S[CoreName_String, nfIn_Integer]
  MassKin[CoreName_String, nfIn_Integer, scaleKin_Real]
  MassPS[CoreName_String, nfIn_Integer, muf_Real]
  MassRS[CoreName_String, nfIn_Integer, scaleRS_Real]
\end{verbatim}
extract from a \Core{} a quark mass value in the specified short-distance scheme. The variable \texttt{nfIn} is again the specifier for the massive quark, referring to the number of dynamical flavors of the associated standard running mass. The variables \texttt{scaleKin},\footnote{Following Refs.~\cite{Fael:2020iea, Fael:2020njb},
the default value for the kinetic mass intrinsic scale (where the default log resummed conversion between the running and the kinetic masses is carried out, see below) is set to be twice its renormalization scale, $2\times$\texttt{scaleKin}.} \texttt{muF} and \texttt{scaleRS} specify the renormalization scales of the kinetic, potential subtracted and renormalon subtracted masses, respectively. Note that the \texttt{1S}, \texttt{Kin}, \texttt{PS} and \texttt{RS} masses are defined in (\texttt{nfIn}$-1$)-flavor schemes, while the \texttt{RGI} mass is defined in the \texttt{nfIn} flavor scheme.

In \Cpp{}, all routines outlined above are member functions of the class \texttt{Mass}, as indicated by the scope resolution prefix \texttt{Mass::}, and are accessed through the \Core{} member function \texttt{Core::masses} (see Sec.~\ref{sec:structure} and the following examples).

Without specifying any additional optional parameters, the extraction of the low-scale short-distance masses is done in the following default way: first, the running mass in the ($\mathtt{nfIn} - 1$)-flavor scheme is determined at the {\it intrinsic scale} of the low-scale short-distance mass using R-evolution. The intrinsic scales of the \texttt{Kin}, \texttt{PS} and \texttt{RS} schemes are the respective values of $2\times$\texttt{scaleKin}, \texttt{muf} and \texttt{scaleRS}, while for the \texttt{1S} scheme it is the inverse Bohr radius $M_{q,B}$ determined with the routine \texttt{Mass::mBohr} described below with default setting.
Subsequently, the running mass is converted to the low-scale short-distance mass using both masses' perturbative relation to the pole mass employing the strong coupling at the intrinsic scale, where the pole mass is then consistently eliminated to the corresponding order. The resulting relation is renormalon-free avoiding large logarithms. Whenever known, finite mass effects stemming from lighter massive quarks are taken into account in this relation, which is up to three loops for the pole mass relations of the \texttt{1S}, \texttt{PS} and \texttt{Kin}\footnote{For the kinetic mass \REvolver{} adopts the definition of lighter massive quark corrections given in Ref.~\cite{Fael:2020njb} where these quark mass corrections are absent in the flavor number scheme below the corresponding threshold and exclusively come from the flavor number decoupling relations of the strong coupling above the threshold.}
schemes. For the \texttt{RS} mass, finite lighter quark mass effects have not been explicitly specified in the literature. Note that the lighter quark mass effects in the perturbative relation between the pole and running ($\MSbar$ or MSR) masses are known to three loops. In the conversion between the running and the \texttt{RS} masses these lighter quark mass corrections are set to zero coherently everywhere to avoid upsetting the renormalon cancellation. Effective methods to simulate finite quark mass effects (e.g.\ by enforcing a change in the flavor number scheme of the strong coupling) are not implemented as regular \REvolver{} functionalities, but can still be realized, as we show in the examples given in Sec.~\ref{sec:examples}.

\REvolver{} provides the routine \texttt{Mass::mBohr}
to calculate the heavy quarkonium inverse Bohr radius $M_{q,B}$ for a massive quark $q$. In the default setting, $M_{q,B}$
is the root of the function $f(x) = C_F \alpha_s^{(n_\ell)}(x) m_q^{\mathrm{MSR},(n_\ell)}(x) - x$, and determined using an algorithm based on a modified version of Dekker's method. The corresponding routine is a global function and always uses machine precision.
The inverse Bohr radius for a massive quark $q$ can be extracted from a given \Core{} through the functions with the following prototypes
\begin{verbatim}
  double Mass::mBohr(int nfIn, double nb = 1.0) const;
\end{verbatim}

\begin{verbatim}
  MBohr[CoreName_String, nfIn_Integer, nb_Real:1.0]
\end{verbatim}
where \texttt{nfIn} is the specifier for the massive quark and \texttt{nb} refers to an optional rescaling factor such that the inverse Bohr radius is determined from the equality $M_{q,B}^\mathtt{nb} = \mathtt{nb}\, C_F \alpha_s^{(n_\ell)}(M_{q,B}^\mathtt{nb}) m_q^{\mathrm{MSR},(n_\ell)}(M_{q,B}^\mathtt{nb})$. This option is useful for scale variation e.g.\ in the context of higher excited heavy quarkonium states. For the computation of
the intrinsic scale used for the \texttt{1S} mass scheme conversions, $\mathtt{nb}$ is used with the default setting $\mathtt{nb}\,$=1.

\subsubsection*{\Cpp{} example}

The instruction
\begin{verbatim}
  core1.masses().m1S(6);
\end{verbatim}
returns the mass value in GeV units of the heaviest of the six quarks defined in the \Core{} named \texttt{core1}, referred to by the flavor number $6$, in the 1S scheme, which is \texttt{(double)\allowbreak 171.517}.

\subsubsection*{\Mathematica{} example}

The \Mathematica{} command
\begin{verbatim}
  In[]:= MassPS["core1", 5, 2.0]
  Out[]= 4.521091787631138
\end{verbatim}
returns the mass value of the next-to-heaviest of the six quarks defined in the \Core{} named \texttt{core1}, referred to by the flavor number $5$, in the potential subtracted scheme at the scale $2.0\,$GeV, i.e.\ $m_b^\mathrm{PS}(2\,\mathrm{GeV})$, referring to
that quark mass as $m_b$.

\subsubsection*{Optional parameters}

The functions responsible for extracting quark mass values in short-distance mass schemes other than the running mass allow for several optional parameters whose nature is tied to the respective schemes. The full prototypes are given by
\begin{verbatim}
  double Mass::mRGI(int nfIn,
                    int order = kMaxRunMSbar) const;
  double Mass::m1S(int nfIn, int nfConv = kDefault,
                   double scale = kDefault,
                   Count1S counting = Count1S::Default,
                   double muA = kDefault,
                   int order = kMaxOrder1s) const;
  double Mass::mKin(int nfIn, double scaleKin,
                    int nfConv = kDefault,
                    double scale = kDefault,
                    double muA = kDefault,
                    int order = kMaxOrderKinetic) const;
  double Mass::mPS(int nfIn, double muF,
                   int nfConv = kDefault,
                   double scale = kDefault,
                   double muA = kDefault,
                   double rIR = 1,
                   int order = kMaxOrderPs) const;
  double Mass::mRS(int nfIn, double scaleRS,
                   int nfConv = kDefault,
                   double scale = kDefault,
                   double muA = kDefault,
                   int order = kMaxRunMSR,
                   int nRS = kMaxRunAlpha - 1,
                   double N12 = kDefault) const;
\end{verbatim}

\begin{verbatim}
  MassRGI[CoreName_String, nfIn_Integer,
          order_Integer:kMaxRunMSbar]
  Mass1S[CoreName_String, nfIn_Integer,
         nfConv_Integer:kDefault,
         scale_Real:kDefault,
         counting_String:"default",
         muA_Real:kDefault,
         order_Integer:kMaxOrder1s]
  MassKin[CoreName_String, nfIn_Integer, scaleKin_Real,
          nfConv_Integer:kDefault,
          scale_Real:kDefault,
          muA_Real:kDefault,
          order_Integer:kMaxOrderKinetic]
  MassPS[CoreName_String, nfIn_Integer, muF_Real,
         nfConv_Integer:kDefault,
         scale_Real:kDefault,
         muA_Real:kDefault,
         rIR_Real:1.0,
         order_Integer:kMaxOrderPs]
  MassRS[CoreName_String, nfIn_Integer, scaleRS_Real,
         nfConv_Integer:kDefault,
         scale_Real:kDefault,
         muA_Real:kDefault,
         order_Integer:kMaxRunMSR,
         nRS_Integer:kMaxRunAlpha - 1,
         N12_Real:kDefault]
\end{verbatim}

where the variables \texttt{kDefault}, \texttt{kMaxRunMSbar}, \texttt{kMaxRunMSR}, \texttt{kMaxOrder1s}, \texttt{kMaxOrderKinetic} and \texttt{kMaxOrderPs} as well as the values \texttt{"default"} and \texttt{Count1S::Default} in \Mathematica{} and \Cpp{}, respectively, are predefined constants representing the respective default values. In \Cpp{} and \Mathematica{}, if one of the optional parameters shown above is explicitly specified, all parameters appearing prior in the argument list must be specified as well.

In the following the meaning of the optional parameters is explained.

The extraction of the RGI mass has only one optional parameter, which is \texttt{order}, specifying the loop order of the $\beta$-function and $\MSbar$ mass anomalous dimensions (which are taken equal) entering the conversion formula from the standard running mass, see Eq.~\eqref{eq:wTilde}. The default value is the highest available order, which is $5$.

Considering the functions extracting low-scale short-distance masses, there are some optional parameters affecting the formulae used for the conversion computations:
\begin{itemize}
\item \texttt{nfConv}: specifies the flavor number scheme of the running mass that is used in the conversion formula. For \texttt{nfConv = nfIn - 1} the MSR scheme is used; for \texttt{nfConv = nfIn} the $\MSbar$ scheme is used. The strong coupling is always employed in the \texttt{nfIn - 1} flavor scheme. Default: \texttt{nfIn - 1}.

\item \texttt{scale}: specifies the scale of the running mass from which the conversion is determined. Default: intrinsic scale of the low-scale mass.

\item \texttt{muA}: specifies the scale of the strong coupling used for the conversion. Default: intrinsic scale of the low-scale mass.

\item \texttt{order}: specifies how many perturbative orders are used in the conversion. Default: Highest available order for the low-scale mass.
\end{itemize}
Note that for very heavy quarks (such as the top quark) specifications of the parameters \texttt{nfConv}, \texttt{scale} and \texttt{muA} that differ from their defaults can lead to large logarithmic corrections in the conversions involving low-scale short-distance masses. On the other hand, for the case of lighter massive quarks (such as the bottom and especially the charm quark) the default settings may lead to unphysically low scales causing perturbative instabilities.
The parameters \texttt{nfConv}, \texttt{scale} and \texttt{muA} should therefore be used with some care.

Furthermore, for some low-scale masses there are additional optional parameters related to their definition and specific properties:
\begin{itemize}
\item \texttt{counting}: specifies the order counting used for the conversion computation to extract the 1S mass. The available options are \texttt{Count1S::\allowbreak Nonrelativistic} and \texttt{Count1S::Relativistic} (enumerations of type \texttt{Count1S}) in \Cpp{}, and \texttt{"nonrelativistic"} and \texttt{"relativistic"} in \Mathematica{}. For non-relativistic counting it is assumed that \texttt{scale} is of the same order as the inverse Bohr radius, which is much smaller than the quark mass value and both are counted as $\mathcal{O}(m_n\alpha_s)$; for the relativistic counting it is assumed that \texttt{scale} is of the same order as the quark mass, and the inverse Bohr radius is counted as $\mathcal{O}(m_n)$, see Sec.~5.2 of Ref.~\cite{Hoang:2017suc}. If one chooses to convert from the $\MSbar$ mass, only the relativistic counting is supported. Note that here the optional parameter \texttt{order} always specifies how many non-zero terms are taken into account in the conversion series, e.g.\ in the case of non-relativistic counting $\mathtt{order}=1$ refers to $\mathcal O(m_n\alpha_s^2)$, while it refers to $\mathcal O(m_n\alpha_s)$ in the relativistic counting. (So for $\mathtt{order}=1$ the conversion formula is the same in both counting schemes.) To simplify language, we refer to a perturbative term in the 1S-pole mass relation as $n$-loop, if it is combined with the $n$-loop coefficient in the relation between the running mass and the pole mass, independent of the employed counting. Default: \texttt{Count1S::Default} and \texttt{"default"} for \Cpp{} and \Mathematica{}, respectively, which corresponds to \texttt{"nonrelativistic"} for an MSR mass input and to \texttt{"relativistic"} for $\MSbar$.

\item \texttt{rIR}: specifies the ratio of the IR subtraction scale $\mu_\mathrm{IR}$ with respect to the scale \texttt{muF} employed in the 4-loop term of the pole-PS mass relation, $\mathtt{rIR} = \mu_\mathrm{IR}/\mu_F$, see Secs.~4.5.3 and 5.1 of Ref.~\cite{Hoang:2017suc}. Default: $1.0$, which corresponds to the definition given in Ref.~\cite{Beneke:2005hg}.

\item \texttt{nRS}: specifies the number of terms (used in the perturbative construction of the Borel function and the normalization \texttt{N12}) for the calculation of the coefficients of the pole-RS mass perturbation series, see Ref.~\cite{Pineda:2001zq}. Default: highest available order which is $4$.

\item \texttt{N12}: specifies the pole mass renormalon normalization constant employed in the pole-RS mass relation. Default: value computed by the sum rule formula employed by the routine \texttt{Mass::N12} (\Cpp{}) or \texttt{N12} (\Mathematica{}), see Sec~\ref{sec:poleambiguity}, summing up \texttt{nRS} terms in the sum rule series using all available information on the QCD $\beta$-function and anomalous dimension of the MSR mass.
\end{itemize}

\subsubsection*{\Cpp{} example}

The instruction
\begin{verbatim}
  core1.masses().mKin(5, 2.0, 4, 5.0, 3.0, 2);
\end{verbatim}
returns the kinetic mass value of the next-to-heaviest of the six quarks defined in the \Core{} named \texttt{core1}, referred to by the flavor number 5 of the corresponding standard running mass flavor scheme. The intrinsic kinetic mass scale is set to $2\,$GeV, the conversion formula is applied to the MSR mass in the $4$ flavor scheme at $5\,$GeV and with the renormalization scale of the strong coupling set to $3\,$GeV. Terms up to $\mathcal O(\alpha_s^2)$ are included in the conversion formula. The returned value is \texttt{(double)4.3105}.

\subsubsection*{\Mathematica{} example}

The \Mathematica{} command
\begin{verbatim}
  In[]:= MassRS["core1", 6, 20.0, 6, 163.0, 60.0, 3]
  Out[]= 170.48477730509723
\end{verbatim}
returns the RS mass value of the heaviest of the six quarks defined in
the \Core{} named \texttt{core1}, referred to by the flavor number 6. The intrinsic scale of the RS mass is set to $20\,$GeV and the conversion formula is applied to the $\MSbar$ mass in the $6$ flavor scheme. The $\MSbar$ mass renormalization scale is specified to be $163\,$GeV, while the renormalization scale of the strong coupling is set to $60\,$GeV. The last parameter shown above specifies that three perturbative orders are used in the conversion. For \texttt{nRS} and \texttt{N12} no inputs are specified, consequently the default setting is applied.

\subsubsection{Pole mass}
\label{sec:pole}

Quark mass values in the pole mass scheme can be extracted from a given \Core{}. The pole mass scheme suffers from a renormalon ambiguity so that there are several options to quote a value. \REvolver{} supports two ways of extracting pole quark mass values, accessible through the functions corresponding to the prototypes
\begin{verbatim}
  double Mass::mPoleFO(int nfIn, int nfConv, double scale,
                       double muA, int order) const;
  double Mass::mPole(int nfIn, double scale) const;
\end{verbatim}

\begin{verbatim}
  MassPoleFO[CoreName_String, nfIn_Integer, nfConv_Integer,
             scale_Real, muA_Real, order_Integer]
  MassPole[CoreName_String, nfIn_Integer, scale_Real]
\end{verbatim}
All parameters shown have to be specified by the user. For both routines the variable \texttt{nfIn} is the specifier for the quark whose pole mass value is extracted, referring to the number of dynamical flavors of the associated standard running mass, while \texttt{scale} specifies the scale of the running mass for which the conversion formula to the pole mass is employed. Furthermore, the loop orders of the various components entering the conversion formula used in the computation of the pole mass value are provided by the settings of the loop order parameters specified during \Core{} creation (\texttt{runAlpha}, \texttt{orderAlpha}, \texttt{runMSbar}, \texttt{orderMSbar}, \texttt{runMSR}, \texttt{orderMSR}), see Sec.~\ref{sec:core}. Here the parameters \texttt{runMSR} and \texttt{runAlpha} are particularly important because they also set the loop order of the exact coefficients accounted for in the perturbative series between the MSR and the pole masses, see Eq.~\eqref{eq:anasymptotic} in Sec.~\ref{sec:asymptotic}. Because in \REvolver{} matching is always applied at flavor thresholds, the latter series are the only ones being affected by the renormalon.

The functions \texttt{Mass::PoleFO} and \texttt{MassPoleFO}, respectively, return the pole mass using the conversion formula up to the order set by \texttt{order} from the running mass in the flavor number scheme specified by \texttt{nfConf}. The input parameter \texttt{muA} specifies the renormalization scale of the strong coupling used in the conversion formula.
The input \texttt{order} allows, in principle, an arbitrarily large integer. For $\mathtt{order} \leq\min(\texttt{runMSR},\,\texttt{runAlpha}-1)$ the exact perturbative coefficients up to this order are utilized in the relation between MSR and pole mass. For $\mathtt{order} > \min(\texttt{runMSR},\,\texttt{runAlpha}-1)$ a renormalon-based asymptotic formula for the coefficients of the asymptotic series is used, see Eq.~\eqref{eq:anasymptotic} in Sec.~\ref{sec:asymptotic} for details.

The functions \texttt{Mass::Pole} and \texttt{MassPole}, respectively, return the value of the asymptotic pole mass, i.e.\ the conversion is done by summing the perturbative series for the conversion formula relating running and pole mass to the order of minimal correction, again potentially employing the already mentioned asymptotic formula if the order of minimal correction is larger than $\min(\texttt{runMSR},\,\texttt{runAlpha}-1)$. The asymptotic pole mass is always determined from the MSR mass in the flavor number scheme where all massive quarks are integrated out regardless of the value of \texttt{scale}. The method used for the calculation of the asymptotic value is the minimal correction method (\texttt{"min"}) described below.

\subsubsection*{\Cpp{} example}

The instruction
\begin{verbatim}
  core1.masses().mPole(5, 2.0);
\end{verbatim}
returns the asymptotic pole mass of the next-to-heaviest of the six quarks defined in the \Core{} named \texttt{core1} in GeV units, referred to by the flavor number $5$, using the minimal correction method, which is \texttt{(double)\allowbreak 4.91150}. The scale employed for the running mass at conversion is $2\,$GeV.

\subsubsection*{\Mathematica{} example}

The \Mathematica{} command
\begin{verbatim}
  In[]:= MassPoleFO["core1", 6, 5, 20.0, 20.0, 16]
  Out[]= 1190.2576448418606
\end{verbatim}
returns the order-dependent pole mass value of the heaviest of the six quarks defined in the \Core{} named \texttt{core1}, referred to by the flavor number 6. The conversion formula is applied to the respective MSR mass with $5$ active flavors at the scale $20\,$GeV, choosing the same scale for the renormalization scale of the strong coupling. The perturbative series terms are summed up to ${\cal O}(\alpha_s^{16})$.\footnote{The value obtained with this command is unphysically large due to the asymptotic (non-convergent) nature of the series that relate the pole mass and short-distance masses. This behavior is the origin of the ambiguity of the pole mass.}

\subsubsection*{Optional parameters}

The functions \texttt{Mass::mPoleFO} and \texttt{MassPoleFO} in \Cpp{} and \Mathematica{}, respectively, are completely general and do not support any additional optional parameters.

The full prototypes of the functions responsible for accessing the asymptotic pole mass are
\begin{verbatim}
  double Mass::mPole(int nfIn, double scale,
                     double muA = kDefault,
                     PoleMethod method = PoleMethod::Default,
                     double f = 1.25,
                     double* ambiguity = nullptr,
                     int* nMin = nullptr) const;
\end{verbatim}

\begin{verbatim}
  MassPole[CoreName_String, nfIn_Integer, scale_Real,
           muA_Real:kDefault,
           method_String:"min",
           f_Real:1.25]
  MassPoleDetailed[CoreName_String, nfIn_Integer, scale_Real,
                   muA_Real:kDefault,
                   method_String,
                   f_Real:1.25]
\end{verbatim}
where \texttt{MassPoleDetailed} in \Mathematica{} is a function returning a list, containing the asymptotic pole mass, the associated renormalon ambiguity and the order of the minimal correction term, providing the functionality corresponding to the \Cpp{} routine \texttt{Mass::mPole} (with specified optional pointer parameters \texttt{ambiguity} and \texttt{nMin}), see the descriptions below. In \Python{} there is the analogous function \texttt{Mass.MassPoleDetailed}.

The meaning of the optional input parameters is as follows:
\begin{itemize}
\item \texttt{muA}: specifies the renormalization scale of the strong coupling in the conversion formula.
The default is the value of \texttt{scale}.
\item \texttt{method}: specifies the method used to obtain the asymptotic pole mass and the ambiguity. In \Cpp{} the available options are \texttt{PoleMethod::Min} for the minimal correction term method, \texttt{PoleMethod::Range} for the range method and \texttt{PoleMethod::DRange} for the corresponding discrete version, see the explanation below. In \Mathematica{} these options correspond to the string inputs \texttt{"min"}, \texttt{"range"} and \texttt{"drange"}, respectively. The input \texttt{method} is optional if only the asymptotic pole mass is returned; the defaults in that case are \texttt{PoleMethod::Min} (\Cpp{}) and \texttt{"min"} (\Mathematica{}). The input \texttt{method} is \textit{not} optional if the pole mass ambiguity is also returned, i.e.\ for
\texttt{MassPoleDetailed} in \Mathematica{} and when the optional pointer to \texttt{ambiguity} is given in \Cpp{}.
\item \texttt{f}: specifies a constant larger than unity multiplying the minimal correction for the methods \texttt{"range"} and \texttt{"drange"}. Default: $1.25$
\item \texttt{ambiguity}: a pointer to \texttt{double}. If specified, the value of the pole mass ambiguity is saved in the variable pointed to (\Cpp{} only).
\item \texttt{nMin}: a pointer to \texttt{int}. If specified, the order of the minimal correction term is saved in the variable pointed to (\Cpp{} only).
\end{itemize}

Note that for the \Mathematica{} routine \texttt{MassPoleDetailed} the input variable \texttt{method} does not have a default and must always be specified (because it always returns a value for the pole mass). Calling the function with $4$ arguments means that
values are specified for the four input parameters \texttt{CoreName}, \texttt{nfIn}, \texttt{scale}, and \texttt{method}, while the variables \texttt{muA} and \texttt{f} are set to their default values.

The \textit{minimal correction} method (\texttt{"min"}) to obtain the asymptotic pole mass value and its ambiguity refers to the method suggested in Ref.~\cite{Beneke:2016cbu} where the ambiguity is determined from the size of the minimal correction based on a quadratic function fitted to the smallest correction and the two neighboring corrections. However, in contrast to the procedure described in Ref.~\cite{Beneke:2016cbu}, \REvolver{} accounts for the mass effects of lighter massive quarks by the exact expressions given in Ref.~\cite{Hoang:2017btd} instead of including them in an approximate way by flavor number scheme modifications of the strong coupling. Also, for coefficients of order higher than $\min(\texttt{runMSR},\,\texttt{runAlpha}-1)$ the asymptotic formula of Eq.~\eqref{eq:anasymptotic} in Sec.~\ref{sec:asymptotic} is employed.

The \textit{drange} (``discrete range'', \texttt{"drange"}) choice refers to the method suggested in Ref.~\cite{Hoang:2017btd}, where the pole mass value and its ambiguity are computed from the range in orders around the minimal term where the corrections are smaller than \texttt{f} times the minimal correction. The method \textit{range} (\texttt{"range"}) refers to a continuous generalization which is analogous but provides smoother results. Here the order-dependent discrete-valued individual perturbative coefficients of the relation between the pole and running masses, as well as the related cumulant, are made continuous by a cubic interpolation. From these functions, the asymptotic pole mass value and the ambiguity are determined in analogy to the \texttt{"drange"} method. For all methods (including \texttt{"range"}), the returned order of the minimal correction term is an integer and refers to the original series without any interpolation.

\subsubsection*{\Cpp{} example}

With the instructions
\begin{verbatim}
  double ambiguity;
  int nMin;
  core1.masses().mPole(6, 10.0, 10.0, PoleMethod::Range, 1.25,
                       &ambiguity, &nMin);
\end{verbatim}
first the variables \texttt{nMin} and \texttt{ambiguity} are initialized and pointers to them are passed in the call of the function \texttt{Mass::mPole}. The instruction returns the asymptotic pole mass value \texttt{(double)173.107} for the heaviest of the $6$ quarks of the \Core{} named \texttt{core1}, obtained from the conversion formula for the running mass at the scale $10.0\,$GeV for $3$ active flavors [\,$m_t^{{\rm MSR},(3)}(10,\mbox{GeV})$\,] using the \textit{range} method. The values of the pole mass ambiguity \texttt{(double)0.179811}, and the order of the minimal correction term \texttt{(int)4} are stored in the variables \texttt{ambiguity} and \texttt{nMin}, respectively.

\subsubsection*{\Mathematica{} example}

The \Mathematica{} command
\begin{verbatim}
  In[]:= MassPoleDetailed["core1", 6, 10.0, "min"]
  Out[]= {173.09681693689498, 0.1307735468951421, 4}
\end{verbatim}
corresponds to the previous commands given in \Cpp{}, however, here the minimal correction method is used. \texttt{muA} is automatically set to default value which is \texttt{10.0} in this case. The output list entries correspond to the asymptotic pole mass value, the ambiguity and the order of the smallest correction term, respectively.

\subsubsection{Norm of the pole mass renormalon ambiguity}
\label{sec:poleambiguity}

The pole mass renormalon normalization constant can be accessed using the functions with the following prototypes
\begin{verbatim}
  double Mass::N12(double lambda = 1.0) const;
  double Mass::P12(double lambda = 1.0) const;
\end{verbatim}

\begin{verbatim}
  N12[CoreName_String, lambda_Real:1.0]
  P12[CoreName_String, lambda_Real:1.0]
\end{verbatim}
where the two functions correspond to the normalization conventions $P_{1/2}$ and $N_{1/2}=\Gamma(1+\hat b_1)\beta_0 P_{1/2}/(2\pi)$ as described in Ref.~\cite{Hoang:2017suc}. The routines employ the renormalon sum rule formula shown in Eq.~(\ref{eq:N12def}) of App.~\ref{sec:sumrule}. The number of massless quarks for which the normalization is determined is tied to the \Core{} used to extract the normalization. The optional input parameter \texttt{lambda} is a scaling parameter to estimate the uncertainty of the output ($1.0$ by default). The number of terms summed up in the sum rule formula is set by \texttt{runMSR} specified at \Core{} construction. The QCD $\beta$-function coefficients entering the sum rule formula are used up to \texttt{runAlpha} loop order, all higher order coefficients are set to zero, so that
$\hat b_1=0$ when \texttt{runAlpha} $=1$.

\subsubsection{Extracting \texorpdfstring{$\Lambda_\mathrm{QCD}$}{Lambda QCD}}

The functions corresponding to the prototypes
\begin{verbatim}
  double Alpha::lambdaQCD(int nf) const;
\end{verbatim}

\begin{verbatim}
  LambdaQCD[CoreName_String, nf_Integer]
\end{verbatim}
return from a given \Core{} the QCD scale $\Lambda_\mathrm{QCD}^{(\mathtt{nf})}$ in the $\MSbar$ definition (see Ref.~\cite{Zyla:2020zbs}) for the QCD coupling in the \texttt{nf} flavor scheme, utilizing the exponential formula given in Eq.~\eqref{eq:lambda} of App.~\ref{sec:strongcoupling}. The parameters \texttt{runAlpha} and \texttt{lambdaAlpha} specified at \Core{} creation set the number of coefficients of the $\beta$-function used.
The possible values for \texttt{nf} range from the number of massless quarks to the total number of quarks in the specified \Core{}.

\subsubsection*{\Cpp{} example}

The instruction
\begin{verbatim}
  core1.alpha().lambdaQCD(3);
\end{verbatim}
returns $\Lambda_\mathrm{QCD}^{(3)}$ in the $\MSbar$ definition in GeV units from the scenario encoded in \texttt{core1}, corresponding to \texttt{(double)0.335547}

\subsubsection*{Optional parameters}

The definition in which $\Lambda_\mathrm{QCD}$ is extracted can be chosen with an optional parameter. The full function prototypes are
\begin{verbatim}
  double Alpha::lambdaQCD(
      int nf, LambdaConvention convention =
                  LambdaConvention::MSbar) const;
\end{verbatim}

\begin{verbatim}
  LambdaQCD[CoreName_String, nf_Integer,
            convention_String:"MSbar"]
\end{verbatim}

In addition to the conventional $\MSbar$ definition according to the PDG 2020 \cite{Zyla:2020zbs}, which is the default, the ``t-scheme'' is supported referring to the definition based on the $t$-variable notations of Ref.~\cite{Hoang:2017suc}. To choose the t-scheme the optional parameter \texttt{convention} has to be set to \texttt{LambdaConvention::tScheme} and \texttt{"tScheme"} in \Cpp{} and \Mathematica{} respectively. See App.~\ref{sec:strongcoupling} for more details on the different conventions.

\subsection{Adding Masses to an Existing \Core{}}
\label{sec:addmass}

The following section presents routines to extend existing \Core{} scenarios by adding one {\it heavier} massive quark. The mass value of the {\it additional} quark can be given as a running mass, in the pole mass scheme (order-dependent as well as asymptotic), or in any of the short-distance schemes described in Sec.~\ref{sec:sdm}, with the same obligatory and optional parameters. Unless the mass of the additional quark is already given in the running mass scheme, the quark mass value is first converted to the running mass (specified by the parameters). The value for the running mass is then added to the scenario. The optional parameter \texttt{fnQ} specifies the matching scale for the new threshold, analogous to the individual entries of \texttt{fMatch} at \Core{} creation, see Sec.~\ref{sec:core}.

The routines employed for converting to the running mass are the \textit{exact inverse} of those used for extracting mass values in the respective scheme. This is achieved by numerically inverting the respective relations with iterative algorithms.

The related function prototypes, including all relevant optional parameters are
\begin{verbatim}
  void Core::addMsMass(int nf, double mass, double scale,
                       double fnQ = 1.0);
  void Core::addPoleMass(double mPole, double scale,
                         double muA = kDefault,
                         PoleMethod method =
                           PoleMethod::Default,
                         double f = 1.25,
                         double fnQ = 1.0);
  void Core::addPoleMassFO(double mPole, int nfConv,
                           double scale, double muA, int order,
                           double fnQ = 1.0);
  void Core::addPSMass(double mPS, double muF,
                       int nfConv = kDefault,
                       double scale = kDefault,
                       double muA = kDefault,
                       double rIR = 1,
                       int order = kMaxOrderPs,
                       double fnQ = 1.0);
  void Core::add1SMass(double m1S,
                       int nfConv = kDefault,
                       double scale = kDefault,
                       Count1S counting = Count1S::Default,
                       double muA = kDefault,
                       int order = kMaxOrder1s,
                       double fnQ = 1.0);
  void Core::addRGIMass(double mRGI,
                        int order = kMaxRunMSbar,
                        double fnQ = 1.0);
  void Core::addRSMass(double mRS, double scaleRS,
                       int nfConv = kDefault,
                       double scale = kDefault,
                       double muA = kDefault,
                       int order = kMaxRunMSR,
                       int nRS = kMaxRunAlpha - 1,
                       double N12 = kDefault,
                       double fnQ = 1.0);
  void Core::addKinMass(double mKin, double scaleKin,
                        int nfConv = kDefault,
                        double scale = kDefault,
                        double muA = kDefault,
                        int order = kMaxOrderKinetic,
                        double fnQ = 1.0);
\end{verbatim}
in \Cpp{} and
\begin{verbatim}
  AddMSMass[CoreName_String, NewCoreName_String, nf_Integer,
            mass_Real, scale_Real]
  AddPoleMass[CoreName_String, NewCoreName_String, mPole_Real,
              scale_Real,
              muA_Real:kDefault,
              method_String:"min",
              f_Real:1.25]
  AddPoleMassFO[CoreName_String, NewCoreName_String,
                mPole_Real, nfConv_Integer, scale_Real,
                muA_Real, order_Integer]
  AddPSMass[CoreName_String, NewCoreName_String, mPS_Real,
            muF_Real,
            nfConv_Integer:kDefault,
            scale_Real:kDefault,
            muA_Real:kDefault,
            rIR_Real:1,
            order_Integer:kMaxOrderPs]
  Add1SMass[CoreName_String, NewCoreName_String, m1S_Real,
            nfConv_Integer:kDefault,
            scale_Real:kDefault,
            counting_String:"default",
            muA_Real:kDefault,
            order_Integer:kMaxOrder1s]
  AddRGIMass[CoreName_String, NewCoreName_String, mRGI_Real,
             order_Integer:kMaxRunMSbar]
  AddRSMass[CoreName_String, NewCoreName_String, mRS_Real,
            scaleRS_Real,
            nfConv_Integer:kDefault,
            scale_Real:kDefault,
            muA_Real:kDefault,
            order_Integer:4,
            nRS_Integer:kMaxRunAlpha - 1,
            N12_Real:kDefault]
  AddKinMass[CoreName_String, NewCoreName_String, mKin_Real,
            scaleKin_Real,
            nfConv_Integer:kDefault,
            scale_Real:kDefault,
            muA_Real:kDefault,
            order_Integer:kMaxOrderKinetic]
\end{verbatim}
in \Mathematica{}, where \texttt{fnQ} can be set by the option parameter syntax (\texttt{fnQ~->~<value>}).

In \Mathematica{}, adding a new heavier quark results in creating a new \Core{} with a (not yet assigned) name specified in the obligatory input argument \texttt{NewCoreName}. This new \Core{} is based on the \Core{} with the name specified in \texttt{CoreName} containing only lighter massive quarks. The scenario parameters of the \Core{} \texttt{CoreName} are passed on to the new \Core{} named \texttt{NewCoreName}. If it turns out that the standard running mass associated to the mass to be added to a \Core{} is not the heaviest one in the new configuration, an error is returned and the new \Core{} is not created.

In \Cpp{} the new heavier quark is added directly to the \Core{} from which the member function is called. This saves resources and copying the old \Core{} can be easily done applying the assignment operator \texttt{=}, see the \Cpp{} example below.

\subsubsection*{\Cpp{} example}

With the instructions
\begin{verbatim}
  RunParV mPar4;
  mPar4.push_back({4, 1.3, 1.3});
  mPar4.push_back({5, 4.2, 4.2});
  Core core4(5, alphaPar, mPar4);
  Core core5 = core4;
  core5.add1SMass(171.51726494075493);
\end{verbatim}
in the first line, the \texttt{std::vector<RunPar>} \texttt{mPar4} is declared, and subsequently filled in the next two lines according to Sec.~\ref{sec:runpar} with information on two massive quarks. Together with \texttt{alphaPar} taken from the example in Sec.~\ref{sec:runpar}, it is the input for creating the \Core{} named \texttt{core4} in the third line, containing a total of $5$ quark flavors. In the fifth line a copy of \texttt{core4}, named \texttt{core5}, is produced. In the last line
a heavier quark with a mass value in the 1S scheme is added to \texttt{core5} which contains a total number of $6$ flavors. The input value for the 1S mass is taken from the \Cpp{} example of Sec.~\ref{sec:sdm} where the 1S mass value was extracted from \texttt{core1}.
Therefore \texttt{core5} and \texttt{core1} contain physically equivalent scenarios. \texttt{core4} has not been modified by adding the additional heavier quark and consequently its scenario still contains $5$ quark flavors.

\subsubsection*{\Mathematica{} example}

The command
\begin{verbatim}
  In[]:= mPar4 = {{4, 1.3, 1.3}, {5, 4.2, 4.2})};
         CoreCreate["core4", 5, alphaPar, mPar4]
         Add1SMass["core4", "core5", 171.51726494075493]
\end{verbatim}
has the same effect as the \Cpp{} example. First a list \texttt{mPar4} is defined, containing the standard running masses of two quarks, which are used as an input for creating \texttt{core4} with a total number of $5$ flavors. In the next line the new \Core{} named \texttt{core5} is created, which is a copy of \texttt{core4}, to which a heavier quark is added with a mass value specified in the 1S scheme.

\section{Applications and Pedagogical Examples}
\label{sec:examples}

In this section we demonstrate some of the features provided by \REvolver{} in a number of concrete examples as they may arise in practical applications. We give the examples in terms of \Mathematica{} code since, due to its interactive nature, it is especially suitable for that purpose. We provide all examples given here as a \Mathematica{} notebook, as \Python{} code in form of a Jupyter Notebook as well as \Cpp{} code, together with the library. For sake of clarity, all digits of the \Mathematica{} output are displayed.

\subsection{\Core{}s without Massive Quarks}

\subsubsection{Strong coupling evolution}

\subsubsection*{Strong coupling from inclusive jet cross sections}
\vspace{2mm}

In this first application we demonstrate one possible way to make use of the strong coupling evolution in \REvolver{} by reproducing and analyzing some results of Ref.~\cite{Khachatryan:2016mlc}, where the CMS collaboration carried out a strong coupling measurement from inclusive jet cross sections in different $p_T$ bins based on $8$\,TeV LHC data. In that publication, the evolution of the strong coupling from different values of $Q$ (the average $p_T$ in the bins) to $m_Z$ was carried out with $n_f=5$ active flavors and 2-loop accuracy. In the following, we focus on the result $\alpha^{(5)}_s(Q)=0.0822^{+0.0034}_{-0.0031}$ for $Q=1508.04$\,GeV shown in Table~5 of that article.

In a first step the relevant values can be defined by
\begin{verbatim}
In[]:= {aQCentral, Q} = {0.0822, 1508.04};
       {aQMin, aQMax} = aQCentral + {-0.0031, 0.0034};
\end{verbatim}
and the respective \Core{}s for the central, upper and lower strong coupling values, setting the strong-coupling evolution to 2-loop order, can be created by
\begin{verbatim}
  In[]:= CoreCreate["central2", {5, aQCentral, Q},
           runAlpha -> 2];
         CoreCreate["min2", {5, aQMin, Q}, runAlpha -> 2];
         CoreCreate["max2", {5, aQMax, Q}, runAlpha -> 2];
\end{verbatim}

The central value of $\alpha^{(5)}_s(m_Z)$ can then be extracted by
\begin{verbatim}
  In[]:= amZCentral2 = AlphaQCD["central2", mZdef]
  Out[]= 0.11616452350227859
\end{verbatim}
coinciding with $\alpha_s^{(5)}(m_Z)=0.1162$, as given in Ref.~\cite{Khachatryan:2016mlc}. Likewise the quoted uncertainties $^{+0.0070}_{-0.0062}$ are easily reproduced by executing
\begin{verbatim}
 In[]:= (AlphaQCD[#, mZdef] & /@ {"max2","min2"}) - amZCentral2
 Out[]= {0.007009865309335922, -0.006168760659529524}
\end{verbatim}
Note that in the \Mathematica{} interface the variable \texttt{mZdef} has the predefined value $91.187$\,GeV.

For comparison, we also employ 5-loop evolution for the strong coupling, resulting in equivalent numbers after rounding:
\begin{verbatim}
  In[]:= CoreCreate["central5", {5, aQCentral, Q},
           runAlpha -> 5];
         CoreCreate["min5", {5, aQMin, Q}, runAlpha -> 5];
         CoreCreate["max5", {5, aQMax, Q}, runAlpha -> 5];

  In[]:= amZCentral5 = AlphaQCD["central5", mZdef]
         AlphaQCD[#, mZdef] & /@ {"max5", "min5"} - amZCentral5
  Out[]= 0.11624523920392597
  Out[]= {0.007030395615252941, -0.006184136846029031}
\end{verbatim}
Note that the specification of the running order \texttt{runAlpha -> 5} at \Core{} creation is not mandatory since 5-loop running is the default. We show the specification to be explicit.

We now have a look at the perturbative uncertainty of these results. One approach to estimate the perturbative uncertainty of the given 2-loop result is to consider the difference of the values obtained by 2-loop and 1-loop evolution. This can be easily computed using \REvolver{} by creating a new \Core{} with specified 1-loop coupling evolution:
\begin{verbatim}
  In[]:= CoreCreate["central1", {5, aQCentral, Q},
           runAlpha -> 1];
         amZCentral2 - AlphaQCD["central1", mZdef]
  Out[]= 0.0017745561167654411
\end{verbatim}
The obtained conservative perturbative error estimate is about $25\%$ of the stated experimental error.
For the highest available perturbative order for running, $5$ loop, this proportion shrinks to $0.005\%$:
\begin{verbatim}
  In[]:= CoreCreate["central4", {5, aQCentral, Q},
           runAlpha -> 4];
         amZCentral5 - AlphaQCD["central4", mZdef]
  Out[]= 3.573226484421266*^-7
\end{verbatim}
\begin{figure}[t!]
\centering
\includegraphics[width=0.49\textwidth]{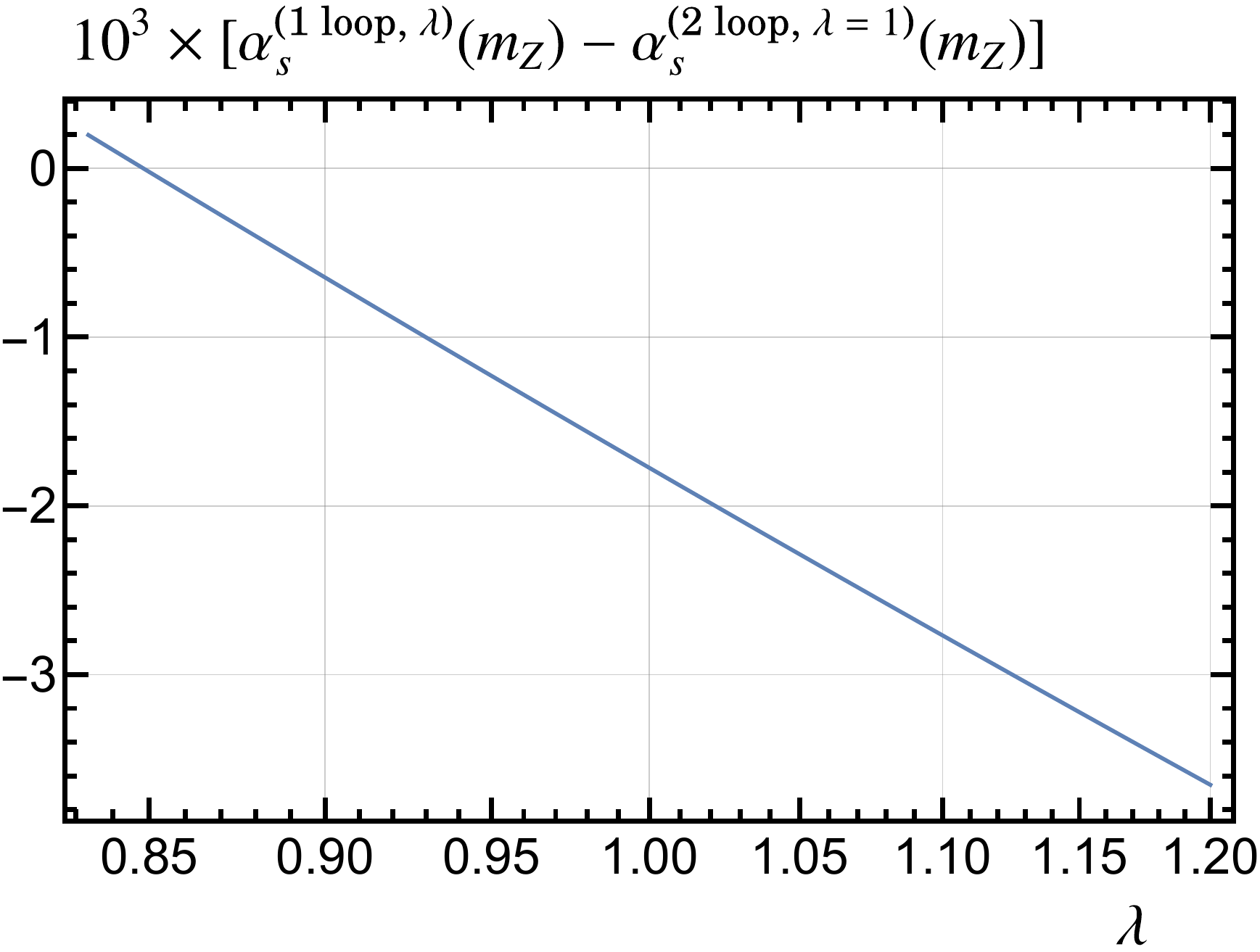}\hfill
\includegraphics[width=0.49\textwidth]{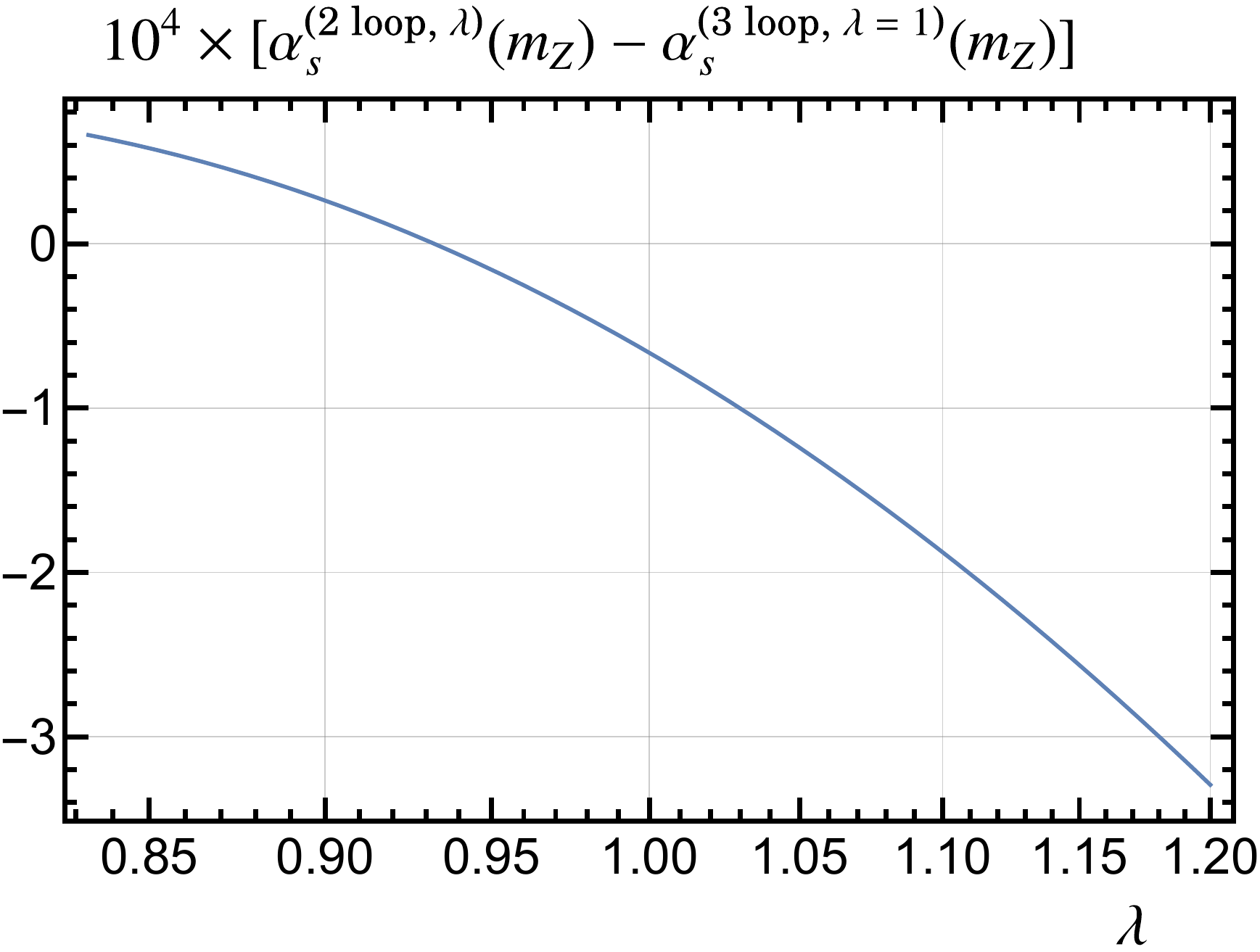}\\[15pt]
\includegraphics[width=0.49\textwidth]{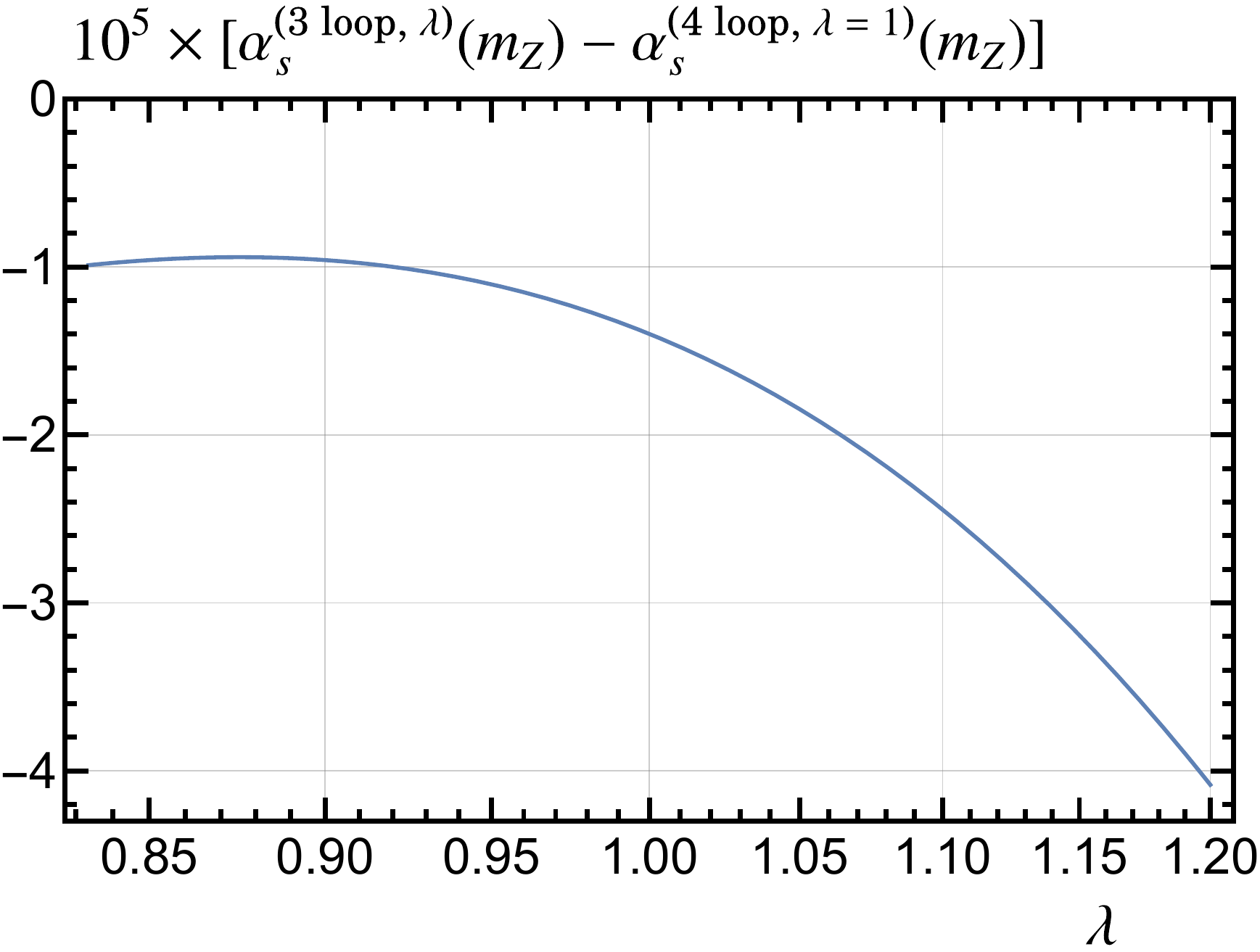}\hfill
\includegraphics[width=0.49\textwidth]{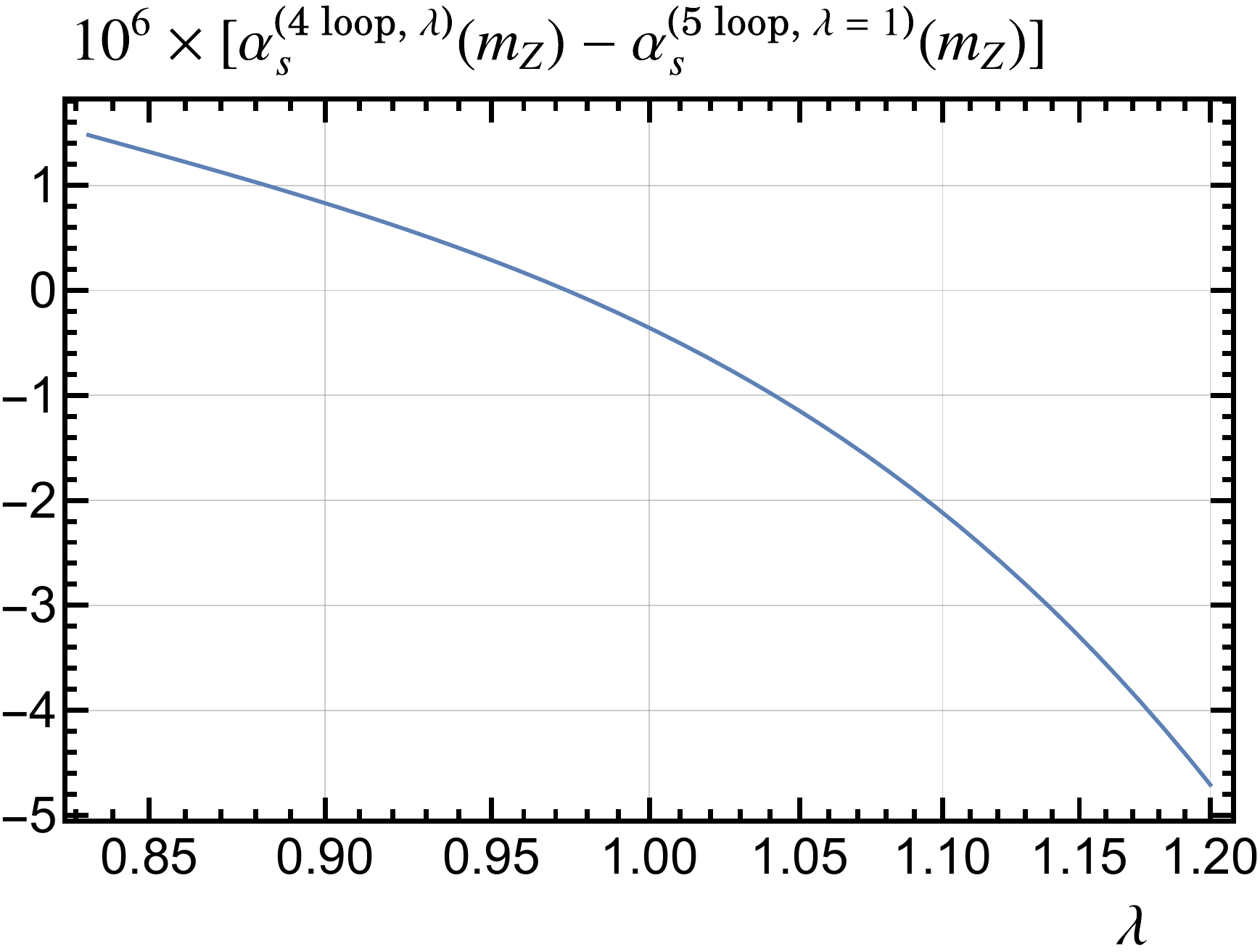}
\caption{\label{fig:lambda1} Values of $\alpha^{(5)}_s(m_Z)$ over the $\beta$-function scaling parameter $\lambda$, obtained by evolving the strong coupling down from $\alpha^{(5)}_s(1508.04\,\mathrm{GeV})$ with different loop orders in the evolution equation. The value of $\alpha^{(5)}_s(m_Z)$ computed by including one more perturbative order in the evolution with $\lambda=1$ is subtracted to highlight for which ranges of $\lambda$ both results for $\alpha^{(5)}_s(m_Z)$ coincide or come close. The results indicate that a variation for $\lambda$ of about $10\%$ around $1$ is a reasonable range to estimate the perturbative uncertainty.}
\end{figure}

A different approach to estimate the perturbative error of $\alpha_s^{(5)}(m_Z)$ is to vary the $\beta$-function scaling parameter $\lambda$ (controlled by the optional parameter \texttt{lambdaAlpha}) as described in Sec.~\ref{sec:core}. In Fig.~\ref{fig:lambda1} the value of $\alpha^{(5)}_s(m_Z)$ is shown, obtained by evolving the strong coupling down from $\alpha^{(5)}_s(1508.04\,\mathrm{GeV})$ with different loop orders in the evolution equation and with varying values of $\lambda$; the value of $\alpha^{(5)}_s(m_Z)$ obtained by including one more perturbative order in the evolution with $\lambda=1$ (which corresponds to the standard form of the $\beta$-function) is subtracted. We observe that, to reproduce a value near to the one obtained with one more perturbative order included in the running, $\lambda$ has to be varied by about $10\%$ around $1$, as already described in Sec.~\ref{sec:core}.

\begin{figure}[t!]
\centering
\includegraphics[width=0.49\textwidth]{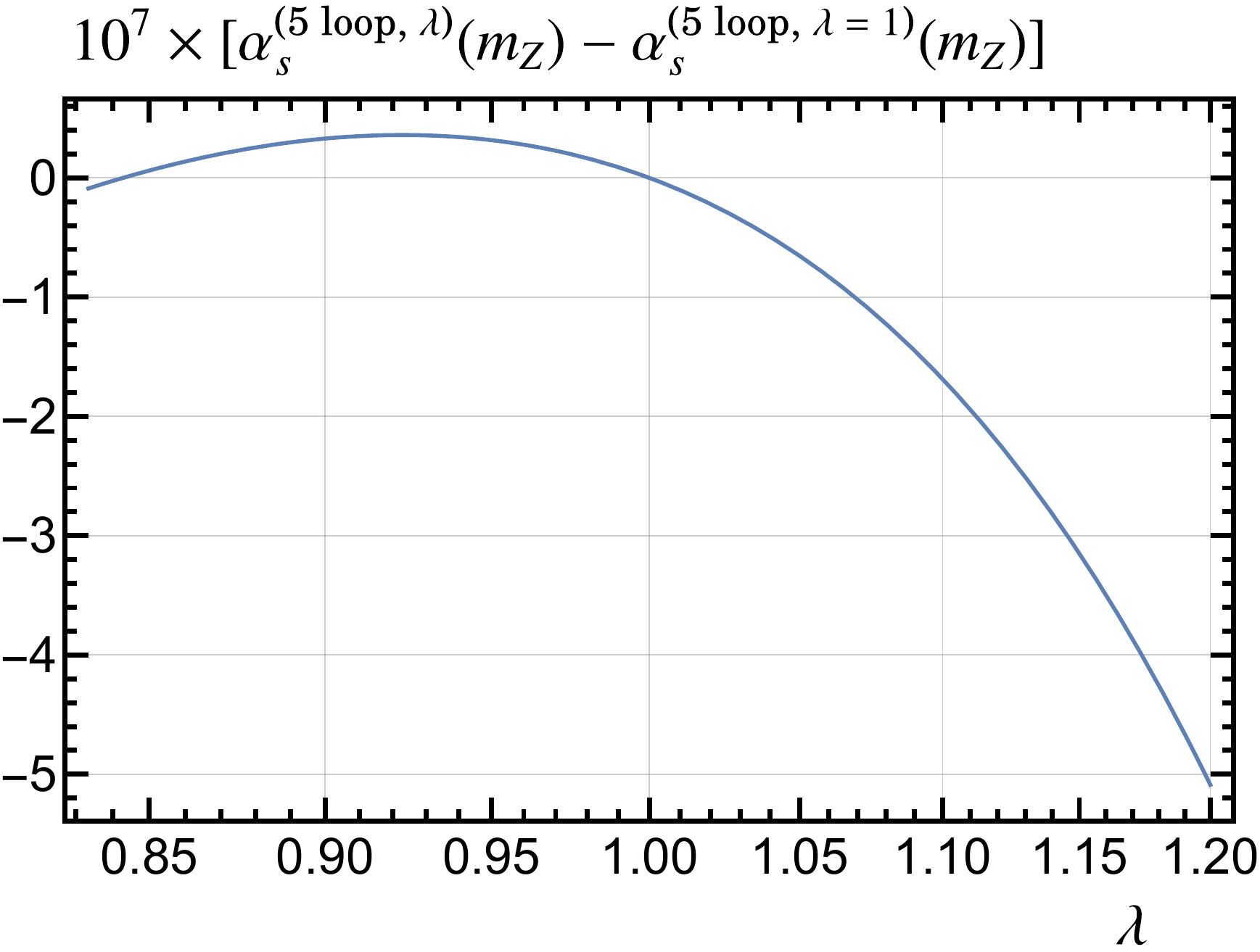}
\caption{\label{fig:lambda2} Variation of $\alpha^{(5)}_s(m_Z)$ with $\lambda$, obtained by evolving the strong coupling down from $\alpha^{(5)}_s(1508.04\,\mathrm{GeV})$ with 5-loop accuracy. The value of $\alpha_s(m_Z)$ for $\lambda=1$ is subtracted. Varying $\lambda$ in a range of about $10\%$ around $1$ the value of $\alpha_s(m_Z)$ changes by $10^{-7}$.}
\end{figure}

Assuming this to be also the appropriate range to estimate the perturbative uncertainty of the 5-loop result, the error can be estimated by scanning over $\lambda$ values. Employing \REvolver{} for this task, we first define a list of $20$ logarithmically distributed values of $\lambda$ in the appropriate range
\begin{verbatim}
  In[]:= lamList = 1.1^Range[-1, 1, 2/19];
\end{verbatim}
and create one \Core{} for each value in $\lambda$. For convenience, the \Core{} names are set to the associated $\lambda$ values
\begin{verbatim}
  In[]:= CoreCreate[ToString[#], {5, aQCentral, Q},
             lambdaAlpha -> #] & /@ lamList;
\end{verbatim}
The span of values of $\alpha^{(5)}_s(m_Z)$ corresponding to the range in $\lambda$ can then be obtained with
\begin{verbatim}
  In[]:= aLamList = AlphaQCD[ToString[#], mZdef] & /@ lamList;
\end{verbatim}
The central value of this range and the associated error are consequently given by
\begin{verbatim}
  In[]:= (Max[aLamList] + {1, -1} * Min[aLamList])/2
  Out[]= {0.11624517272452983, 1.0224197660724244*^-7}
\end{verbatim}
giving the same order of magnitude as the conservative approach. The variation in $\lambda$ of $\alpha^{(5)}_s(m_Z)$ when employing 5-loop running is also depicted in Fig.~\ref{fig:lambda2}.

\subsubsection{Complex renormalization scales}
\label{sec:alphacomplex}

In this application we illustrate the \REvolver{} functionality to determine the strong coupling $\alpha_s(\mu)$ at complex scales $\mu$. To this end we consider the analyses in Refs.~\cite{Beneke:2008ad,Boito:2016pwf}, where the perturbative QCD corrections $\delta^{(0)}$ to the inclusive $\tau$ hadronic width were considered in fixed-order perturbation theory (FOPT) as well as contour-improved perturbation theory (CIPT). In the second reference a scheme different from the usual $\MSbar$ definition is used for $\alpha_s$. We exploit this fact to demonstrate the ability of \REvolver{} to deal with user-defined $\beta$-function coefficients.

\subsubsection*{Hadronic $\tau$ decay}
\vspace{2mm}

In CIPT, determining $\delta^{(0)}$ involves integrating over powers of the strong coupling $\alpha_s^{(3)}(\mu)$ multiplied with a kinematic weight function in the complex $\mu^2$ plane along a circle with radius $m_\tau^2$
\begin{align}
\delta^{(0)}_\mathrm{CI} &= \sum_{n=1}^\infty c_{n,1} J_n^a(m_\tau^2)\,,\nonumber\\
J_n^a(m_\tau^2) &= \frac{1}{2\pi i} \oint_{|x|=1}\! \frac{\dd x}{x} (1-x)^3
(1+x) \biggl[\frac{\alpha_s^{(3)}(\sqrt{-m_\tau\, x})}{\pi}\biggr]^n\,,
\end{align}
where the coefficients $c_{n,1}$ are given in Eqs.~(2.13), (2.15) and (3.10) of Ref.~\cite{Beneke:2008ad}.
In that article, 4-loop running for the strong coupling was used, as well as the input value $\alpha_s^{(3)}(m_\tau)=0.34$. To reproduce the numbers for $\delta^{(0)}_\mathrm{CI}$ given in Eq.~(3.9) of Ref.~\cite{Beneke:2008ad} we first define variables accounting for the employed input values
\begin{verbatim}
 In[]:= {nfa, aTau, mTau} = {3, 0.34, 1.77686};
        {c11, c21, c31, c41, c51} =
          {1, 1.640, 6.371, 49.076, 283};
\end{verbatim}
and a function to compute $J_n^a(m_\tau^2)$
\begin{verbatim}
  In[]:= Ja[n_, core_] := 1/(2 Pi)
            NIntegrate[(1 - E^(I phi))^3 (1 + E^(I phi))
              (AlphaQCD[core, Sqrt[-mTau^2 E^(I phi)]]/Pi)^n,
             {phi, 0, 2 Pi}];
\end{verbatim}
where we change variables via $x=\exp(i\phi)$. Furthermore, we define a \Core{} with 4-loop running for the coupling
\begin{verbatim}
  In[]:= CoreCreate["CI4", {nfa, aTau, mTau}, runAlpha -> 4]
\end{verbatim}

All requirements are now set to reproduce the numbers given in Eq.~(3.9) of Ref.~\cite{Beneke:2008ad}
\begin{verbatim}
  In[]:= Ja[1, "CI4"]*c11
         Ja[2, "CI4"]*c21
         Ja[3, "CI4"]*c31
         Ja[4, "CI4"]*c41
         Ja[5, "CI4"]*c51
  Out[]= 0.14789839179248082 + 0. I
  Out[]= 0.02968556861749222 - 1.3583618952083769*^-18 I
  Out[]= 0.0121854520515486 + 4.947098114161149*^-18 I
  Out[]= 0.008592183988059407 - 8.468364884304664*^-19 I
  Out[]= 0.0037863076541974016` - 9.919281576456942*^-19 I
\end{verbatim}
with full agreement.

We can now easily inspect the corrections induced by the 5-loop running of the strong coupling by creating an appropriate \Core{}
\begin{verbatim}
  In[]:= CoreCreate["CI5", {nfa, aTau, mTau}, runAlpha -> 5];
\end{verbatim}
and evaluating
\begin{verbatim}
  In[]:= Ja[1, "CI5"]*c11
         Ja[2, "CI5"]*c21
         Ja[3, "CI5"]*c31
         Ja[4, "CI5"]*c41
         Ja[5, "CI5"]*c51
  Out[]= 0.14775354880331787 - 8.834874115176436*^-18 I
  Out[]= 0.02960219692750483 - 1.811149193611169*^-18 I
  Out[]= 0.012122455789576415 + 3.4080009230887917*^-18 I
  Out[]= 0.008521503950040327 - 1.6936729768609328*^-18 I
  Out[]= 0.0037376807938461127 - 1.5260433194549141*^-18 I
\end{verbatim}

We observe a small negative $\mathcal O(1\%)$ shift in the individual coefficients. Note that the explicit specification of the running order \texttt{runAlpha -> 5} at \Core{} creation is not mandatory since 5-loop running is the default. We included it to be explicit.

\subsubsection*{Hadronic $\tau$ decay with $C$-scheme strong coupling}
\vspace{2mm}

We now turn to Ref.~\cite{Boito:2016pwf} for which we reproduce the value quoted for $\delta^{(0)}_\mathrm{CI}$ in Eq.~(22). To obtain this number the 3-flavor strong coupling, the related $\beta$-function coefficients and the coefficients $c_{n,1}$ were converted to a class of schemes for the strong coupling where the $\beta$-function adopts the following exact all order form
\begin{equation}
\hat\beta(\hat\alpha_s) = -2\hat\alpha_s
\frac{\frac{\hat\alpha_s}{4\pi}\beta_0}{1-\frac{\hat\alpha_s}{4\pi}\frac{\beta_1}{\beta_0}}
\,=-2\hat\alpha_s \!\sum_{i=0} \hat \beta_i\biggl(\frac{\hat\alpha_s}{4\pi}\biggr)^{\!\!i+1}
\,,
\end{equation}
where $\hat \beta_i = \beta_0 (\beta_1/\beta_0)^i$.
Within that class of schemes one needs to fix
a parameter $C$ to uniquely specify the strong coupling. This parameter was set to $C=-1.246$ in Eq.~(22) of Ref.~\cite{Boito:2016pwf} with the argument that the unknown 5-loop coefficient can then be neglected
resulting in the value \mbox{$\delta^{(0)}_\mathrm{CI}(\hat\alpha(m_\tau), C=-1.246)=0.1840\pm0.0062$}. The quoted uncertainty refers to the size of the 4-loop correction term. Solving Eq.~(6) of the reference paper to convert the quoted value of the $\MSbar$ strong coupling \mbox{$\alpha_s^{(3)}(m_\tau)=0.316$} to the scheme described above leads to $\hat\alpha_s(m_\tau, C=-1.246)=0.477$. The transformed coefficients $c_{n,1}$ can be extracted from Eq.~(12) therein. \REvolver{} allows the order-by-order specification of user-defined $\beta$-functions, so we expand $\hat\beta(\hat\alpha_s)$ up to $\mathcal O (\hat\alpha_s^{11})$.

We define the relevant input values related to $\hat\alpha_s$, the coefficients $c_{n,1}$ and $\hat\beta_i$ for $n_f=3$ with
\begin{verbatim}
  In[]:= {nfa, aTau, mTau} = {3, 0.477, 1.77686};
         {c11, c21, c31, c41} =
           {1, 1.640 + 2.25 c, 7.682 + 11.38 c + 5.063 c^2, 
             61.06 + 72.08 c + 47.4 c^2 + 11.39 c^3} /.
            c -> -1.246;
         betaHat = 9. * (64/9)^Range[0, 9];
\end{verbatim}
and the related \Core{}
\begin{verbatim}
  In[]:= CoreCreate["Hat", {nfa, aTau, mTau}, betaHat]
\end{verbatim}

Reusing the function \verb|Ja[n_, core_]| as defined in the previous example we obtain
\begin{verbatim}
  In[]:= Ja[1, "Hat"]*c11 + Ja[2, "Hat"]*c21 +
          Ja[3, "Hat"]*c31 + Ja[4, "Hat"]*c41
  Out[]= 0.18403340012158337 - 1.0177988675705949*^-17 I
\end{verbatim}
for $\delta^{(0)}_\mathrm{CI}(\hat\alpha(m_\tau), C=-1.246)$ and
\begin{verbatim}
  In[]:= Ja[4, "Hat"]*c41
  Out[]= 0.006222359703784199 - 2.951274747776552*^-19 I
\end{verbatim}
for the last correction term. Both the central value and the size of the last correction term are in perfect agreement with Ref.~\cite{Boito:2016pwf}.

\subsubsection*{Cauchy integral theorem for the strong coupling}
\vspace{2mm}

It is worth mentioning that Cauchy's integral formula can be utilized to check the numerical quality of the coupling evolution routine implemented in \REvolver{}. For example, valuating $\alpha_s^{(3)}(2\,\mathrm{GeV})$ directly as well as by employing Cauchy's integral formula with a radius of $1$\,GeV gives equivalent results up to machine precision
\begin{verbatim}
  In[]:= mu0 = 2;
         aDirect = AlphaQCD["CI5", mu0]
         aResidue =
          1/(2 Pi) NIntegrate[AlphaQCD["CI5", mu0 + E^(I phi)],
            {phi, 0, 2 Pi}, PrecisionGoal -> 10]
         aResidue - aDirect
  Out[]= 0.3169005366613899
  Out[]= 0.31690053666139 + 1.1043592643970545*^-17 I
  Out[]= 1.1102230246251565*^-16 + 1.1043592643970545*^-17 I
\end{verbatim}

For all practical purposes the solution for the strong coupling evolution provided by \REvolver{} based on a given QCD $\beta$-function can be considered as exact.

\subsection{\Core{}s with one Massive Quark}

In the previous sample applications the impact of flavor thresholds was not considered. In the following we discuss examples where threshold effects associated to one massive quark are accounted for.

\subsubsection{Strong coupling with a flavor threshold}

\subsubsection*{Strong coupling from multijet events}
\vspace{2mm}

We consider Ref.~\cite{Aaboud:2017fml}, where the strong coupling value was determined by the ATLAS collaboration from transverse energy-energy correlations in multijet events based on $8$\,TeV LHC data. The measurements were made for different values of the transverse momentum sum $H_\mathrm{T2}$ of the two leading jets.
All values of $H_\mathrm{T2}$ considered in that analysis were much larger than the top quark mass, consequently the associated respective strong coupling values $\alpha_s^{(6)}(Q)$ are defined in the 6-flavor scheme. In Tabs.~2 and 3 of Ref.~\cite{Aaboud:2017fml} values for $\alpha_s^{(6)}(Q)$ and the associated results for $\alpha_s^{(5)}(m_Z)$ are quoted. For the strong coupling running an approximate analytic 2-loop solution of the evolution equation was used accounting for continuous matching at the top quark mass (which is correct for 1-loop matching when the matching scale is at the top quark mass).
In the following, we focus on the associated results $\alpha_s^{(5)}(m_Z)=0.1186^{+0.0090}_{-0.0047}$ and $\alpha_s^{(6)}(810\,\mathrm{GeV})=0.0907^{+0.0052}_{-0.0026}$, where for simplicity, we added the respective upper and lower uncertainties in quadrature.
We reproduce the asymmetric uncertainties for $\alpha_s^{(6)}(810\,\mathrm{GeV})$ for the given $\alpha_s^{(5)}(m_Z)$ range
and start by defining the relevant parameters for the 5-flavor coupling $\alpha_s^{(5)}(m_Z)$, where we set the top quark standard running mass to $\overline m_t = 163$\,GeV
\begin{verbatim}
  In[]:= {amZCentral, Q, mtmt} = {0.1186, 810, 163};
         {amZMax, amZMin} = amZCentral + {0.0090, -0.0047};
\end{verbatim}
Next we create the \Core{}s for our evaluation (including a \verb|CoreDeleteAll[]| to remove older cores)
\begin{verbatim}
  In[]:= CoreDeleteAll[]
         CoreCreate["central2", 6, {5, amZCentral, mZdef},
          {{6, mtmt, mtmt}}, runAlpha -> 2]
         CoreCreate["max2", 6, {5, amZMax, mZdef},
          {{6, mtmt, mtmt}}, runAlpha -> 2]
         CoreCreate["min2", 6, {5, amZMin, mZdef},
          {{6, mtmt, mtmt}}, runAlpha -> 2]
\end{verbatim}

To obtain the $6$-flavor strong coupling values at $Q=810$\,GeV including the error range we simply evaluate
\begin{verbatim}
  In[]:= aQCentral2 = AlphaQCD["central2", Q, 6]
         AlphaQCD[#, Q, 6] & /@ {"min2", "max2"} - aQCentral2
  Out[]= 0.09079931707610696
  Out[]= {-0.002757013433150532, 0.005129916277931454}
\end{verbatim}
agreeing very well with the result given in Ref.~\cite{Aaboud:2017fml}.

We investigate this setup with two-loop evolution further by determining the uncertainty related to varying the matching scale accounting for the top threshold corrections at one loop. This is easily done using \REvolver{} by creating a set of \Core{}s with a range of \texttt{f}-parameters. To this end we first define a table containing $20$ logarithmically scaled \texttt{f}-parameters in the range $[1/2, 2]$, which corresponds to matching scales between
one half and twice the standard running top mass. Subsequently the table is used to create the corresponding set of \Core{}s
\begin{verbatim}
  In[]:= list2 = 2^Range[-1, 1, 2/19];
  In[]:= CoreCreate[ToString[#] <> "run2", 6,
             {5, amZCentral, mZdef}, {{6, mtmt, mtmt}},
             runAlpha -> 2, orderAlpha -> 1,
             fMatch -> {#}] & /@ list2;
\end{verbatim}
Finally, we can determine a list of the corresponding strong coupling values $\alpha_s^{(6)}(810\,\mathrm{GeV})$ and compute the central value as well as the error range
\begin{verbatim}
  In[]:= aQList2 = AlphaQCD[ToString[#] <> "run2", Q, 6] & /@
            list2;
         (Max[aQList2] + {1, -1} * Min[aQList2])/2
  Out[]= {0.09081555469961156, 0.00009936201139525841}
\end{verbatim}
The perturbative uncertainties associated to the threshold corrections are about $3\%$ of the quoted experimental error.

Employing 5-loop instead of 2-loop running for the strong coupling and \mbox{4-loop} matching corrections we obtain
\begin{verbatim}
  In[]:= CoreCreate["central5", 6, {5, amZCentral, mZdef},
          {{6, mtmt, mtmt}}, runAlpha -> 5]
         CoreCreate["max5", 6, {5, amZMax, mZdef},
          {{6, mtmt, mtmt}}, runAlpha -> 5]
         CoreCreate["min5", 6, {5, amZMin, mZdef},
          {{6, mtmt, mtmt}}, runAlpha -> 5]
  In[]:= aQCentral5 = AlphaQCD["central5", Q, 6]
         AlphaQCD[#, Q, 6] & /@ {"min5", "max5"} - aQCentral5
  Out[]= 0.09078701609518454
  Out[]= {-0.00275523573924355, 0.0051259314534172346}
\end{verbatim}
for $\alpha_s^{(6)}(810\,\mathrm{GeV})$ and its upper and lower uncertainty and
\begin{verbatim}
  In[]:= CoreCreate[ToString[#] <> "run5", 6,
             {5, amZCentral, mZdef}, {{6, mtmt, mtmt}},
             runAlpha -> 5, fMatch -> {#}] & /@ list2;
  In[]:= aQList5 = AlphaQCD[ToString[#] <> "run5", Q, 6] & /@
            list2;
         (Max[aQList5] + {1, -1} * Min[aQList5])/2
  Out[]= {0.09078682756270738, 1.9445996186917558*^-7}
\end{verbatim}
for the central value and error estimate derived from varying the matching scale.
Within the experimental uncertainties as quoted in Ref.~\cite{Aaboud:2017fml} using 2-loop evolution and continuous matching is perfectly adequate.

\subsubsection*{Strong coupling from inclusive jet cross sections}
\vspace{2mm}

We return to the analysis of Ref.~\cite{Khachatryan:2016mlc} and investigate the impact of a top quark threshold on the values given there, staying with 2-loop running as employed in that reference. In the following we create \Core{}s with the given range of values of $\alpha_s^{(5)}(m_Z)=0.1162^{+0.0070}_{-0.0062}$
and determine the strong coupling $\alpha_s^{(6)}(1508.04\,\mathrm{GeV})$ in the 6-flavor scheme instead of the 5-flavor coupling $\alpha_s^{(5)}(1508.04\,\mathrm{GeV})$ determined in Ref.~\cite{Khachatryan:2016mlc}.
After defining the parameters
\begin{verbatim}
  In[]:= {amZCentral, Q, mtmt} = {0.1162, 1508.04, 163};
         {amZMax, amZMin} = amZCentral + {0.007, -0.0062};
\end{verbatim}
and creating the respective \Core{}s, accounting for the standard running top quark mass $\overline m_t = 163$\,GeV
and default 4-loop matching
\begin{verbatim}
  In[]:= CoreCreate["central2", 6, {5, amZCentral, mZdef},
           {{6, mtmt, mtmt}}, runAlpha -> 2];
         CoreCreate["min2", 6, {5, amZMin, mZdef},
           {{6, mtmt, mtmt}}, runAlpha -> 2];
         CoreCreate["max2", 6, {5, amZMax, mZdef},
           {{6, mtmt, mtmt}}, runAlpha -> 2];
\end{verbatim}
we obtain
\begin{verbatim}
  In[]:= aQCentral2 = AlphaQCD["central2", Q, 6]
         (AlphaQCD[#, Q, 6] & /@ {"min2","max2"}) - aQCentral2
  Out[]= 0.08406136347372324
  Out[]= {-0.003262229612787812, 0.0035626775539831235}
\end{verbatim}
for the central value and the upper and lower uncertainties, respectively. Comparing these values to $\alpha_s^{(5)}(1508.04\,\mathrm{GeV})=0.0822^{+0.0034}_{-0.0031}$, quoted in Ref.~\cite{Khachatryan:2016mlc}, we observe a positive shift of about $0.002$ in the central value. This already amounts to about $60\%$ of the given experimental error, illustrating that flavor-thresholds effects can lead to significant changes.

\subsubsection{Asymptotic pole and low-scale MSR mass}

To demonstrate the \REvolver{} functionalities related to mass conversions accounting for flavor threshold effects, we investigate how much the asymptotic top quark pole mass $m_t^{\rm pole}$ as well as the top MSR mass at $2$\,GeV $m_t^{\rm MSR}(2\,\mbox{GeV})$ would change if the strong coupling value quoted in Ref.~\cite{Khachatryan:2016mlc}
$\alpha_s(1508.04\,\mbox{GeV})=0.0822^{+0.0034}_{-0.0031}$ would be interpreted as a $6$-flavor compared to a $5$-flavor result.
After defining the relevant constants and \Core{}s (we keep using the values of \texttt{Q} and \texttt{mtmt} already defined in the previous section)
\begin{verbatim}
  In[]:= aQ = 0.0822;
         CoreCreate["5", 6, {5, aQ, Q}, {{6, mtmt, mtmt}}];
         CoreCreate["6", 6, {6, aQ, Q}, {{6, mtmt, mtmt}}];
\end{verbatim}
where we have employed default highest-order precision for coupling and mass evolution and flavor threshold matching,
we first extract $m_{t}^\mathrm{MSR}(2\,\mathrm{GeV})$ from both \Core{}s and then determine their difference:
\begin{verbatim}
  In[]:= mMSR25 = MassMS["5", 6, 2]
         mMSR26 = MassMS["6", 6, 2]
         mMSR25 - mMSR26
  Out[]= 172.45858180514585
  Out[]= 172.10074770514723
  Out[]= 0.3578340999986267
\end{verbatim}

For the asymptotic pole masses we obtain
\begin{verbatim}
  In[]:= mPole5 = MassPole["5", 6, mtmt]
         mPole6 = MassPole["6", 6, mtmt]
         mPole5 - mPole6
  Out[]= 172.8732843267365
  Out[]= 172.4868549739421
  Out[]= 0.3864293527944085
\end{verbatim}
where we adopted $\overline m_t=163$\,GeV as the conversion scale for the asymptotic pole mass determination.
The differences we obtain amount to $358$\,MeV and $386$\,MeV, both of which exceed the uncertainty of the current world average for direct top mass measurements~\cite{Zyla:2020zbs}.

\subsubsection{Bottom and charm quark short-distance masses}
\label{sec:bcshort}

\subsubsection*{Bottom $\MSbar$ mass at high scales}
\vspace{2mm}

In Ref.~\cite{Duhr:2020kzd} the partonic cross section for Higgs production via bottom quark fusion was presented. This involves the bottom running mass at the Higgs scale $m_H=125.09$\,GeV, quoted to be \mbox{$m_b^{(5)}(m_H)\equiv\overline m_b^{(5)}(m_H)=2.79$\,GeV} using 4-loop $\MSbar$-mass running, $\alpha_s^{(5)}(m_Z)=0.118$ and the bottom quark standard running mass $\overline m_b=4.18$\,GeV, see Table~2 therein. We can reproduce this relation in \REvolver{} by defining the relevant parameters
\begin{verbatim}
  In[]:= {mH, amZ, mbmb} = {125.09, 0.118, 4.18};
\end{verbatim}
and creating a corresponding \Core{}
\begin{verbatim}
  In[]:= CoreDeleteAll[]
         CoreCreate["b", 5, {5, amZ, mZdef}, {5, mbmb, mbmb},
            runMSbar -> 4]
\end{verbatim}

The value of $m_b^{(5)}(m_H)$ can now be extracted with
\begin{verbatim}
  In[]:= MassMS["b", 5, mH]
  Out[]= 2.78854676339097
\end{verbatim}
showing perfect agreement with Ref.~\cite{Duhr:2020kzd}.

\subsubsection*{Bottom quark PS and 1S masses}
\vspace{2mm}

In Ref.~\cite{Marquard:2015qpa} the conversion between the $\MSbar$ and various low-scale short-distance masses was carried out with 4-loop fixed-order formulae. To illustrate the functionalities of \REvolver{} for mass conversions we now consider the values of the PS and 1S bottom quark masses given in Table~II of that article.
We start defining the input parameters $\alpha_s^{(5)}(m_Z)=0.1185$, $\overline m_b=4.163$\,GeV and $\mu_f=2.0$\,GeV for the strong coupling, the standard running bottom mass and the bottom PS mass renormalization scale, respectively, as specified in Ref.~\cite{Marquard:2015qpa},
\begin{verbatim}
  In[]:= {amZ, mbmb, mufB} = {0.1185, 4.163, 2.0};
\end{verbatim}
and the related \Core{}
\begin{verbatim}
  In[]:= CoreCreate["b", 5, {5, amZ, mZdef},
          {{5, mbmb, mbmb}}, fMatch -> {2.0}]
\end{verbatim}

Following Ref.~\cite{Marquard:2015qpa} we set the bottom quark matching scale to twice the standard running bottom mass.
Comparing the 4-loop value for the bottom PS mass $m_b^{\rm PS}(2\,\mbox{GeV})=4.483$\,GeV quoted in Ref.~\cite{Marquard:2015qpa} and obtained by \REvolver{} we obtain
\begin{verbatim}
  In[]:= mbPSRef = 4.483;
         mbPSREvo = MassPS["b", 5, mufB, 5, mbmb, mbmb]
         mbPSREvo - mbPSRef
  Out[]= 4.484037803646092
  Out[]= 0.0010378036460920725
\end{verbatim}
where, following Ref.~\cite{Marquard:2015qpa}, the conversion is carried out using the fixed-order relation between the standard running and PS masses using the standard running mass $\overline m_b$ as the renormalization scale for the strong coupling.

The small difference of about $1$\,MeV results from the fact that in \REvolver{} {\it all conversions} between mass schemes are based on formulae starting from the value of the running mass, while in Ref.~\cite{Marquard:2015qpa} the conversion was obtained the other way around, i.e.\ the running mass was computed starting from a value in the PS scheme. The difference is naturally covered by the perturbative uncertainty and not relevant for practical purposes.

In the previous example we have used \REvolver{} to determine the PS from the standard running mass $\overline m_b$. The conversion in the opposite direction can be achieved using the \REvolver{} functionality to add a heavier mass to an existing \Core{}, see Sec.~\ref{sec:addmass}. To determine the standard running mass
we first create a \Core{} containing $4$ massless flavors
\begin{verbatim}
  In[]:= anf4mu3 = AlphaQCD["b", 3.0, 4];
         CoreCreate["O4", {4, anf4mu3, 3.0}]
\end{verbatim}
which requires the 4-flavor strong coupling $\alpha_s^{(4)}(3\,\mathrm{GeV})$ as an input, which here we obtain from the previously created \Core{} named \texttt{b}.

We now create a new \Core{} with name \texttt{b2} by adding the bottom PS mass \mbox{$m_b^{\rm PS}(2\,\mbox{GeV})=4.483$\,GeV} to the \Core{} named \texttt{O4}, reusing the parameters \texttt{mbPSRef} and \texttt{mufB} defined in the previous example as well as the matching scale factor $2.0$, specified by the option parameter \texttt{fnQ}
\begin{verbatim}
  In[]:= AddPSMass["O4", "b2", mbPSRef, mufB, 5, 4.2,
          4.2, fnQ -> 2.0]
\end{verbatim}
Internally, first the 5-flavor running mass is determined and subsequently used to create the 5-flavor \Core{} named \texttt{b2}. To match precisely the conversion method used in Ref.~\cite{Marquard:2015qpa} we set the running mass scale and the renormalization scale of the strong coupling for that conversion to $4.2$\,GeV here.

Reading out the resulting bottom quark standard running mass results in
\begin{verbatim}
  In[]:= MassMS["b2", 5]
  Out[]= 4.1621220500397325
\end{verbatim}
with the expected $1$\,MeV difference to the value $\overline m_b=4.163$\,GeV quoted in the reference paper.
This difference arises here again because the numerical conversion formulae employed in \REvolver{} are exactly invertible, i.e.\ they produce the same numerical mass differences regardless in which way the conversion is carried out.

Note that the default \REvolver{} routines convert from the MSR mass to a low-scale short-distance mass at the {\it intrinsic scale} of the low-scale short-distance mass (see Sec.~\ref{sec:sdm}) to resum potentially large logarithms involving the ratio of the quark mass and the renormalization scale. In the example above, we have, however, explicitly set the input parameter \texttt{nfConv} of the function \texttt{MassPS} to $5$ to enforce conversion at the scale of the $\MSbar$ mass (which is the approach used in Ref.~\cite{Marquard:2015qpa} and which does not resum these logarithms).
The result including log-resummation via R-evolution differs by around $15$\,MeV and can be extracted by using the corresponding \REvolver{} commands with default parameter settings. For example, converting from the standard running mass to the PS mass we obtain
\begin{verbatim}
  In[]:= MassPS["b", 5, mufB]
  Out[]= 4.468247788253957
\end{verbatim}
where \texttt{nfConv} is automatically set to $4$ and $R$ as well as the renormalization scale of the strong coupling are set to \texttt{mufB}.

The corresponding (log-resummed) conversion from the PS mass to the standard running mass is achieved by adding to the \Core{} named \texttt{O4} the bottom quark PS mass \mbox{$m_b^{\rm PS}(2\,\mbox{GeV})=4.483$\,GeV}, specifying log resummation via \mbox{R-evolution}. The associated new \Core{} named \texttt{b3} is created by
\begin{verbatim}
  In[]:= AddPSMass["O4", "b3", mbPSRef, mufB, fnQ -> 2.0]
\end{verbatim}
resulting in a bottom quark standard running mass of
\begin{verbatim}
  In[]:= MassMS["b3", 5]
  Out[]= 4.176206116625182
\end{verbatim}

At this point it should be mentioned that for bottom quarks large scale hierarchies cannot arise, such that the log-resummed conversion is not superior and the difference between the fixed-order and the log-resummed conversions may be better considered as a scheme variation. To illustrate this we show, order by order, the bottom quark PS mass computed in the fixed-order expansion as well as by utilizing R-evolution:
\begin{verbatim}
  In[]:= MassPS["b", 5, mufB, 5, mbmb, mbmb, 1.0, #] & /@
          {1, 2, 3, 4}
         MassPS["b", 5, mufB, 4, mufB, mufB, 1.0, #] & /@
          {1, 2, 3, 4}
  Out[]= {4.371486051575019, 4.452182487749768,
          4.484416838176134, 4.484037803646092}
  Out[]= {4.456398324160689, 4.487678320598545,
          4.494296775616467, 4.468247788253957}
\end{verbatim}

Next, we consider the 1S scheme. Comparing the 1S bottom quark mass $m_b^{\rm 1S}=4.670$\,GeV quoted Ref.~\cite{Marquard:2015qpa} and obtained in \REvolver{} using the relativistic counting and conversion at the high scale $\overline m_b$ (which again agrees with the approach used in Ref.~\cite{Marquard:2015qpa} and does not account for the resummation of logarithms involving the bottom mass and inverse Bohr radius) we obtain
\begin{verbatim}
  In[]:= mb1SRef = 4.670;
         mb1SRevo = Mass1S["b", 5, 5, mbmb, "relativistic"]
         mb1SRevo - mb1SRef
  Out[]= 4.671281708897272
  Out[]= 0.0012817088972720825
\end{verbatim}

Again, the $1$\,MeV discrepancy emerges from the perturbative difference between converting from or to the standard running mass. For comparison, we also present the corresponding 1S mass value accounting for log-resummation via R-evolution. This can be achieved by calling the routine \texttt{Mass1S} without any optional parameters. This implies that the default setting is used, namely that the 1S mass is converted from the MSR mass at the inverse Bohr radius $M_{q,B}$ (the intrinsic scale of the 1S mass, see Sec.~\ref{sec:sdm}) and that non-relativistic counting is applied. The bottom 1S mass is, however, sensitive to the ultra-soft scale $M^2_{q,B}/m_b$, related to logarithms of the strong coupling. This entails that for the conversion from the MSR mass (at the inverse Bohr radius) a scale larger than $M_{q,B}$ may be adopted as the renormalization scale for the strong coupling to avoid perturbative instabilities.\footnote{This issue does not arise for the top quark due to is large mass value.} Adopting $2M_{q,B}$ for the renormalization scale of the strong coupling, the 1S mass accounting for log-resummation is obtained by
\begin{verbatim}
  In[]:= MbB = MBohr["b", 5]
         Mass1S["b", 5, 4, MbB, "nonrelativistic", 2*MbB]
  Out[]= 1.8792648142678285
  Out[]= 4.6676688462595175
\end{verbatim}

The result differs by $4$\,MeV to the one using fixed-order conversion. As for the bottom PS mass, there is no conceptual improvement using log-resummation. This can again be seen from comparing the 1S mass values at lower orders. Order by order, the values of the bottom quark 1S mass, derived in fixed-order and with R-evolution, respectively, are given by
\begin{verbatim}
  In[]:= Mass1S["b", 5, 5, mbmb, "relativistic", mbmb, #] & /@
          {1, 2, 3, 4}
         Mass1S["b", 5, 4, MbB, "nonrelativistic", 2*MbB,
          #] & /@ {1, 2, 3, 4}
  Out[]= {4.516545706062886, 4.64055194539697,
          4.680062686477036, 4.671281708897272}
  Out[]= {4.609340303152403, 4.666301702298438,
          4.6809114692112495, 4.6676688462595175}
\end{verbatim}

Of course one can also convert from the 1S scheme to the standard running mass using the routine \texttt{Add1SMass}, shortly demonstrated in the following.

Reusing the \Core{} named \texttt{O4}, which contains $4$ massless quarks, we create a new \Core{} with name \texttt{b4} by adding the bottom 1S mass using fixed-order conversion
\begin{verbatim}
  In[]:= Add1SMass["O4", "b4", mb1SRef, 5, mbmb,
          "relativistic", mbmb, fnQ -> 2.0]
\end{verbatim}
which results in the following bottom quark standard running mass:
\begin{verbatim}
  In[]:= MassMS["b4", 5]
  Out[]= 4.161807379452769
\end{verbatim}
again with the expected $1$\,MeV difference to the value $\overline m_b=4.163$\,GeV quoted in Ref.~\cite{Marquard:2015qpa}, related to invertible numerical conversion algorithm employed in \REvolver{}. The analogous procedure for converting the 1S mass to the standard running mass using R-evolution can be performed by executing
\begin{verbatim}
  In[]:= Add1SMass["O4", "b5", mb1SRef, 4, MbB,
          "nonrelativistic", 2*MbB, fnQ -> 2.0]
  In[]:= MassMS["b5", 5]
  Out[]= 4.165169213761735
\end{verbatim}

\subsubsection*{Charm and bottom quark RS masses}
\vspace{2mm}

In Ref.~\cite{Peset:2018ria} the charm quark RS mass $m_c^{\rm RS}(\nu_f)=1.202$\,GeV at the scale \mbox{$\nu_f=1$\,GeV} was extracted from charmonium bound states masses, given in Eq.~(2.15) of that paper. The RS mass was converted to the standard running charm mass $\overline m_c=1.217$\,GeV with 4-loop fixed-order formulae for \mbox{$\alpha_s^{(5)}(m_Z)=0.1184$} and using $\mu_c=1.27$\,GeV as the charm threshold matching scale. The renormalization scale of the strong coupling was set to $\nu_c=1.5$\,GeV and the pole mass renormalon normalization constant for $n_f=3$ active flavors was quoted to be $N_m=0.5626$. This relation can be easily reproduced with \REvolver{}.
First, we set the relevant variables
\begin{verbatim}
  In[]:= {amZ, Nm} = {0.1184, 0.5626};
         {mcmc, mcmcMatch, nufc, nuc} =
          {1.217, 1.27, 1.0, 1.5};
         {mbmb, mbmbMatch} = {4.185, 4.2};
\end{verbatim}
and create the respective \Core{}
\begin{verbatim}
  In[]:= CoreDeleteAll[]
         CoreCreate["cb", 5, {5, amZ, mZdef},
          {{4, mcmc, mcmc}, {5, mbmb, mbmb}}, 
          fMatch -> {mcmcMatch/mcmc, mbmbMatch/mbmb}]
\end{verbatim}

We include the standard running bottom mass $\overline m_b=4.185$\,GeV with the corresponding threshold matching scale $\mu_b=4.2$\,GeV to enable automatic matching of the strong coupling constant which is given at $\mu=m_Z$. The charm quark RS mass computed by \REvolver{} and the difference to the value quoted in the reference paper are given by
\begin{verbatim}
  In[]:= mcRSRef = 1.202;
         mcRSREvo = MassRS["cb", 4, nufc, 4, mcmc, nuc,
           4, 4, Nm]
         mcRSRef - mcRSREvo
  Out[]= 1.201144351209422
  Out[]= 0.0008556487905779786
\end{verbatim}
showing agreement at the sub-MeV level.

In Ref.~\cite{Peset:2018ria} also the bottom quark RS mass value $m_b^{\rm RS}(\nu_f)=4.379$\,GeV at the scale $\nu_f=2$\,GeV was extracted from bottomonium meson masses [\,see Eq.~(2.3) of that paper\,] and subsequently converted to the standard running bottom mass $\overline m_b=4.379$\,GeV with 4-loop fixed-order formulae using $\mu_b=4.2$\,GeV as the bottom threshold matching scale and $\nu_b=2.5$\,GeV as the renormalization scale of the strong coupling. The charm mass corrections in the perturbative relation between the RS and the $\MSbar$ mass have been implemented in an effective way
by evaluating the conversion series that relates the $\MSbar$ and the RS masses for the bottom quark and $n_\ell$ massless quarks for $n_\ell=3$ dynamical flavors, but using the 3-flavor strong coupling
$\alpha_s^{(3)}$ computed with the charm mass threshold properly accounted for. This treats the charm quark as a decoupled flavor. The \REvolver{} RS mass routines do not directly provide this functionality, but in the following we show how this evaluation can still be carried out in \REvolver{}.
We first set the necessary parameters not yet defined before
\begin{verbatim}
  In[]:= {nufb, nub} = {2.0, 2.5};
\end{verbatim}
and compute the 3-flavor strong coupling $\alpha_s^{(3)}(\mu_c)$ at the charm matching scale $\mu_c=1.27$\,GeV, taking into account the bottom as well as charm quark thresholds in the usual way
\begin{verbatim}
  In[]:= as3 = AlphaQCD["cb", mcmcMatch, 3];
\end{verbatim}

Next, we define a \Core{} with decoupled charm quarks by using the precomputed $n_f=3$ strong coupling and specifying a total number of $4$ flavors, including $3$ massless flavors and the massive bottom quark as the 4-th flavor
\begin{verbatim}
  In[]:= CoreCreate["b", 4, {3, as3, mcmcMatch},
          {4, mbmb, mbmb}, fMatch -> {mbmbMatch/mbmb}]
\end{verbatim}

Setting all previously defined scales we get
\begin{verbatim}
  In[]:= mbRSRef = 4.379;
         mbRSREvo = MassRS["b", 4, nufb, 4, mbmb, nub,
           4, 4, Nm]
         mbRSREvo - mbRSRef
  Out[]= 4.379013026999629
  Out[]= 0.00001302699962923981
\end{verbatim}
for the bottom quark mass in the RS scheme with a negligible difference to the value quoted in the reference article.

We can also convert from the bottom RS to the standard running mass using the \REvolver{} routine \texttt{AddRSMass}. To do that, we first create a \Core{} containing the $3$ massless non-decoupled flavors with
\begin{verbatim}
  In[]:= CoreCreate["O3", {3, as3, mcmcMatch}]
\end{verbatim}
and subsequently create a new \Core{} with name \texttt{b2} by adding the bottom RS mass to the \Core{} with name \texttt{O3} setting the various input parameters in analogy to the example above, i.e.\ using fixed-order conversion
\begin{verbatim}
  In[]:= AddRSMass["O3", "b2", mbRSRef, nufb, 4, 4.2, nub, 4,
          4, Nm]
\end{verbatim}
Internally, first the 5-flavor running mass is determined and subsequently used to create the \Core{} named \texttt{b2}. We set the running mass scale for that conversion to $4.2$\,GeV.

Extracting the bottom quark standard running mass from the newly created \Core{} results in
\begin{verbatim}
  In[]:= MassMS["b2", 4]
  Out[]= 4.184987620121892
\end{verbatim}

\subsubsection*{Bottom quark kinetic masses}
\vspace{2mm}

In Ref.~\cite{Fael:2020iea} the 3-loop corrections to the perturbative relation between the pole and kinetic quark mass schemes have been determined for the case of one massive quark and $n_\ell$ massless quarks. Here we show how to employ \REvolver{} to reproduce the values for the bottom quark kinetic mass $m_b^{\rm kin}(\mu)$ at the scale $\mu=1$\,GeV given in Eq.~(8) of that paper, obtained by converting from the standard running mass $\overline m_b=4.163$\,GeV.

The conversion from $\overline m_b$ to the bottom kinetic mass was considered using the perturbative series computed with the massive bottom quark and $n_\ell$ massless quarks for $n_\ell=4$ (i.e.\ with the charm quark treated as massless) with the result $m_b^{\rm kin}(\mu)=4.523$\,GeV, as well as for $n_\ell=3$ (i.e.\ with the charm quark treated as decoupled) with the result $m_b^{\rm kin}(\mu)=4.521$\,GeV. The first case can be reproduced in a straightforward way with \REvolver{} as it corresponds to a realistic physical scenario.

First we define the relevant parameters
\begin{verbatim}
  In[]:= {mbmb, amZ, muCutb} = {4.163, 0.1179, 1.0};
\end{verbatim}
and create a \Core{} in which the charm quark is treated as massless
\begin{verbatim}
  In[]:= CoreDeleteAll[];
         CoreCreate["b4", 5, {5, amZ, mZdef},
          {{5, mbmb, mbmb}}]
\end{verbatim}

The value obtained by \REvolver{} and the difference to the value quoted in Ref.~\cite{Fael:2020iea} is
\begin{verbatim}
  In[]:= mKinb4Ref = 4.523;
         mKinb4REvo = MassKin["b4", 5, muCutb, 5, mbmb, mbmb]
         mKinb4REvo - mKinb4Ref
  Out[]= 4.523457226246508
  Out[]= 0.0004572262465085686
\end{verbatim}
i.e.\ there is perfect agreement.

The second case is in close analogy to the treatment of the RS mass just discussed above and requires that the conversion series, which relates the $\MSbar$ and the kinetic mass for the bottom quark and $n_\ell$ massless quarks, is evaluated for $n_\ell=3$ dynamical flavors, but using the 3-flavor strong coupling
$\alpha_s^{(3)}$ computed with the charm mass threshold properly accounted for. \REvolver{} does not provide functionality to carry out this conversion directly, but in the following we show how this evaluation can still be performed.

To obtain the $n_f=3$ strong coupling in the usual way, we first create a new \Core{} involving a massive bottom as well as charm quark
\begin{verbatim}
  In[]:= mcmc = 1.263;
         CoreCreate["bc", 5, {5, amZ, mZdef},
          {{4, mcmc, mcmc}, {5, mbmb, mbmb}}]
\end{verbatim}
and extract $\alpha_s^{(3)}(\overline m_b)$
\begin{verbatim}
  In[]:= a3mbmb = AlphaQCD["bc", mbmb, 3];
\end{verbatim}

Next, we create the \Core{} to be employed for the mass scheme conversion. We use the computed $n_f=3$ strong coupling and specify a total number of $4$ flavors with $3$ massless flavors and the massive bottom quark
\begin{verbatim}
  In[]:= CoreCreate["b3", 4, {3, a3mbmb, mbmb},
          {{4, mbmb, mbmb}}]
\end{verbatim}

The value of the kinetic bottom quark mass computed by \REvolver{} and the difference to the corresponding result quoted in Ref.~\cite{Fael:2020iea} can now be obtained by
\begin{verbatim}
  In[]:= mKinb3Ref = 4.521;
         mKinb3REvo = MassKin["b3", 4, muCutb, 4, mbmb, mbmb]
         mKinb3Ref - mKinb3REvo
  Out[]= 4.520784332346221
  Out[]= 0.00021566765377922792
\end{verbatim}
The numbers are in perfect agreement.

In Ref.~\cite{Fael:2020njb} the lighter flavor mass corrections to the relation between the pole and kinetic masses were computed explicitly up to ${\cal O}(\alpha_s^3)$. These corrections are implemented in \REvolver{}. They have the property that they exclusively come from the flavor number decoupling relations of the strong coupling, and they are also referred to as ``scheme B'' in that reference.
In Eq.~(79) of that reference the bottom quark kinetic mass $m_b^\mathrm{kin}(1\,\mathrm{GeV}) = 4.526$\,GeV was obtained
using \mbox{$\alpha_s^{(5)}(m_Z)=0.1179$} for the strong coupling, the charm quark running mass $\overline m_c(2\,\mathrm{GeV})=0.993$\,GeV and the bottom quark standard running mass \mbox{$\overline m_b=4.136$\,GeV}. For the calculation, fixed-order conversion was applied using the standard running bottom mass $\overline m_b$ as the renormalization scale of the strong coupling. To reproduce the result with \REvolver{} we define the input values with
\begin{verbatim}
  In[]:= {amZ, mc3, mbmb} = {0.1179, 0.993, 4.163};
\end{verbatim}
and the related \Core{} with
\begin{verbatim}
  In[]:= CoreDelete["bc"]
         CoreCreate["bc", 5, {5, amZ, mZdef},
           {{4, mc3, 3}, {5, mbmb, mbmb}}]
\end{verbatim}
The bottom quark kinetic mass extracted by \REvolver{} and its difference to the reference value are returned by
\begin{verbatim}
  In[]:= mbKinRef = 4.526;
         mbKinREvo = MassKin["bc", 5, 1.0, 5, mbmb, mbmb]
         mbKinRef - mbKinREvo
  Out[]= 4.527084826155039
  Out[]= -0.0010848261550391314
\end{verbatim}
The small deviation of $1$\,MeV originates from the fact that in Ref.~\cite{Fael:2020njb} the light massive flavor corrections related to the charm quark are parametrized in terms of $\overline m_c(3\,\mathrm{GeV})$, while in \REvolver{} they are parametrized in terms of the charm standard running mass $\overline m_c$.

Performing the same conversion with R-evolution to resum logarithms of the intrinsic physical scales of the mass schemes gives
\begin{verbatim}
  In[]:= MassKin["bc", 5, 1.0]
  Out[]= 4.53472077059131
\end{verbatim}
i.e.\ the bottom quark kinetic mass with log-resummation is larger by about $7$\,MeV. We note that, following Ref.~\cite{Fael:2020njb},
the default value for the kinetic mass intrinsic scale (where the default log-resummed conversion between the running and the kinetic mass is carried out) is set to be twice its renormalization scale, that is, $2$\,GeV in the example above.

\subsubsection{Top quark mass at low scales}
\label{sec:topsmallscales}

We return to Ref.~\cite{Marquard:2015qpa} and investigate the top quark mass conversion between the $\MSbar$ and various short-distance schemes with 4-loop accuracy. This concerns Table~I of that article.

Following Ref.~\cite{Marquard:2015qpa} we first define the values for the strong QCD coupling \mbox{$\alpha_s^{(5)}(m_Z)=0.1185$}, the top quark standard running mass $\overline m_t= 163.643$\,GeV and the renormalization scale of the PS mass $\mu_f=20$\,GeV
\begin{verbatim}
  In[]:= {amZ, mtmt, mufT} = {0.1185, 163.643, 20.0};
\end{verbatim}
and a \Core{} with a top threshold matching scale of twice the top standard running mass
\begin{verbatim}
  In[]:= CoreDeleteAll[]
         CoreCreate["t", 6, {5, amZ, mZdef}, {{6, mtmt, mtmt}},
          fMatch -> {2.0}]
\end{verbatim}

In analogy to Sec.~\ref{sec:bcshort}, we evaluate the PS and 1S top quark mass employing \REvolver{} and determine the difference to the values $m_t^{\rm 1S}=172.227$\,GeV and $m_t^{\rm PS}(20\,\mbox{GeV})=171.792$\,GeV quoted in the reference paper by converting directly from the $\MSbar$ scheme, leading to
\begin{verbatim}
  In[]:= mtPSRef = 171.792;
         mtPSRevo = MassPS["t", 6, mufT, 6, mtmt, mtmt]
         mtPSRevo - mtPSRef
  Out[]= 171.7950829141975
  Out[]= 0.0030829141975061702
\end{verbatim}
for the PS scheme, and
\begin{verbatim}
  In[]:= mt1SRef = 172.227;
         mt1SRevo = Mass1S["t", 6, 6, mtmt, "relativistic"]
         mt1SRevo - mt1SRef
  Out[]= 172.2298841816133
  Out[]= 0.0028841816132967324
\end{verbatim}
for the 1S scheme, respectively.

As already described in Sec.~\ref{sec:bcshort}, the small difference (of about $3$\,MeV) results from the
fact that in \REvolver{} conversion is based on formulae where the (standard) running mass is taken as the input, while in
Ref.~\cite{Marquard:2015qpa} the conversion was carried out starting from the PS and 1S masses.

We now investigate the influence of large logarithms of the ratio between the top quark mass and the intrinsic scale of the respective short-distance schemes. For the top quark the impact of the summation of these logarithms can be significant. To this end we consider the convergence of the perturbative series relating the standard running top mass $\overline m_t$ to the PS mass $m_t^{\rm PS}(20\,\mbox{GeV})$, the 1S mass $m_t^{\rm 1S}$ as well as the running mass $m_t^{(5)}(2\,\mathrm{GeV})$.

Using fixed-order conversion (without log-resummation) from $\overline m_t$ and using $\overline m_t$ as the renormalization scale, the top PS mass at $4$ loops, and the corresponding corrections of order $\mathcal{O}(\alpha_s^n)$ with $1\leq n\leq 4$ amount to
\begin{verbatim}
  In[]:= MassPS["t", 6, mufT, 6, mtmt, mtmt]
         Table[
          MassPS["t", 6, mufT, 6, mtmt, mtmt, 1.0, n] - 
           MassPS["t", 6, mufT, 6, mtmt, mtmt, 1.0, n - 1],
          {n, 1, 4}]
  Out[]= 171.7950829141975
  Out[]= {6.636196514422409, 1.199280138792858,
          0.26779429813998945, 0.048811962842222556}
\end{verbatim}

On the other hand, employing R-evolution to first evolve down to the PS mass renormalization scale $\mu_f=20$\,GeV and then converting to the PS mass at that scale, which is \REvolver{}s default procedure, we obtain
\begin{verbatim}
  In[]:= MassPS["t", 6, mufT]
         Table[
          MassPS["t", 6, mufT, 5, mufT, mufT, 1.0, n] - 
           MassPS["t", 6, mufT, 5, mufT, mufT, 1.0, n - 1],
          {n, 1, 4}]
  Out[]= 171.79912206134247
  Out[]= {0., 0.06895809427228983,
          0.001838267705210228, -0.01976315025407871}
\end{verbatim}

We observe that the corrections are considerably smaller when R-evolution is accounted for and that there is a difference of around \mbox{$4$\,MeV} in the final 4-loop converted values for the two approaches.
The $\mathcal O(\alpha_s)$ correction term is zero in the case of the log-resummed conversion since the 1-loop perturbative coefficients of the PS and MSR masses coincide.

For the conversion to the 1S mass the observation regarding fixed-order versus log-resummed conversion is similar. In the fixed-order case we get
\begin{verbatim}
  In[]:= Mass1S["t", 6, 6, mtmt, "relativistic", mtmt]
         Table[
          Mass1S["t", 6, 6, mtmt, "relativistic", mtmt, n] -
           Mass1S["t", 6, 6, mtmt, "relativistic",
            mtmt, n - 1], {n, 1, 4}]
  Out[]= 172.2298841816133
  Out[]= {7.129280949172028, 1.227243021981991,
          0.21914263057425387, 0.01121757988499894}
\end{verbatim}
while the log-resummed evaluation gives again substantially smaller corrections:
\begin{verbatim}
  In[]:= mBohr = MBohr["t", 6]
         Mass1S["t", 6, 5]
         Table[
          Mass1S["t", 6, 5, mBohr, "nonrelativistic",
            mBohr, n] - Mass1S["t", 6, 5, mBohr,
            "nonrelativistic", mBohr, n - 1], {n, 1, 4}]
  Out[]= 32.11684612826197
  Out[]= 172.20797480544945
  Out[]= {1.1666583987086199, 0.18536848855083576,
          -0.047109547951663444, -0.010351024770614004}
\end{verbatim}
where the inverse Bohr radius, i.e.\ the intrinsic scale of the 1S mass, is shown as the first output for completeness.

Finally, we investigate the effect of log-resummation on the computation of the top quark $5$-flavor running mass at \mbox{$2$\,GeV}, $m_t^{(5)}(2\,\mathrm{GeV})$, when converting from $m_t^{(5)}(\overline m_t)$. For simplicity we create a new \Core{} named \texttt{t}, this time where for the top threshold matching scale the default value, the top quark standard running mass, is adopted:
\begin{verbatim}
  In[]:= CoreDelete["t"]
         CoreCreate["t", 6, {5, amZ, mZdef},
           {{6, mtmt, mtmt}}];
\end{verbatim}
Now we extract the value of the 5-flavor MSR mass $m_t^{(5)}(\overline m_t)$ at the scale of the standard top running mass from this \Core{} as a reference
\begin{verbatim}
  In[]:= m5mt = MassMS["t", 6, mtmt, 5]
  Out[]= 163.67571467667918
\end{verbatim}

\REvolver{} does not provide fixed-order conversions of the running masses between two different renormalization scales. It is, however, possible to access these fixed-order corrections through the pole mass routine \texttt{MassPoleFO}. Care has to be taken that the two calls of the \texttt{MassPoleFO} routines involve the same renormalization scale (which here is the standard running mass) to ensure that the pole mass renormalon is properly canceled:
\begin{verbatim}
  In[]:= FOTab =
           MassMS["t", 6, 2.0, 5] -
              MassPoleFO["t", 6, 5, 2.0, mtmt, #] +
              MassPoleFO["t", 6, 5, mtmt, mtmt, #] & /@
            {0, 1, 2, 3, 4};
         FOTab[[5]]
         Table[FOTab[[n]] - FOTab[[n - 1]], {n, 2, 5}]
  Out[]= 173.2337130005733
  Out[]= {7.46778271952536, 1.5242807051151317,
          0.4246371655552821, 0.14129773369833742}
\end{verbatim}
The difference of the two calls of the \texttt{MassPoleFO} routine removes the log-resummed corrections from the output of \texttt{MassMS} and replaces it by the corresponding fixed-order terms.
As for the above examples, we first display the $m_t^{(5)}(2\,\mathrm{GeV})$ value at $\mathcal O(\alpha_s^4)$ and the fixed-order perturbative correction terms at order $\mathcal{O}(\alpha_s^n)$ with $1\leq n\leq 4$.

For the investigation of the correction terms in the case of R-evolution we create four \Core{}s, setting the respective loop order $n$ of the R-evolution equation to $1\leq n\leq 4$
\begin{verbatim}
  In[]:= CoreCreate["t-R" <> ToString[#], 6, {5, amZ, mZdef},
             {{6, mtmt, mtmt}}, runMSR -> #] & /@ {1, 2, 3, 4};
\end{verbatim}
The four \Core{}s differ only by the loop order used for \mbox{R-evolution}, but employ the default 4-loop matching to determine
$m_t^{(5)}(\overline m_t)$ from the standard running mass $\overline m_t$.
The value of $m_t^{(5)}(2\,\mathrm{GeV})$ corresponding to \mbox{$4$-loop} running and the corrections coming from the individual running orders can be extracted with
\begin{verbatim}
  In[]:= MassMS["t-R4", 6, 2.0]
         RevoTab = MassMS["t-R" <> ToString[#], 6, 2.0] & /@
            {1, 2, 3, 4};
         RevoTab = Prepend[RevoTab, m5mt];
         Table[RevoTab[[n]] - RevoTab[[n - 1]], {n, 2, 5}]
  Out[]= 173.36186564820926
  Out[]= {8.841888795217045, 0.8531892592812653,
          0.035471795296302844, -0.044398878264530595}
\end{verbatim}
where the first entry in the curly brackets is the difference between $m_t^{(5)}(\overline m_t)$ and $m_t^{(5)}(2\,\mathrm{GeV})$ obtained with 1-loop R-evolution, while the subsequent entries refer to the differences to the previous order in $m_t^{(5)}(2\,\mathrm{GeV})$ obtained when adding the 2-, 3- and 4-loop terms to the \mbox{R-evolution} anomalous dimension.
Once again we observe that the convergence is much better when using R-evolution, and that the corresponding results are more precise than using fixed-order conversion.
Using 4-loop R-evolution leads to a value for $m_t^{(5)}(2\,\mathrm{GeV})$ that is about $130$\,MeV higher than when using 4-loop fixed-order conversion. This difference is consistent with the size of the 4-loop fixed-order correction of $141$\,MeV, when adopting the latter as an uncertainty for the \mbox{4-loop} fixed-order conversion.

We see that the higher order corrections to the R-evolution anomalous dimension lead to very small effects, so that one could worry that their size may not reflect the perturbative uncertainty at the corresponding loop order. A different way to estimate perturbative uncertainties is to perform $\lambda$-variation in the R-evolution equation, which we shortly demonstrate in the following.

We create a set of \Core{}s with $\lambda$ values in the range $1/2 \le \lambda \le 2$ and for various loop orders in the R-evolution equation. The \Core{}s are created with
\begin{verbatim}
  In[]:= list2 = 2^Range[-1, 1, 2/49];
         Outer[
           CoreCreate["t-l" <> ToString[#2] <> "-R" <>
              ToString[#1], 6, {5, amZ, mZdef},
             {{6, mtmt, mtmt}}, lambdaMSR -> #2,
             runMSR -> #1] &, {1, 2, 3, 4}, list2];
\end{verbatim}
and the table containing the corresponding values of $m_t^{(5)}(2\,\mathrm{GeV})$ is generated by
\begin{verbatim}
  In[]:= mListRevo = 
           Outer[
            MassMS["t-l" <> ToString[#2] <> "-R" <>
               ToString[#1], 6, 2.0, 5] &, {1, 2, 3, 4},
            list2];
\end{verbatim}
Order by order, the values of the $\lambda$ variations divided by two are then given by
\begin{verbatim}
  In[]:= Table[(Max[mListRevo[[n]]] - Min[mListRevo[[n]]])/2,
          {n, 1, 4}]
  Out[]= {1.082493387031036, 0.2525991201191715,
          0.041115232629010734, 0.02465406235192802}
\end{verbatim}
The values of $\lambda$ variations are consistent with the size of the corrections to the \mbox{R-evolution} equation and substantially smaller than the size of the fixed-order corrections, illustrating that resumming logarithms via \mbox{R-evolution} leads to a more precise mass conversion than using fixed-order corrections.

\subsubsection{Running masses for complex renormalization scales}

\REvolver{} provides the functionality to determine running masses at complex renormalization scales.
In analogy to Sec.~\ref{sec:alphacomplex}, we demonstrate the quality of the running mass evolution in the complex plane by showing numerical consistency for the high-scale ($\MSbar$) running top quark mass concerning the Cauchy's integral formula, where we adopt $\mathtt{muMS}=350$\,GeV for the central scale and $174$\,GeV for the radius of the Cauchy integral.
We set up a \Core{} with the top quark standard running mass \mbox{$\overline m_t=163$\,GeV} and subsequently determine the central running mass and the corresponding value using the residue theorem
\begin{verbatim}
  In[]:= CoreCreate["O", 6, {5, amZdef, mZdef},
          {{6, 163, 163}}]
  In[]:= {muMS, rMS} = {350, 174};
         mMSDirect = MassMS["O", 6, muMS, 6]
         mMSResidue =
          1/(2 Pi)
           NIntegrate[MassMS["O", 6, muMS + rMS E^(I phi), 6],
            {phi, 0, 2 Pi}, PrecisionGoal -> 10]
         mMSResidue - mMSDirect
  Out[]= 154.072755231406
  Out[]= 154.07275523140612 + 4.240739575284689*^-16 I
  Out[]= -1.1368683772161603*^-13 - 4.240739575284689*^-16 I
\end{verbatim}
Both results are in perfect agreement reflecting the high numerical precision of the evolution routines implemented in \REvolver{}.

This high numerical precision is also maintained at much smaller renormalization scales (where perturbation theory in general is less reliable).
We demonstrate this for the top quark running (MSR) mass at a central scale of $2$\,GeV and a radius of $1$\,GeV for the Cauchy integral:
\begin{verbatim}
  In[]:= muR = 2;
         mRDirect = MassMS["O", 6, muR, 5]
         mRResidue =
          1/(2 Pi)
           NIntegrate[MassMS["O", 6, muR + E^(I phi), 5],
            {phi, 0, 2 Pi}, PrecisionGoal -> 10]
         mRResidue - mRDirect
  Out[]= 172.64605095372784
  Out[]= 172.64605095372804 + 4.417437057588218*^-18 I
  Out[]= 1.9895196601282805*^-13 + 4.417437057588218*^-18 I
\end{verbatim}

\subsubsection{Top quark pole masses}
\label{sec:expoleml}

The pole mass scheme suffers from an ${\cal O}(\Lambda_{\rm QCD})$ renormalon, which entails that its perturbative relation to a short-distance mass involves a factorially divergent perturbative series and an associated ambiguity in its value \cite{Bigi:1994em,Beneke:1994sw,Beneke:2016cbu,Hoang:2017suc}. The pole mass (and its value) can either be treated as an order-dependent concept or one can assign its value to be in the region where the perturbative series reaches its minimal correction term (and the
partial sum of the perturbative series increases linearly), sometimes called the asymptotic region. We call the latter value the ``asymptotic pole mass''. The associated ambiguity arises from the principle ignorance where precisely to truncate the partial sum within the asymptotic region. Different lines of reasoning have been proposed concerning the determination of the asymptotic value and the associated ambiguity, see Refs.~\cite{Hoang:2017suc,Beneke:2016cbu} for recent analyses.

\REvolver{} provides functionalities to extract the order-dependent as well as the asymptotic pole mass value from a \Core{}. In case of the asymptotic value and the associated ambiguity, \REvolver{} provides routines employing a number of different methods and allowing for various options as explained in Sec.~\ref{sec:pole}. In the following we demonstrate some of these functionalities reproducing results quoted in Refs.~\cite{Beneke:2016cbu} and \cite{Hoang:2017suc} focusing on the scenario of the top quark with massless bottom and charm quarks. Scenarios with massive bottom and charm quarks are treated in Sec.~\ref{sec:polelmq}.

\subsubsection*{Minimal correction approach for the pole mass ambiguity}
\vspace{2mm}

First, we consider Ref.~\cite{Beneke:2016cbu} where the top quark pole mass renormalon ambiguity quoted in Eq.~(4.7) is determined from the perturbative series between the pole mass and the standard running mass for the case where the bottom and charm mass effects are ignored.
To reproduce the result we define the top standard running mass value used in Ref.~\cite{Beneke:2016cbu}
\begin{verbatim}
  In[]:= mtmt = 163.508;
\end{verbatim}
and create an associated \Core{}
\begin{verbatim}
  In[]:= CoreDeleteAll[]
         CoreCreate["t", 6, {5, amZdef, mZdef},
          {{6, mtmt, mtmt}}]
\end{verbatim}

We execute the function \texttt{MassPoleDetailed} to extract the asymptotic pole mass value, the ambiguity, and the order of the smallest correction term. The method adopted in Ref.~\cite{Beneke:2016cbu} to determine the pole mass ambiguity was associated to the size of the minimal correction and is accessed in \REvolver{} specifying the \texttt{min} method:
\begin{verbatim}
  In[]:= MassPoleDetailed["t", 6, mtmt, mtmt, "min"]
  Out[]= {173.60839591880665, 0.06443134543422957, 8}
\end{verbatim}

Furthermore the third and fourth arguments are set to the standard running top mass \texttt{mtmt} to mimic the choices adopted in Ref.~\cite{Beneke:2016cbu}. The choice of arguments when calling the function \texttt{MassPoleDetailed} ensures that the analyzed series is the one for the difference between pole and the (5-flavor) MSR mass at the scale \texttt{mtmt} and that the renormalization scale for the strong coupling is \texttt{mtmt} as well.
The values for the asymptotic pole mass and its ambiguity quoted in Eq.~(4.8) in Ref.~\cite{Beneke:2016cbu} are $173.608$\,GeV and $67$\,MeV, respectively. The asymptotic value is in perfect agreement. The small discrepancy of $3$\,MeV in the ambiguity is related to the fact that \REvolver{} uses the perturbative series for the relation between the pole and the MSR mass for the routine \texttt{MassPoleDetailed} while in Ref.~\cite{Beneke:2016cbu} the relation between the pole and the standard running mass was considered.

\subsubsection*{Asymptotic series for the pole-$\MSbar$ mass relation}
\vspace{2mm}

The method of estimating higher order coefficients in the pole-$\MSbar$ relation employed by \REvolver{} is described in Sec.~4.4 of Ref.~\cite{Hoang:2017suc} and relies on an asymptotic formula that can, depending on the specified options, reproduce the exactly known coefficients, see Eq.~(\ref{eq:anasymptotic}) in Sec.~\ref{sec:asymptotic}.

The values of the asymptotic coefficients can be easily read out using the function \texttt{MassPoleFO}. For the orders $5$--$9$ \REvolver{} returns
\begin{verbatim}
  In[]:= (MassPoleFO["t", 6, 5, mtmt, mtmt, #] - 
              MassPoleFO["t", 6, 5, mtmt, mtmt, # - 1])/
            (mtmt * (AlphaQCD["t", mtmt, 5]/(4 Pi))^#) & /@
          Range[5, 9]
  Out[]= {1.4248584469453057*^7, 1.1661884055877335*^9,
          1.1323781055413771*^11, 1.2729796430260885*^13,
          1.6260702903901682*^15}
\end{verbatim}
for $a_n^\mathrm{MSR\,\prime}$ (i.e.\ the series coefficients for the MSR and pole mass difference) and
\begin{verbatim}
  In[]:= (MassPoleFO["t", 6, 6, mtmt, mtmt, #] - 
              MassPoleFO["t", 6, 6, mtmt, mtmt, # - 1])/
            (mtmt * (AlphaQCD["t", mtmt, 6]/(4 Pi))^#) & /@
          Range[5, 9]
  Out[]= {1.429074531848818*^7, 1.1687368433901165*^9,
          1.1344063409072752*^11, 1.2749318929376549`*^13,
          1.6282469370459092*^15}
\end{verbatim}
for $a_n^\mathrm{\MSbar\,\prime}$ (i.e.\ the series coefficients for the standard running and pole mass difference), agreeing well with Table~2 of Ref.~\cite{Hoang:2017suc} within errors.
The deviation from the central values quoted in that table arises because in Ref.~\cite{Hoang:2017suc} the central values of asymmetric uncertainty intervals have been quoted.

\subsubsection*{Asymptotic series for the pole-$\MSbar$ mass relation and the asymptotic pole mass}
\vspace{2mm}

Having access to the pole mass at in principle arbitrary order using conversion from $\MSbar$ and MSR masses
at arbitrary renormalization scales, \REvolver{} allows to easily reproduce the numbers utilized to produce Fig.~6 of Ref.~\cite{Hoang:2017btd}, where the order-dependent top quark pole mass was shown. We create a new \Core{} to switch to the top standard running mass value $\overline m_t = 163$\,GeV in accordance with Ref.~\cite{Hoang:2017btd}
\begin{verbatim}
  In[]:= mtmt = 163;
         CoreDeleteAll[]
         CoreCreate["t", 6, {5, amZdef, mZdef},
          {{6, mtmt, mtmt}}]
\end{verbatim}

Converting directly from the $\MSbar$ mass at the scale $\overline m_t$, including scale variation in the range $\overline m_t/2\leq\mu\leq2 \overline m_t$ the pole mass, order by order $0\leq n\leq 12$, is given by
\begin{verbatim}
  In[]:= Table[{{ord, MassPoleFO["t", 6, 6, mtmt, mtmt, ord]},
           {MassPoleFO["t", 6, 6, mtmt, mtmt/2, ord] - 
             MassPoleFO["t", 6, 6, mtmt, mtmt, ord], 
            MassPoleFO["t", 6, 6, mtmt, 2*mtmt, ord] - 
             MassPoleFO["t", 6, 6, mtmt, mtmt, ord]}},
          {ord, 0, 12}]
  Out[]= {{{0, 163.}, {0., 0.}},
          {{1, 170.5097731904309},
           {0.7122980457679375, -0.5972354956839183}},
          {{2, 172.11277687226905},
           {0.27618516105792423, -0.3086769323029728}},
          {{3, 172.6079790748946},
           {0.12091854036228256, -0.15116686757511388}},
          {{4, 172.80297651883814},
           {0.06337072738099891, -0.08027361964516899}},
          {{5, 172.9150350500228},
           {0.04822992022724293, -0.05421541398965246}},
          {{6, 172.99420261176041},
           {0.04210709824803871, -0.043768064396573436}},
          {{7, 173.06058298165766},
           {0.042838521236177485, -0.04090645714344987}},
          {{8, 173.1250293570062},
           {0.04963473302441912, -0.04338206485874707}},
          {{9, 173.19612979388887},
           {0.0644919645042421, -0.05147133915082236}},
          {{10, 173.28398617671283},
           {0.09283257854019666, -0.06758882965976909}},
          {{11, 173.40418238677069},
           {0.14657118914180955, -0.09737521610099975}},
          {{12, 173.58454177427788},
           {0.2518136273276639, -0.1527834480848469}}}
\end{verbatim}

\begin{figure}[t!]
\centering
\includegraphics[width=0.49\textwidth]{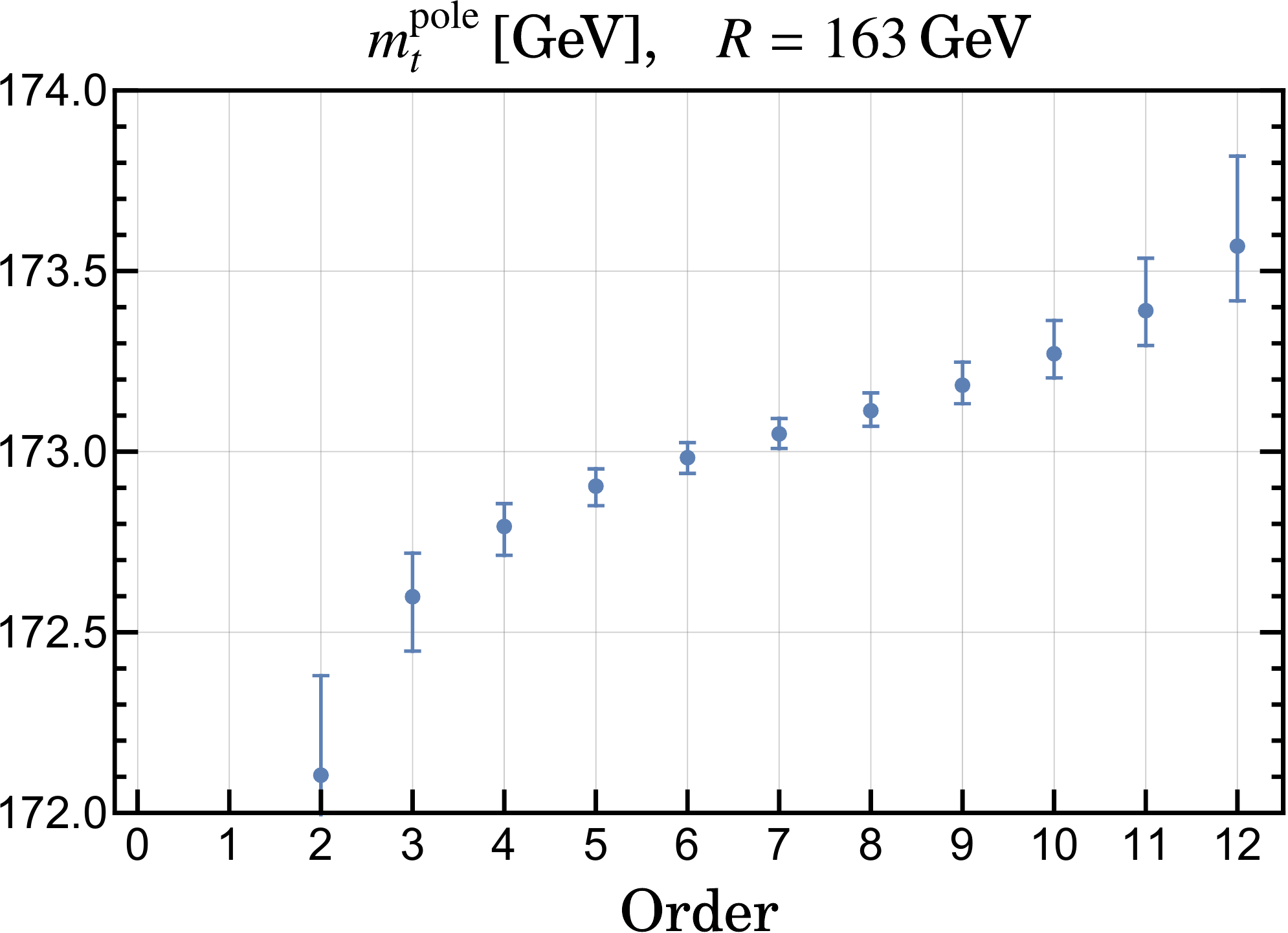}\hfill
\includegraphics[width=0.49\textwidth]{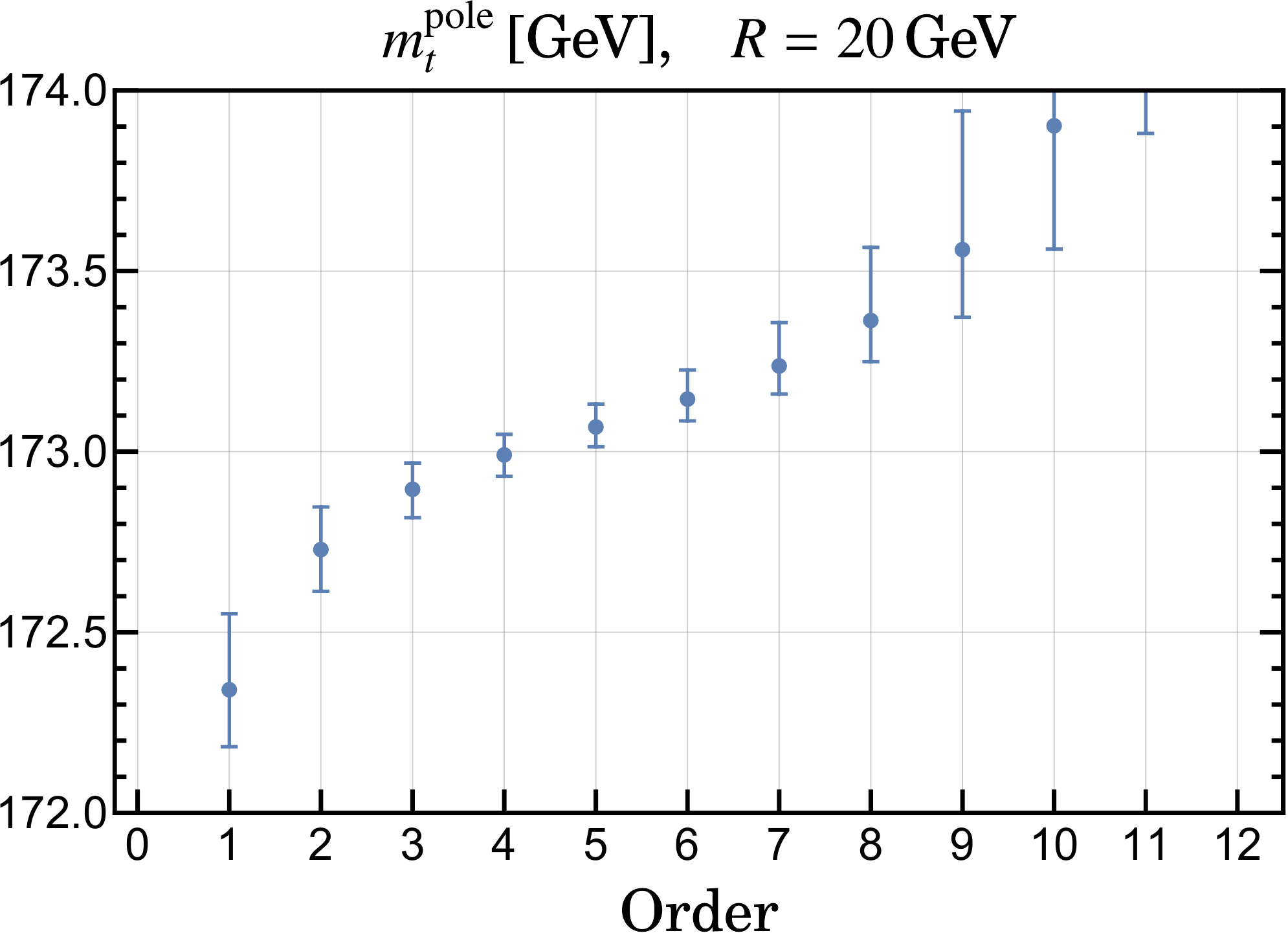}\\[15pt]
\includegraphics[width=0.49\textwidth]{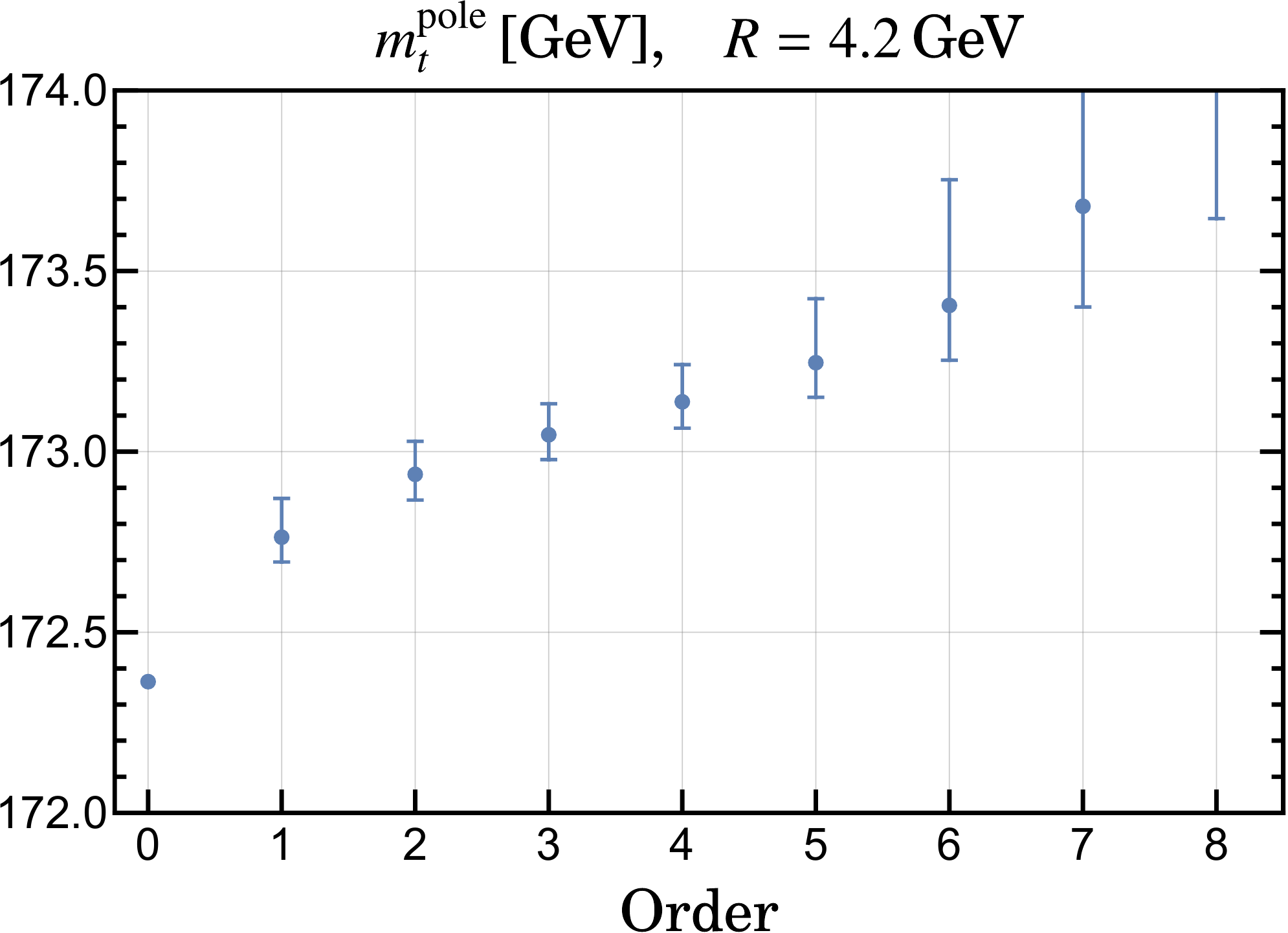}\hfill
\includegraphics[width=0.49\textwidth]{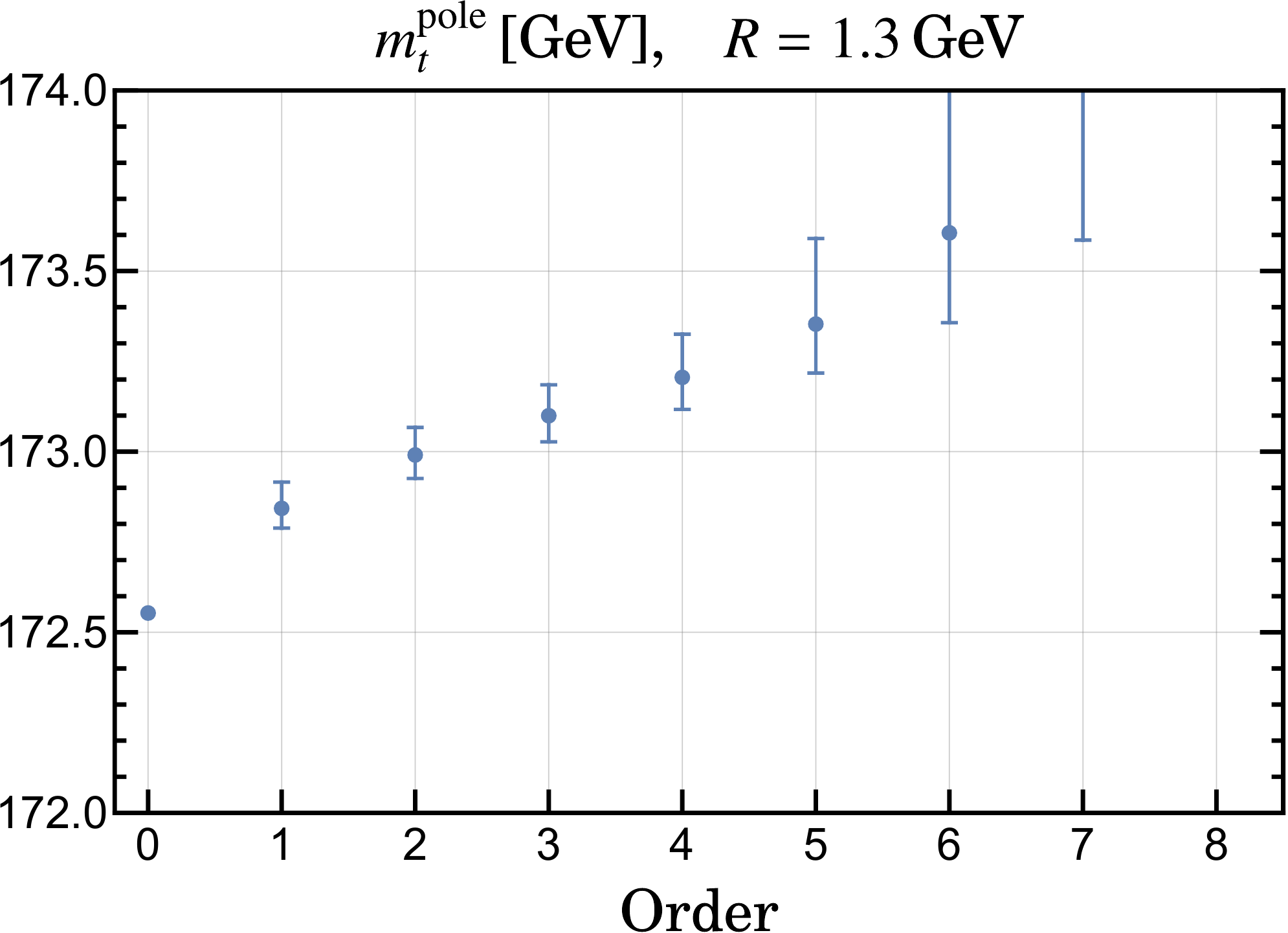}
\caption{\label{fig:poleml}Top quark pole mass $m^\mathrm{pole}_t$ as a function of the perturbative order for massless bottom and charm quarks, converted from the $\MSbar$ mass (upper left panel) or the MSR mass (other panels) at different scales of $R$. The central dots are obtained for the default renormalization scales for the strong coupling, the error bars represent the scale variation. This corresponds to the black dots and error bars in Fig.~6 of Ref.~\cite{Hoang:2017btd}.}
\end{figure}

Analogous commands can be used to extract the relevant values of the pole mass when converting from the MSR scheme at various scales of $R$. Utilizing these numbers it is now straightforward to reproduce Fig.~6 of Ref.~\cite{Hoang:2017btd} for vanishing bottom and charm mass (black dots in Ref.~\cite{Hoang:2017btd}), see Fig.~\ref{fig:poleml}.\footnote{Note that the upper left panel of Fig.~\ref{fig:poleml} (corresponding to the numbers given above) shows the conversion directly from the standard $\MSbar$ mass instead of the MSR mass at the high scales.} The asymptotic value which can be assigned to the pole mass lies in the region where the series grows linearly, which is in Fig.~\ref{fig:poleml}, roughly in the region around $173$\,GeV. This is confirmed by the output of the command \texttt{MassPoleDetailed}, which gives
\begin{verbatim}
  In[]:= MassPoleDetailed["t", 6, mtmt, mtmt, "min"]
  Out[]= {173.07508395890844, 0.06445302708283407, 8}

  In[]:= MassPoleDetailed["t", 6, mtmt, mtmt, "range"]
  Out[]= {173.07976320487433, 0.12605058936964042, 8}

  In[]:= MassPoleDetailed["t", 6, mtmt, mtmt, "drange"]
  Out[]= {173.1081654841447, 0.16284602132711257, 8}
\end{verbatim}
for the three methods to estimate the asymptotic pole mass supported by \REvolver{} from the relation between the pole and the MSR mass $m_t^{(5)}(\overline m_t)$
at the scale $163$\,GeV. While the estimation of the ambiguity can vary by more than a factor of $2$ depending on the employed strategy, the estimated asymptotic pole mass is fairly stable.

\begin{figure}
\centering
\includegraphics[width=0.6\textwidth]{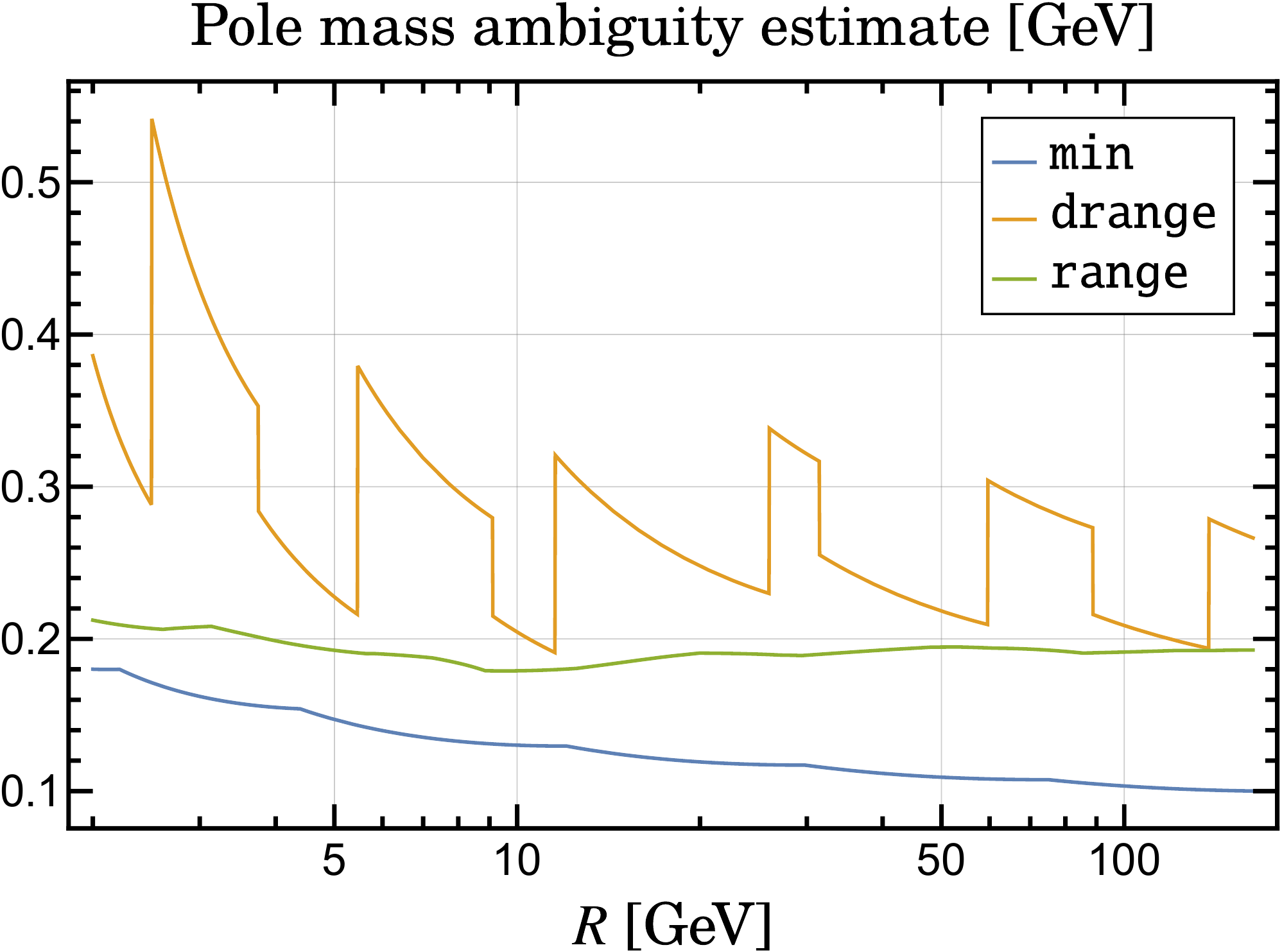}
\caption{\label{fig:ambest} Comparison of the pole mass ambiguity estimating strategies available in \REvolver{} obtained from the series between the pole mass and the running mass \mbox{$m_t^{(5)}(R)=m_t^{\rm MSR}(R)$} for the case that all quarks lighter than the top quark are considered to be massless. The figure shows the size of the ambiguity of the top quark pole mass in GeV as estimated by the \texttt{min} (blue lower curve), \texttt{drange} (orange upper curve) and \texttt{range} (green middle curve) strategies over the scale $R$.}
\end{figure}

Figure~\ref{fig:ambest} shows a comparison of the three strategies of estimating the pole mass ambiguity implemented in \REvolver{} from the perturbative series between the pole mass and the running (MSR) mass $m_t^{(5)}(R)=m_t^{\rm MSR}(R)$ at the scale $R$. The plot is produced with the command
\begin{verbatim}
  In[]:= LogLinearPlot[
          {MassPoleDetailed["t", 6, R, "min"][[2]],
           MassPoleDetailed["t", 6, R, "drange"][[2]],
           MassPoleDetailed["t", 6, R, "range"][[2]]},
          {R, 2, mtmt}, PlotRange -> {0.05, 0.26}]
\end{verbatim}
omitting all options related to plot styling and annotations for brevity.

The figure shows the size of the ambiguity of the top quark pole mass in GeV as estimated by the strategies over the scale $R$, specifying the scale in the perturbative relation between the pole and running mass. All quarks lighter than the top quark are considered to be massless. Due to its discrete nature, the \texttt{drange} method produces a discontinuous curve, while the continuous version \texttt{range} provides a very stable result. The \texttt{min} method leads to an ambiguity estimate that is logarithmically decreasing with $R$.

\subsection{\Core{}s with Multiple Massive Quarks}

\subsubsection{Strong coupling and \texorpdfstring{$\Lambda_{\rm QCD}$}{Lambda QCD}}

\vspace{1mm}
\subsubsection*{Strong coupling evolution through multiple flavor thresholds}
\vspace{2mm}

In Ref.~\cite{Boito:2018yvl} the 3-flavor strong coupling value $\alpha_s^{(3)}(m_\tau)$ was determined from hadronic $e^+ e^-$ $R$-ratio data for c.m.\ energies below the charm production threshold. For this determination, fixed-order (FO) as well as contour-improved (CI) perturbation theory were applied, and the corresponding values $\alpha_s^{(3)}(m_\tau)_{\rm FO}=0.298$ and $\alpha_s^{(3)}(m_\tau)_{\rm CI}=0.304$ were subsequently converted to $\alpha_s^{(5)}(m_Z)$ employing the \mbox{4-loop} beta function as well as the \mbox{3-loop} matching relations at the charm and bottom thresholds. The respective values for the strong coupling at the Z-scale were quoted in Eq.~(4.7) of that reference as $\alpha_s^{(5)}(m_Z)_{\rm FO}=0.1158$ and $\alpha_s^{(5)}(m_Z)_{\rm CI}=0.1166$, respectively.

For the conversion, the values $\overline m_c=1.28$\,GeV and $\overline m_b=4.2$\,GeV were used for the charm and bottom standard running masses, respectively, and the charm and bottom threshold matching scales were set to \mbox{$\mu_c=2.0$\,GeV} and $\mu_b=4.0$\,GeV, respectively. To reproduce the conversion with \REvolver{} we define these input values with
\begin{verbatim}
  In[]:= mTau = 1.77686;
         {mcmc, mbmb} = {1.28, 4.2};
         matchList = {2.0/mcmc, 4.0/mbmb};
         {amTauFO, amTauCI} = {0.298, 0.304};
\end{verbatim}
and define one \Core{} for each value of $\alpha_s^{(3)}(m_\tau)$
\begin{verbatim}
 In[]:= CoreCreate["FO", 5, {3, amTauFO, mTau},
         {{4, mcmc, mcmc}, {5, mbmb, mbmb}}, runAlpha -> 4,
         orderAlpha -> 3, fMatch -> matchList]
        CoreCreate["CI", 5, {3, amTauCI, mTau},
         {{4, mcmc, mcmc}, {5, mbmb, mbmb}}, runAlpha -> 4,
         orderAlpha -> 3, fMatch -> matchList]
\end{verbatim}
where the parameters relevant for running and matching were set with the optional parameters \texttt{runAlpha}, \texttt{orderAlpha} and \texttt{fMatch}.

The values of $\alpha_s^{(5)}(m_Z)$ can now be easily extracted by
\begin{verbatim}
  In[]:= AlphaQCD["FO", mZdef, 5]
         AlphaQCD["CI", mZdef, 5]
  Out[]= 0.11581049250660494
  Out[]= 0.11662117512902259
\end{verbatim}
in full agreement with Eq.~(4.7) of Ref.~\cite{Boito:2018yvl}.

Similarly, in Ref.~\cite{Baikov:2008jh} $\alpha_s^{(3)}(m_\tau)$ was determined from $\tau$ decay data and converted to $\alpha_s^{(5)}(m_Z)$ employing 4-loop evolution and matching. We update the values for the standard running bottom and charm masses to the ones in the reference paper and define the value of $\alpha_s^{(3)}(m_\tau)$ as given in Eq.~(19) therein as well as the associated \Core{}
\begin{verbatim}
  In[]:= {mcmc, mbmb} = {1.286, 4.164};
         amTau = 0.332;

  In[]:= CoreDeleteAll[]
         CoreCreate["O", 5, {3, amTau, mTau},
          {{4, mcmc, mcmc}, {5, mbmb, mbmb}}, runAlpha -> 4]
\end{verbatim}
where, following Ref.~\cite{Baikov:2008jh}, the strong coupling evolution is set to 4-loop precision. We focus on the central value here for brevity. The value of $\alpha_s^{(5)}(m_Z)$ can now easily be extracted with
\begin{verbatim}
  In[]:= AlphaQCD["O", mZdef, 5]
  Out[]= 0.12019776978833413
\end{verbatim}
agreeing with Eq.~(20) of Ref.~\cite{Baikov:2008jh}.

\subsubsection*{Flavor number dependence of the QCD scale $\Lambda_\mathrm{QCD}^{(n_f)}$}
\vspace{2mm}

To demonstrate the capability of \REvolver{} to extract $\Lambda_\mathrm{QCD}^{(n_f)}$ in various flavor number schemes we consider Ref.~\cite{Bruno:2017lta}, where $\alpha_s^{(5)}(m_Z)=0.1179$ has been determined by the ALPHA collaboration and values for $\Lambda_\mathrm{QCD}^{(nf)}$ with \mbox{$3\leq n_f\leq 5$} were determined, see Eqs.~(4.14) as well as (5.3) and (5.4) of that reference.

To reproduce the $\Lambda_\mathrm{QCD}^{(n_f)}$ values with \REvolver{}
we create a \Core{} with massive bottom and charm quarks (with standard running mass values taken from the PDG \cite{Zyla:2020zbs}) and define the value of $\alpha_s^{(5)}(m_Z)$ to be the one given in the reference paper
\begin{verbatim}
  In[]:= {amZ, mcmc, mbmb} = {0.1179, 1.27, 4.18};

  In[]:= CoreDeleteAll[]
         CoreCreate["O", 5, {5, amZ, mZdef},
          {{4, mcmc, mcmc}, {5, mbmb, mbmb}}]
\end{verbatim}

We now extract $\Lambda_\mathrm{QCD}^{(n_f)}$ in the $n_f=5\,,4\,,3$ flavor schemes in the so-called $\MSbar$ scheme [\,see Eq.~\eqref{eq:lambdaMSbar} in App.~\ref{sec:strongcoupling}\,] with
\begin{verbatim}
  In[]:= LambdaQCD["O", 5]
         LambdaQCD["O", 4]
         LambdaQCD["O", 3]
  Out[]= 0.20745573124097272
  Out[]= 0.288991643859333
  Out[]= 0.33153068987761314
\end{verbatim}
The numbers are in exact agreement with the values shown in Ref.~\cite{Bruno:2017lta}.

\subsubsection{Top quark running mass at low scales}

In Ref.~\cite{Butenschoen:2016lpz} a calibration analysis was provided which suggested that the top quark running mass $m_t(1\,\mathrm{GeV})$ agrees within theoretical uncertainties with the MC top mass parameter $m_t^\mathrm{MC}$ of the \texttt{Pythia} event generator. Specifically, the authors quote a value of $m_t(1\,\mathrm{GeV}) = 172.82\pm0.19$\,GeV for a calibration with $m_t^\mathrm{MC}=173$\,GeV, given in Table~1. That analysis was carried out in the approximation of massless bottom and charm quarks. Here we demonstrate how \REvolver{} can be used to investigate how finite charm and bottom quark masses affect the value of the standard running mass $\overline m_t$ calculated from $m_t(1\,\mathrm{GeV})$. As an estimate of the uncertainties we quote the difference of the values computed using $4$- and $3$-loop R-evolution.

To this end we create six \Core{}s: two with massless bottom and charm quarks, two with a massless charm quark and a massive bottom quark, and two with massive bottom and charm quarks, employing $\overline m_c=1.27$\,GeV and \mbox{$\overline m_b=4.18$\,GeV} for the standard running charm and bottom quark masses. Among the two respective \Core{}s one employs 3-loop and the other 4-loop precision for the R-evolution equation:
\begin{verbatim}
  In[]:= {mcmc, mbmb, mt1} = {1.27, 4.18, 172.82};

  In[]:= CoreDeleteAll[]
         CoreCreate["t4", 6, {5, amZdef, mZdef}, {{5, mt1, 1}},
          runMSR -> 4]
         CoreCreate["t3", 6, {5, amZdef, mZdef}, {{5, mt1, 1}},
          runMSR -> 3]
         CoreCreate["bt4", 6, {5, amZdef, mZdef},
          {{5, mbmb, mbmb}, {4, mt1, 1}}, runMSR -> 4]
         CoreCreate["bt3", 6, {5, amZdef, mZdef},
          {{5, mbmb, mbmb}, {4, mt1, 1}}, runMSR -> 3]
         CoreCreate["cbt4", 6, {5, amZdef, mZdef},
          {{4, mcmc, mcmc}, {5, mbmb, mbmb}, {3, mt1, 1}},
          runMSR -> 4]
         CoreCreate["cbt3", 6, {5, amZdef, mZdef},
         {{4, mcmc, mcmc}, {5, mbmb, mbmb}, {3, mt1, 1}},
         runMSR -> 3]
\end{verbatim}

The \Core{} setups specify that $m_t^{\rm MSR}(1\,\mbox{GeV})= 172.82$\,GeV is the running top quark mass
in the scheme in which all massive flavors above the scale of $1$\,GeV are integrated out. It is easy to extract the top quark standard running masses and their uncertainties from the \Core{}s above with
\begin{verbatim}
  In[]:= MassMS["t4", 6]
         MassMS["t4", 6] - MassMS["t3", 6]
  Out[]= 163.00308510782256
  Out[]= 0.05105270648016358

  In[]:= MassMS["bt4", 6]
         MassMS["bt4", 6] - MassMS["bt3", 6]
  Out[]= 162.94352028117146
  Out[]= 0.050962554040836494

  In[]:= MassMS["cbt4", 6]
         MassMS["cbt4", 6] - MassMS["cbt3", 6]
  Out[]= 162.92268002391415
  Out[]= 0.049529435610025985
\end{verbatim}
where the first numbers refer to the setup with massless bottom and charm quarks and the last ones to the one where bottom and charm quarks have mass.

We observe that the finite bottom mass lowers the standard top running mass by around $60$\,MeV and the finite charm mass decreases the standard top running mass by about another $20$\,MeV. The perturbative uncertainty of the conversion amounts to around $50$\,MeV in all cases. Together, the finite charm and bottom masses lower the standard running top mass by about $80$\,MeV, which is larger than the perturbative uncertainty.

\subsubsection{Charm mass effects for the bottom quark mass}

To demonstrate the effect of lighter massive flavors in conversions between short-distance mass schemes we consider Ref.~\cite{Hoang:1999us}, where effects of the finite charm quark mass on bottom quark mass determinations from $\Upsilon$ mesons were (calculated and) examined. In Eq.~(99) of that paper the bottom standard running mass is computed from the 1S mass $m_b^\mathrm{1S}=4.7$\,GeV with and without the contribution of the charm standard running quark mass $\overline m_c=1.5$\,GeV. The strong coupling value was set to $\alpha_s^{(4)}(m_b^\mathrm{1S})=0.216$. We define these parameters with
\begin{verbatim}
  In[]:= {mb1S, mcmc, amb1S} = {4.7, 1.5, 0.216};
\end{verbatim}
and create two 4-flavor \Core{}s, one with a massless charm quark and one with the charm mass given above
\begin{verbatim}
  In[]:= CoreDeleteAll[]
         CoreCreate["c0", {4, amb1S, mb1S}]
         CoreCreate["cm", 4, {4, amb1S, mb1S},
          {{4, mcmc, mcmc}}]
\end{verbatim}

We then create two additional \Core{}s by adding the bottom quark 1S mass $m_b^\mathrm{1S}=4.7$\,GeV to the existing \Core{}s, where,
following Ref.~\cite{Hoang:1999us}, the perturbative order of its relation to the standard running mass is $\mathcal O(\alpha_s^3)$ and the relativistic (``upsilon-expansion'') counting scheme is adopted.
\begin{verbatim}
  In[]:= Add1SMass["c0", "bc0", mb1S, 5, mb1S, "relativistic",
          mb1S, 3]
         Add1SMass["cm", "bcm", mb1S, 5, mb1S, "relativistic",
          mb1S, 3]
\end{verbatim}
We have furthermore set the running mass scale and the renormalization scale of the strong coupling to $m_b^\mathrm{1S}$ and the parameter \texttt{nfConv} to $5$ to specify direct conversion without R-evolution to be in accordance with the computation carried out in Ref.~\cite{Hoang:1999us}.

Extracting the standard mass from the massless charm \Core{} gives
\begin{verbatim}
  In[]:= MassMS["bc0", 5]
  Out[]= 4.190311348710497
\end{verbatim}
in perfect agreement with $\overline m_b = (4.7 - 0.382 - 0.098 - 0.030)\,\mathrm{GeV} = 4.190$\,GeV as quoted in Ref.~\cite{Hoang:1999us}. For the standard mass from the massive charm \Core{} we get
\begin{verbatim}
  In[]:= MassMS["bcm", 5]
  Out[]= 4.178691644159289
\end{verbatim}
with a sub-MeV difference to the reference value $\overline m_b = (4.7 - 0.382 - (0.098 + 0.0072) - (0.03 + 0.0049))\,\mathrm{GeV} = 4.178$\,GeV.

\subsubsection{Top quark pole masses}
\label{sec:polelmq}

In the following we investigate the influence of massive bottom and charm quarks on the perturbative series relating the top quark standard running mass and the pole mass.

\subsubsection*{Pole-$\MSbar$ mass series dependence on lighter massive quarks}
\vspace{2mm}

In Ref.~\cite{Hoang:2017btd} the dependence of the high-order asymptotic series for the top quark pole-$\MSbar$ mass relation on finite bottom and charm quark mass was examined and an algorithm to determine explicit analytic formulae (accounting for the semianalytical results given in Ref.~\cite{Bekavac:2007tk}) were provided, which are implemented in \REvolver{}. Explicit results for the asymptotic series coefficients beyond ${\cal O}(\alpha_s^4)$ were provided in Ref.~\cite{Hoang:2017btd} for the scenarios of massless bottom and charm quarks, massive bottom and massless charm quarks as well as massive bottom and charm quarks.
To compare the results of \REvolver{} to those of Ref.~\cite{Hoang:2017btd} we set the appropriate relevant parameters for the standard running charm, bottom and top quark masses as well as for the strong coupling and create the \Core{}s for the three scenarios
\begin{verbatim}
  In[]:= {mcmc, mbmb, mtmt} = {1.3, 4.2, 163};
         {amZ, mZ} = {0.118, 91.187};

  In[]:= CoreDeleteAll[]
         CoreCreate["tbc", 6, {5, amZ, mZ},
          {{4, mcmc, mcmc}, {5, mbmb, mbmb}, {6, mtmt, mtmt}}]
         CoreCreate["tb", 6, {5, amZ, mZ},
          {{5, mbmb, mbmb}, {6, mtmt, mtmt}}]
         CoreCreate["t", 6, {5, amZ, mZ}, {{6, mtmt, mtmt}}]
\end{verbatim}

\begin{figure}
\centering
\includegraphics[width=0.6\textwidth]{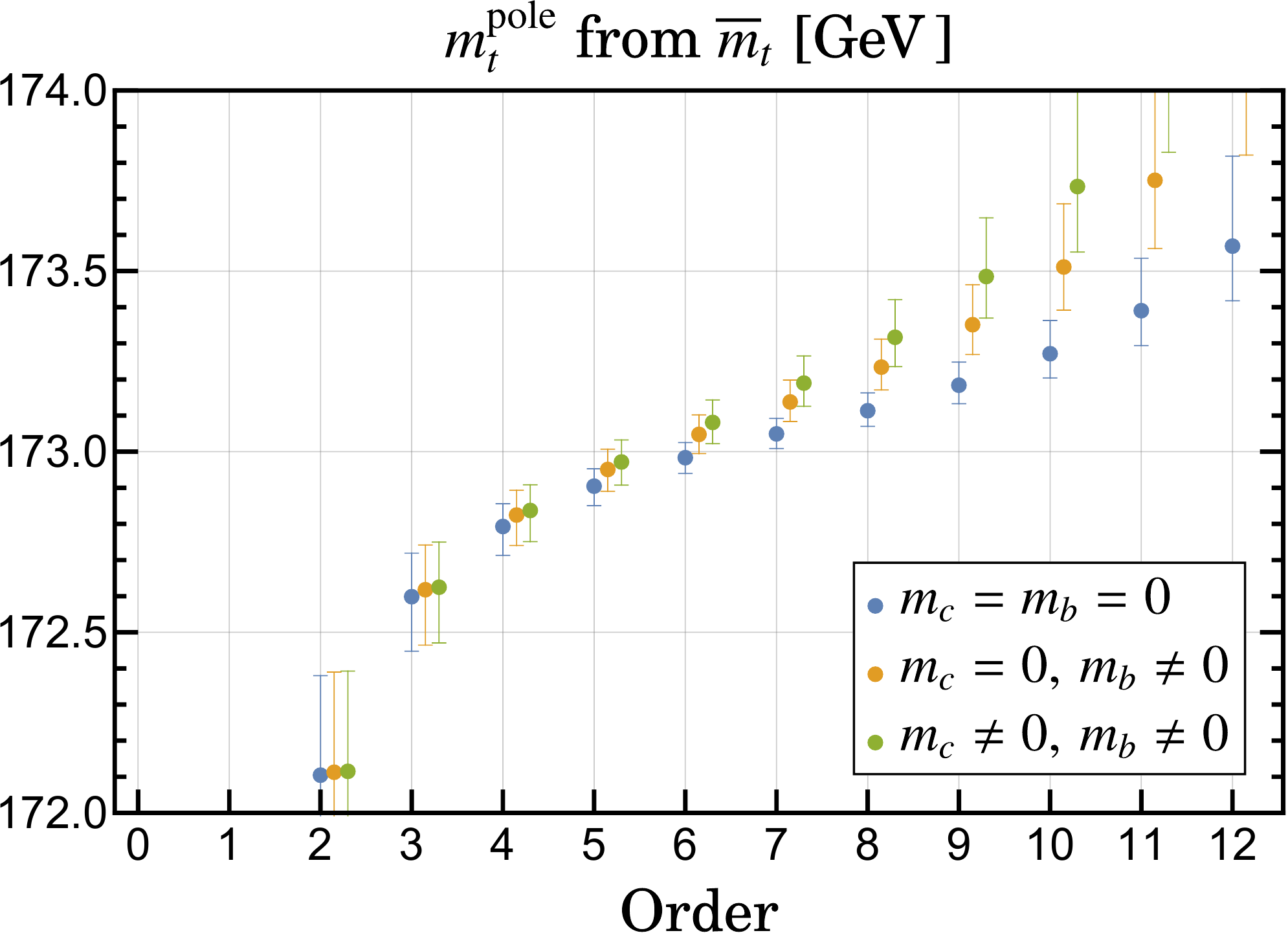}
\caption{\label{fig:polelmf}Top quark pole mass $m^\mathrm{pole}_t$ as a function of the perturbative order with bottom and charm quark masses set to zero (left blue), massless charm quark (middle orange) and massive bottom and charm masses (right green). The conversion is performed from the $\MSbar$ mass. The central dots are obtained for the default renormalization scales for the strong coupling $\mu=\overline m_t$, the error bars represent the scale variation $\overline m_t/2\leq\mu\leq2\,\overline m_t$.}
\end{figure}

The perturbative series relating the standard running mass and the pole mass can be generated as shown in Sec.~\ref{sec:expoleml} and depicted in the upper left panel of Fig.~\ref{fig:poleml} for massless bottom and charm quarks. In Fig.~\ref{fig:polelmf} we show the result of repeating this procedure for each of the \Core{}s created above.
It is clearly visible that for each additional accounted massive quark flavor the series diverges faster and consequently the respective pole mass ambiguity increases.

In accordance with Ref.~\cite{Hoang:2017btd} we read out the value of the asymptotic top quark pole mass, the associated ambiguity and the order of the minimal correction term from each \Core{} employing the \texttt{drange} approach, which is based on all series terms in the superasymptotic region,
by executing
\begin{verbatim}
  In[]:= MassPoleDetailed["t", 6, mtmt, "drange"]
         MassPoleDetailed["tb", 6, mtmt, "drange"]
         MassPoleDetailed["tbc", 6, mtmt, "drange"]
  Out[]= {173.09649964256326, 0.16178589168021063, 8}
  Out[]= {173.17525229067104, 0.2336983055654116, 7}
  Out[]= {173.17973196293923, 0.26632540751468525, 7}
\end{verbatim}
The three results agree with the numbers given in Table~5 of Ref.~\cite{Hoang:2017btd} within uncertainties\footnote{The $10$\,MeV discrepancy in the asymptotic pole mass value is because in Ref.~\cite{Hoang:2017btd}, an additional $\lambda$ variation was applied in the estimate of the asymptotic coefficients and the quoted coefficients were the central values of the obtained intervals. \REvolver{}
does not support this functionality.}.

We now return to Ref.~\cite{Beneke:2016cbu}. There, finite charm and bottom quark mass effects on the value of the top quark asymptotic pole mass were quantified by an approximate and heuristic method utilizing order-dependent light flavor decoupling in the strong coupling. A numerical value for the resulting ambiguity of the asymptotic pole mass was quoted in Eq.~(5.1) of that paper using size of the minimal correction of the asymptotic series.

Fixing the value of the top, bottom and charm quark standard running masses to the values given in Ref.~\cite{Beneke:2016cbu} and creating an associated \Core{}
\begin{verbatim}
  In[]:= {mtmt, mbmb, mcmc} = {163.508, 4.2, 1.3};

  In[]:= CoreCreate["tbc2", 6, {5, amZdef, mZdef},
          {{4, mcmc, mcmc}, {5, mbmb, mbmb}, {6, mtmt, mtmt}}]
\end{verbatim}
we obtain
\begin{verbatim}
  In[]:= MassPoleDetailed["tbc2", 6, mtmt, mtmt, "min"]
  Out[]= {173.61531203486197, 0.10050070355543994, 7}
\end{verbatim}
where we employ the \texttt{min} method to match the approach used in Ref.~\cite{Beneke:2016cbu}. The asymptotic pole mass value and the associated ambiguity are quoted as $173.667$\,GeV and $108$\,MeV, respectively, in that reference and differs by around $50$ and $8$\,MeV, respectively, to the result provided by \REvolver{}. The difference is mainly related to the exact treatment of light massive flavor corrections provided by \REvolver{}.

\subsubsection*{Normalization of the pole mass renormalon}
\vspace{2mm}

Finally, we demonstrate \REvolver{}'s capability to compute the normalization $N_{1/2}^{(n_\ell)}$ of the pole mass renormalon for a given number $n_\ell$ of massless quark flavors.
Reusing the \Core{}s created in one of the previous examples, the value of the normalization for $n_\ell=3, 4, 5$ can be extracted with
\begin{verbatim}
  In[]:= N12[#] & /@ {"tbc","tb","t"}
  Out[]= {0.5371120004494861, 0.5055628679671248,
          0.4612967890971816}
\end{verbatim}
in excellent agreement with the corresponding values $N_{1/2}^{(n_\ell=3,4,5)} = \{0.5370\pm 0.0011,\, 0.5056\pm 0.0015,\, 0.4616\pm 0.0020\}$ given for $\mu/\overline m = 1$ in Table~1 of Ref.~\cite{Beneke:2016cbu}.

In Eq.~(2.21) of Ref.~\cite{Hoang:2017btd} the $\lambda$-parameter was varied in the range \mbox{$1/2\leq\lambda\leq 2$}, and the associated central value and uncertainty of the renormalon normalization $N_{1/2}^{(n_\ell)}$ were determined. The corresponding values for $n_\ell=3,4,5$ can be reproduced by \REvolver{} with

\begin{verbatim}
  In[]:= list2 = 2^Range[-1, 1, 2/19];

  In[]:= N3Tab = N12["tbc", #] & /@ list2;
         (Max[N3Tab] + {1, -1} * Min[N3Tab])/2
  Out[]= {0.5256087104548577, 0.011913515642471062}

  In[]:= N4Tab = N12["tb", #] & /@ list2;
         (Max[N4Tab] + {1, -1} * Min[N4Tab])/2
  Out[]= {0.4916415824077933, 0.016342715842279004}

  In[]:= N5Tab = N12["t", #] & /@ list2;
         (Max[N5Tab] + {1, -1} * Min[N5Tab])/2
  Out[]= {0.44606744334202203, 0.023867353429116556}
\end{verbatim}
in perfect agreement with the values $N_{1/2}^{(n_\ell=3,4,5)} = \{0.526\pm 0.012,\, 0.492\pm 0.016,\, 0.446\pm 0.024\}$, quoted in Eq.~(2.21) of Ref.~\cite{Hoang:2017btd}.

\section{Theoretical input}
\label{sec:theoryinput}

The perturbative coefficients of the QCD $\beta$-function, that is the renormalization group equation of the $\overline{\rm MS}$ strong coupling $\alpha_s^{(n_f)}(\mu)$ in a given $n_f$ flavor number scheme, are known to five loops and can be found in Refs.~\cite{Tarasov:1980au,Larin:1993tp,vanRitbergen:1997va,Czakon:2004bu,Baikov:2016tgj,Chetyrkin:2017bjc,Luthe:2017ttg,Herzog:2017ohr}. If \texttt{REvolver} is used for scenarios with massless quarks only, it is recommended to cite those references. The perturbative coefficients for the strong coupling flavor matching relation at the quark thresholds are known up to four loops and can be found in Refs.~\cite{Chetyrkin:1997un,Chetyrkin:2005ia,Schroder:2005hy,Kniehl:2006bg}. If \texttt{REvolver} is used for strong coupling evaluations in different flavor number schemes involving massive quark threshold matching, these references should be cited.

The perturbative coefficients of the relation between the $\overline{\rm MS}$ mass $\overline m_q^{(n_f)}(\mu)$, for the case where $n_f-1$ is the number of flavors lighter than the massive quark $q$, and the pole mass $m^{\rm pole}_q$ for the case that all lighter quarks are massless, are known to four loops and can be found in Refs.~\cite{Tarrach:1980up,Gray:1990yh,Melnikov:2000qh,Marquard:2016dcn}. If REvolver is used to compute relations between the $\overline{\rm MS}$ mass and the pole mass when all lighter quarks are massless, these references should be included. The two- and three-loop corrections coming from the masses of lighter massive quarks can be found in Refs.~\cite{Gray:1990yh} and \cite{Bekavac:2007tk,Fael:2020bgs}, respectively. It would be appropriate to also cite these references, if REvolver is used to compute lighter massive flavor effects in the relation between the $\overline{\rm MS}$ mass and the pole mass of a heavy quark. The perturbative coefficients of the $\overline{\rm MS}$ mass renormalization group equations (relevant for renormalization scales above the mass of quark $q$) are known to five loops and can be found in Refs.~\cite{Vermaseren:1997fq,Chetyrkin:1997dh,Baikov:2014qja,Luthe:2016xec}. These references should be cited whenever the code is used to compute the running of the $\overline{\rm MS}$ mass in any flavor scheme that includes the heavy quark itself. The perturbative coefficients of the flavor matching relations needed when the running $\overline{\rm MS}$ mass evolves to scales beyond even heavier massive quarks
are known up to four loops and can be found in Refs.~\cite{Chetyrkin:1997un,Liu:2015fxa}. These references should be included, if the $\overline{\rm MS}$ mass is evaluated in different flavor number schemes involving threshold matching associated to even heavier massive quarks.

The MSR mass $m_q^{{\rm MSR},(n_\ell)}(\mu)$ is derived from the perturbative relation between the pole mass and the standard running mass $\overline m_q$ by integrating out the quark flavor $q$, see Sec.~\ref{sec:MSRevol}. Thus, all perturbative properties of the MSR mass are derived from expressions known for the $\overline{\rm MS}$ scheme from the references mentioned in the previous paragraph. The MSR mass concept was first suggested in Ref.~\cite{Hoang:2008yj} and allows to consistently consider renormalization group evolution for scales $\mu$ below the quark mass $m_q$, called ``R-evolution''. The perturbative coefficients of (i) the relation between the MSR mass $m_q^{{\rm MSR},(n_\ell)}(\mu)$, where $n_\ell$ is smaller or equal to the number of flavors lighter than the massive quark $q$, and the pole mass $m^{\rm pole}_q$, (ii) the matching relation of the MSR mass to the $\overline{\rm MS}$ mass, and (iii) its $n_\ell$ flavor-number-dependent renormalization group equation for the case that all lighter quarks are massless (which are all known to four loops) can be found in Ref.~\cite{Hoang:2017suc}. The MSR mass scheme implemented in REvolver is called the ``natural'' MSR mass scheme in Ref.~\cite{Hoang:2017suc}.
The two- and three-loop corrections coming from the masses of lighter massive quarks including the flavor matching relations, when the MSR mass evolution crosses a lighter massive quark threshold, can be found in Ref.~\cite{Hoang:2017btd}. The results for the lighter massive quark effects given in Ref.~\cite{Hoang:2017btd} involve a parametrization of the results provided in Ref.~\cite{Bekavac:2007tk} (in terms of coefficients $\delta_2$ and $\delta_{Q,n}^{(q,q^\prime,\ldots)}$) that accounts for theoretical interrelations of the different contributions not considered in Ref.~\cite{Bekavac:2007tk}, and allows for a straightforward generalization for the case of multiple lighter massive quark flavors. The parametrization is also fully compatible with the recent analytic updates on the three-loop corrections given Ref.~\cite{Fael:2020bgs} within a few MeV.
The expression for $\delta_2$ used in REvolver has been given in Ref.~\cite{Hoang:2017btd}. The concrete expressions for the other coefficients $\delta_{Q,n}^{(q,q^\prime,\ldots)}$ have been taken from Ref.~\cite{Mateu:2017hlz}.
If REvolver is used to calculate the MSR mass of a massive quark in the case that all lighter quarks are massless, Refs.~\cite{Hoang:2017suc,Hoang:2017btd} should be cited.
If REvolver is used to calculate the MSR mass in different flavor number schemes involving flavor threshold matching as light massive quark thresholds Ref.~\cite{Hoang:2017btd} should be cited. The possibility to resum (large) logarithms in the relation between short-distance quark mass definitions involving scales below the quark mass scale (e.g.\ concerning the ratio of the heavy quark mass and a lighter quark mass) is currently only provided through the MSR mass, which is therefore the primary short-distance mass scheme for scales below the quark masses contained in the REvolver core objects.
If REvolver is used to resum such logarithms in the relation of short-distance masses (even not involving the MSR directly), this MSR mass functionality is used as an intermediate step, and it is appropriate to cite Ref.~\cite{Hoang:2017suc,Hoang:2008yj}.

The quark mass dependence of the formulae of the flavor threshold matching relations for the flavor-number-dependent strong coupling $\alpha_s^{(n_f)}(\mu)$, the $\overline{\rm MS}$ mass $\overline m_q^{(n_f)}(\mu)$ and the MSR mass $m_q^{{\rm MSR},(n_\ell)}(\mu)$ implemented in \REvolver{} are expressed in terms of the standard running mass $\overline m_q$. Changing the scheme of the quark mass in these matching relations is not supported in \REvolver{} because the corresponding numerical impact is tiny and negligible for practical applications where large logarithmic corrections are properly resummed.

The perturbative coefficients of the relation between the kinetic and pole mass, assuming lighter massive quarks to be massless, can be found in Refs.~\cite{Czarnecki:1997sz,Fael:2020iea}. The 1S mass scheme was first suggested in Refs.~\cite{Hoang:1998hm,Hoang:1998ng}. The perturbative coefficients for the 1S mass scheme have been derived in Refs.~\cite{Hoang:1998hm,Hoang:1998ng,Kiyo:2014uca,Beneke:2005hg,Brambilla:2001qk,Penin:2002zv,Kiyo:2013aea}.
The perturbative coefficient for the PS mass scheme, first suggested in Ref.~\cite{Beneke:1998rk}, can be obtained from Refs.~\cite{Beneke:1998rk,Beneke:2005hg} using analytic results for the static potential as given in Refs.~\cite{Fischler:1977yf,Billoire:1979ih,Schroder:1998vy,Smirnov:2008pn,Smirnov:2009fh,Anzai:2009tm,Lee:2016cgz}. The relation connecting the RS and pole mass schemes is described in Refs.~\cite{Peset:2018ria,Pineda:2001zq} and references therein. Whenever values for any of these masses are obtained through \texttt{REvolver}, the corresponding references should be mentioned.
The lighter massive flavor corrections in the relation to the pole mass are known for the kinetic \cite{Fael:2020njb}, 1S \cite{Eiras:2000rh, Hoang:2000fm} and PS masses \cite{Beneke:2014pta}.
For the 1S mass \REvolver{} currently only accounts for the fixed-order corrections in the relation to the pole mass. Therefore, if lighter quarks are considered massive,
the appropriate references should be cited.

To compute the RGI mass \cite{Floratos:1978jb} the standard formula is evaluated as described below Eq.~\eqref{eq:dtilde}.

\section{Summary}
\label{sec:conclusions}

In this article we have presented \REvolver{}, a \Cpp{} library for carrying out state-of-the-art renormalization group evolution and flavor matching for the QCD coupling and quark masses, and conversion between the most common quark mass renormalization schemes. For short-distance quark masses the achievable precision is at the level of $10$ to $20$\,MeV. In addition to similar libraries that are already available \REvolver{} offers the \Core{} concept, that allows to define and manage different physical scenarios for coupling and quark mass values, and to carry out renormalization group summation of logarithms from scales above and below the quark masses. \REvolver{} supports, in particular, the summation of logarithms described by the \mbox{R-evolution} equation, which describes the linear scale evolution characteristic to low-scale short-distance masses, and it accounts for the flavor threshold corrections that arise in this linear scale evolution. Furthermore, \REvolver{} provides access to the asymptotic perturbative relations to the pole mass to in principle any perturbative order, and it provides quasi-exact solutions to the renormalization group equations (i.e. exact up to machine-precision) for complex renormalization scales. The \REvolver{} library can be also accessed through \Mathematica{} and \Python{} interfaces. We have provided a large number of examples for the most common applications, which are shipped along with the library in the form of \Mathematica{} and Jupyter notebooks, as well as a \texttt{C++} program.

\section*{Acknowledgments}

This work was supported in part by FWF Austrian Science Fund under the Project No.\ P28535-N27, the Spanish MINECO Ram\'on y Cajal program
(RYC-2014-16022), the MECD grants FPA2016-78645-P and PID2019-105439GB-C22, the IFT Centro de Excelencia Severo Ochoa Program under Grant SEV-2012-0249, the EU STRONG-2020 project under the program H2020-INFRAIA-2018-1, grant agreement No.\ 824093 and the COST Action CA16201 PARTICLEFACE. CL is supported by the FWF Doctoral Program ``Particles and Interactions'' No.\ W1252-N27. We thank D.\ Boito and M.\ Steinhauser for explanations concerning the content of their respective articles.

\appendix
\renewcommand*{\thesection}{\Alph{section}}

\section{Algorithms to Solve the Differential Equations}
\label{sec:algo}
In this appendix some details on the {algorithms used in \REvolver{} are provided.

\subsection{\Core{} Creation}
For the creation of a \Core{}, values for the strong coupling and the running masses of the $n_q$ massive quarks at specified scales and in specified flavor-number schemes, as well as the total number of flavors $n_T$, are provided by the user.\footnote{The \Core{} contains $n_T-n_q$ massless quarks, and the standard running mass of the $n$-th massive quark is defined with
$n_T + n -n_q$ active flavors. In the \Core{} the strong coupling and the running masses can be evaluated in a $n_f$ flavor number scheme with $n_T -n_q\le n_f\le n_T$, i.e.\ for $n_q+1$ different flavor number schemes.} Let us call the corresponding tuples $\{n_\alpha, \alpha_s^{(n_\alpha)}(\mu_\alpha)$, $\mu_\alpha\}$ for the strong coupling and $\{k_n, m_{q_n}^{(k_n)}(R_n), R_n\}$,
with $\max[k_n,n_\alpha]\le n_T$, \mbox{$1\leq n\leq n_q$}, for the $n_q$ massive quarks, where $q_1$ refers to the lightest and $q_{n_q}$ to the heaviest massive quark. In a first step \REvolver{} determines the values of the standard running masses $\overline m_{q_n}$. Subsequently, the following strong coupling and running mass values at each flavor threshold matching scale $\mu_n$ are obtained: $\alpha_s^{(n_T + n -1 -n_q)}(\mu_n)$, $\alpha_s^{(n_T + n -n_q)}(\mu_n)$ and $m_n^{(n_T + n -1 -n_q)}(\mu_n)$, $m_n^{(n_T + n -n_q)}(\mu_n)$. These numbers serve as the initial conditions to compute $\alpha_s$ and the running masses $m_{q_n}$ at any renormalization scale and in any specified flavor-number scheme.
We have implemented into our code a fast multi-dimensional recursive algorithm that solves the coupled system of equations to the specified precision.

The determination of the standard running masses $\overline m_{q_n}$ in the algorithm is particularly important since the flavor matching scales $\mu_n$ are specified as dimensionless coefficients times $\overline m_{q_n}$. If at the $i$-th iteration the standard running masses are $[\overline m_{q_n}]_i$, the numerical values $[\overline m_{q_n}]_{i+1}$ at step $(i+1)$
are computed evolving (and matching if necessary) the quark masses $m_{q_n}^{(k_n)}(R_n)$ from the user-specified $\mu=R_n$, for which the numerical value of the mass was provided, to $\mu=[\overline m_{q_n}]_i$.
To do that, before running the masses, the program computes the values of $\alpha_s$ at the various thresholds $\mu_n$ using $[\overline m_{q_n}]_i$ as well. After each iteration is carried out, the new standard running masses are compared to the previous ones, and if their largest relative deviation is smaller than the required precision, the process stops. In the initial
step it is assumed that $[\overline m_{q_n}]_{i=0}=m_{q_n}^{(k_n)}(R_n)$. We checked that the algorithm is very robust and converges quickly even for very extreme starting
conditions such as a charm quark mass defined with $k_1=6$ active flavors taking for $R_1$ a very large scale, together with a top quark mass defined at $R_3=1\,$GeV with $k_3=3$ active flavors.
There is the possibility that the algorithm does not converge, if a particular sequence of $[\overline m_{q_n}]_i$-values repeats itself. This is avoided by adopting, after the $20$-th iteration ($i\ge 21$) the linear combination $(1-\mathtt{damp})[\overline m_{q_n}]_i+\mathtt{damp}[\overline m_{q_n}]_{i-1}$, with \texttt{damp} being a random number between $0$ and $0.2$, as the outcome of the $i$-th iteration.

To compute the strong coupling flavor matching \REvolver{} takes as exact the upward relation (at the user-specified loop order)
\begin{equation}\label{eq:matching}
\alpha^{(n_{\ell} + 1)}_s (\mu_n) = \alpha^{(n_{\ell})}_s (\mu_n) \Biggl\{ 1 +
\sum_{i = 1} \biggl[ \frac{\alpha^{(n_{\ell})}_s (\mu_n)}{4 \pi} \biggr]^i \xi_i(n_\ell) \Biggr\},
\end{equation}
where $\xi_i$ may depend on $\log(\overline m_{q_n}/\mu_n)$. This means that the values for $\alpha_s^{\!(n_\ell)}\!(\mu_n)$ and $\alpha_s^{(n_\ell+1)}(\mu_n)$ in \REvolver{} always satisfy exactly Eq.~\eqref{eq:matching}. If $\alpha_s^{(n_\ell)}(\mu_n)$ is given, the value of $\alpha_s^{(n_\ell+1)}(\mu_n)$ is computed directly from Eq.~\eqref{eq:matching}. If, on the contrary, $\alpha_s^{(n_\ell+1)}(\mu_n)$ is given and $\alpha_s^{(n_\ell)}(\mu_n)$ shall be obtained, our program numerically inverts Eq.~\eqref{eq:matching} in an iterative way as described in the following. For simplicity we omit the argument $\mu_n$ and write $[\alpha^{(n_\ell)}]_{k}$ for the $k$-th iteration value. The value at the $(k+1)$-th step the reads
\begin{equation}\label{eq:threshold-inverse}
[\alpha^{(n_\ell)}]_{k+1} = \frac{\alpha_s^{(n_\ell+1)}(\mu_n)}{1 + \sum_{i = 1} \Bigl( \frac{[\alpha^{(n_\ell)}]_k}{4 \pi} \Bigr)^{\!i}\, \xi_i(n_\ell)}\,,
\end{equation}
where for the first iteration $[\alpha^{(n_\ell)}]_{k=0}=\alpha_s^{(n_\ell+1)}(\mu_n)$ is adopted. The algorithm converges very quickly, even for $\alpha_s^{(n_\ell+1)}(\mu_n)$ values
as large as $0.69$. For smaller values of the strong coupling, and specially if $\mu_n\simeq \overline m_{q_n}$, the RHS of
Eq.~\eqref{eq:threshold-inverse} is very weakly depending on $\alpha^{(n_\ell)}_k$ (since the 1-loop term is very small or vanishing) and only a few iterations are necessary. If $\alpha_s^{(n_\ell+1)}(\mu_n)$ is larger than $0.69$, there is no solution to Eq.~\eqref{eq:matching} and the iterative procedure will fail. In those cases the
program will return the leading order solution $\alpha_s^{(n_\ell)}(\mu_n)=\alpha_s^{(n_\ell+1)}(\mu_n)$. If the solution exists, the algorithm converges as well, because
the slope of the left-hand-side is smaller than unity in absolute value.

\subsection{Strong Coupling}
\label{sec:strongcoupling}
For the QCD $\beta$-function at $N$-loop order (where $N$ is the integer value specified in the variable \texttt{runAlpha}), see Eq.~\eqref{eq:alphaRGE}, the coefficients $b_i$ are defined by
\begin{equation}\label{eq:betaN}
\beta^N_{\rm{QCD}} (\alpha_s) \equiv -\frac{\alpha^2_s}{2 \pi} \beta_0
\biggl[ 1 + \sum_{i = 1}^{N-1} \Bigl( \frac{\alpha_s}{4 \pi} \Bigr)^{\!\!i} \,b_i \biggr],
\qquad b_i \equiv \frac{\beta_i}{\beta_0} \,.
\end{equation}
The algorithms implemented in \REvolver{} provide solutions for the strong coupling differential equation based on the $N$-loop $\beta$-function that are numerically exact (within the specified precision) and do not involve any additional approximation related to $N$-loop precision.
The associated coefficients $c^N_n$, which parametrize the associated inverse QCD $\beta$-function, are defined by
\begin{equation}\label{eq:betaInv}
\frac{1}{\beta^N_{\rm{QCD}} (\alpha_s)} = -\frac{2 \pi}{\beta_0}
\frac{1}{\alpha^2_s} \biggl[ 1 + \sum_{i = 1} c^N_i \Bigl( \frac{\alpha_s}{4
\pi} \Bigr)^{\!\!i}\, \biggr] ,
\end{equation}
where, except for $N=1$, the sum extends to infinity\footnote{The notation where the upper limit of the sum is omitted signifies that the \REvolver{} algorithms employ the number of terms mandatory to reach numerical results that are exact within the specified precision.
The analogous notation is used for other formulae shown below.} even though there is a finite limit $N$ in the sum defining $\beta^N_{\rm{QCD}}$ (for $N=1$ one has $c^N_i=0$ for all $i\ge 1$). The (infinite set of) $c^N_i$ coefficients can be computed in terms of $N-1$ coefficients $b_1,\ldots b_{N-1}$ using the recursive relation shown in Eq.~\eqref{eq:cRed}. The convergence radius of the series is set by the distance from the origin to the nearest pole in the complex plane, which is equivalent to the smallest module of the $\beta$-function roots (excluding the double pole at the origin), which are in general complex numbers. Hence the convergence radius depends both on the number of flavors and the loop order $N$. It increases with the number of flavors and decreases with the loop order $N$, and in all cases is much larger than any physical value of $\alpha_s$ that may arise in phenomenological applications.

For convenience we define $b_0=c^N_0=1$, which allows to write
\begin{equation}\label{eq:cRed}
c^N_{n + 1} = - \!\!\!\!\sum_{i = 1}^{\min(N-1, n + 1)}\!\! c^N_{n + 1 - i} \,b_i\, .
\end{equation}
Many analytic formulae implemented in \REvolver{} use the \mbox{$t$-variable} formalism~\cite{Hoang:2008yj,Hoang:2017suc} which is based on the definition
\begin{equation}
t = - \frac{2 \pi}{\beta_0} \frac{1}{\alpha_s}\,,
\end{equation}
which gives the relation
\begin{equation}
\label{eq:dlnmudt}
{\rm d}\ln(\mu) = - \hat{b}^N\!(t) \,{\rm d}t \,,
\end{equation}
for the renormalization scale $\mu$ where
\begin{equation}
\label{eq:hatbetaN}
\hat{b}^N\!(t) = - \frac{2 \pi}
{\beta_0\,t^2 \beta^N_{\rm{QCD}}\bigl(- \frac{2 \pi}{\beta_0} \frac{1}{t}\bigr) }
\equiv 1 + \sum_{i = 1} \hat{b}^N_i \,t^{- i}\,,
\end{equation}
with $\hat{b}^N_n = (- 1)^n c^N_n/(2 \beta_0)^n$, which implies $\hat{b}^N_0=1$.\footnote{The $\hat{b}^N_i$ should not be confused with the $b_i$ coefficients defined in Eq.~\eqref{eq:betaN}.}
The $c_n^N$ and $\hat b_n^N$ coefficients are obtained numerically whenever needed and stored in a member vector (whose length is extended if necessary) such that they are not re-computed again when used later. For the integral of Eq.~\eqref{eq:dlnmudt} it is useful to define the following polynomial in $1/t$ expressed in terms of $\hat b_i^N$:
\begin{equation}\label{eq:Gtilde}
\tilde{G}^N\!(t) \equiv \int_{-\infty}^t \text{d} t' \biggl[ \hat{b}^N(t') - 1 -
\frac{\hat{b}^N_1}{t'}\, \biggr] = - \sum_{n = 1} \frac{\hat{b}^N_{n + 1}}{n}\, t^{-n}\,.
\end{equation}
For $N=1$ one has $\tilde{G}^N\!(t) = 0$ while for $N\ge 2$ the sum in $\tilde{G}^N\!(t)$ always extends to infinity.
The exponential of this expression defines the coefficients $g_\ell^N$ upon reexpansion in powers of $1/t$, $e^{\tilde{G}^N(t)} \equiv \sum_{\ell = 0} g^N_{\ell} (- t)^{-\ell}$, which can be computed with the recursive relation
\begin{equation}
g^N_{n + 1} = \frac{1}{n + 1} \sum_{i = 0}^n (- 1)^i\, \hat{b}^N_{i + 2}\, g^N_{n -i}\,,
\end{equation}
where ${g}^N_{0}=1$.
It is also necessary to define the coefficients that result from expanding $e^{-\tilde{G}^N(t)} \equiv \sum_{\ell = 0} \tilde{g}^N_{\ell} (- t)^{-\ell}$ in an analogous way. They can be computed with a very similar recursive algorithm
\begin{equation}
\tilde{g}^N_{n + 1} = - \frac{1}{n + 1} \sum_{i = 0}^n (- 1)^i \, \hat{b}^N_{i+ 2} \, \tilde{g}^N_{n - i} \,,
\end{equation}
where $\tilde{g}^N_{0}=1$. Once again one has $g_n^N=\tilde g_n^N=0$ for $n\ge 1$ when $N=1$.
Both $g$ and $\tilde g$ coefficients are computed when first needed and then are conveniently stored in member vectors for later use.

The QCD scale $\Lambda_{\rm QCD,t}$ in the t-scheme~\cite{Hoang:2017suc} is defined by
\begin{equation}
\label{eq:lambda}
\Lambda^N_{\rm{QCD,t}} = \mu\,e^{G^N\!\bigl(- \frac{2 \pi}{\beta_0} \frac{1}{\alpha_s(\mu)}\bigr)}\,,
\end{equation}
where $G^N\!(t) = t + \hat{b}^N_1 \log (- t) + \tilde{G}^N(t)$ (where we recall that $\hat b_{i>0}^N=0$ if $N=1$).
The summation for the function $\tilde{G}^N$ in Eq.~\eqref{eq:Gtilde} is terminated after the relative size of the last term compared to the associated partial sum is smaller than the specified precision.
Since in the evaluation of Eq.~\eqref{eq:lambda} the exact solution of the strong coupling differential equation with $N$-loop $\beta$-function $\beta^N_{\rm QCD}$ (see next section) for $\alpha_s(\mu)$ is used, the numerical value of $\Lambda^N_{\rm QCD}$ does (within the specified precision) not depend on $\mu$.
For the evaluation in \REvolver{} the value $\mu=100$\,GeV is always used.
The values for the QCD scale are computed for each number of active flavors and stored as members of the class. For cases in which $\alpha_s$ is very large, the sum $G^N$ may not converge within the specified precision after adding $200$ terms and \REvolver{} will return \texttt{NaN} (see below why the limit is set to $200$). Again, this only happens for unphysical $\alpha_s$ values.

For the QCD scale in the $\MSbar$ definition, $\Lambda_{\rm QCD,\MSbar}$ (following \cite{Zyla:2020zbs}), we use the formula
\begin{equation}
\label{eq:lambdaMSbar}
\Lambda^N_{\rm{QCD,\MSbar}} = 2^{\hat b^N_1}\,\Lambda^N_{\rm{QCD,t}}\,,
\end{equation}
therefore both schemes coincide for $N=1$.

The algorithm used by \REvolver{} to solve the differential equation for the strong coupling is based on variable separation.
Defining $\ell_{\mu}\equiv \log(\mu/\mu_0)$, $\alpha_\mu\equiv\alpha_s(\mu)$, $\alpha_0\equiv\alpha_s(\mu_0)$ and $a\equiv \alpha_s/(4\pi)$ one
can expand $1/\beta_{\rm QCD}^N$ as in Eq.~\eqref{eq:betaInv} and integrate term by term. This gives
\begin{equation}
\ell_{\mu} = \!\!\int_{\alpha_0}^{\alpha_\mu}\!\!\! \frac{\text{d} \alpha}{\beta^N_{\rm{QCD}}(\alpha)}
= \! - \frac{1}{2 \beta_0} \biggl[\frac{1}{a_0} - \frac{1}{a_\mu} + c_1^N \log \!\biggl( \frac{a_\mu}{a_0} \biggr) +\!
\sum_{i = 1} \frac{c^N_{i + 1}}{i} ( a_\mu^i - a_0^i ) \biggr] .\label{eq:masterAlpha}
\end{equation}
where the sum on the RHS is carried out
until the last term added is smaller than the specified precision.

We now describe the algorithm to determine $a_\mu$ from the algebraic equation~(\ref{eq:masterAlpha}). At leading-log (LL), i.e.\ using the 1-loop $\beta$-function ($N=1$), one has $\beta_{n>0}=0$ and the solution is $a^{\rm LL}_\mu = a_0/(1 + 2 \beta_0 a_0 \ell_\mu)$. For $N\ge 2$
the LL solution is used to rewrite Eq.~\eqref{eq:masterAlpha} in the form
\begin{equation}
\frac{1}{a_{\mu}} = \frac{1}{a^{\rm{LL}}_{\mu}} + c_1^N \log \biggl(
\frac{a_{\mu}}{a_0} \biggr) + \sum_{i = 1} \frac{c^N_{i + 1}}{i} (a_{\mu}^i -
a_0^i)\,,
\end{equation}
which is solved recursively.
In this iterative procedure, the \mbox{$(n+1)$-th} iteration value for the $N$-loop strong coupling is obtained from the $n$-th iteration one by the relation
\begin{equation}
\label{eq:asmuasLL}
[a_{\mu}]_{n + 1} = \frac{1}{\frac{1}{a^{\rm LL}_{\mu}} + c_1^N \log \Bigl(
\frac{[a_{\mu}]_n}{a_0} \Bigr) + \sum_{i = 1} \frac{c^N_{i + 1}}{i}
[([a_{\mu}]_n)^i - a_0^i]} \,.
\end{equation}
With the initial choice $[a_{\mu}]^N_0 \equiv a^{\rm{LL}}_{\mu}$ it is ensured that the correct solution is obtained in all cases. The iterative method converges provided that (i)~the solution exists and (ii)~the (absolute value of the) slope of the right-hand-side function is smaller than $1$. The second condition is always satisfied, and the former will be discussed later in this paragraph. Since \mbox{$\beta_1>0$} we have that $a_\mu^{\rm LL} < a_\mu$, and the solution beyond 1-loop order is approached from below. The iterative procedure is carried out until the numerical value of $a_\mu$ does not change within the specified precision, with a maximum allowed number of $200$ iterations.

The convergence radius for the infinite sum over $c_{i+1}^N$ in Eq.~\eqref{eq:asmuasLL} is the same as for the inverse $\beta$-function in Eq.~\eqref{eq:betaInv}.
However, due to technical limitations related to double-precision floating numbers, \REvolver{} cannot carry out the sum for $\alpha_s$ values arbitrarily close to the convergence radius within the specified precision. If after adding $200$ terms this precision is not met, the recursive procedure to obtain $\alpha_s$ will stop and \texttt{NaN} will be returned. This happens for $\alpha_s\gtrsim0.7$, hence outside the range of any physical application.

\subsection{\texorpdfstring{$\overline{\rm MS}$}{MSbar} and RGI masses}
The $K$-loop renormalization group equation (where $K$ is the integer value specified in the variable \texttt{runMSbar}) for the $\overline{\rm MS}$ quark mass $\overline m_q(\mu)$ has the form
\begin{equation}
\frac{\mathrm{\text{d}} \,\overline m_q(\mu)}{\mathrm{\text{d}} \ln(\mu)}
= 2 \overline m_q(\mu) \gamma_m^K(\alpha_s) =
2 \overline m_q(\mu) \sum_{n = 0}^{K-1} \gamma_n \biggr[
\frac{\alpha_s(\mu)}{4 \pi} \biggl]^{\!n+1}\,,
\label{eq:mass-adim}
\end{equation}
which implies logarithmic scale evolution. The \REvolver{} algorithm provides (within the user specified precision) the exact solution of Eq.~\eqref{eq:mass-adim} employing the $N$-loop strong coupling described in Sec.~\ref{sec:strongcoupling}. Changing the integration variable from $\ln(\mu)$ to $\alpha_s$ followed by separation of variables, and using the mass $\overline m_q(\mu_0)$ at the scale $\mu_0$ as the initial value,
the analytic solution reads
\begin{equation}\!\!\!\!
\overline m_q (\mu)\! =\! \overline m_q(\mu_0) \exp\! \biggr[2 \!\!\int^{\alpha_\mu}_{\alpha_0} \!\!\text{d} \alpha \,\frac{\gamma_m^K\!
(\alpha)}{\beta^N_{\rm{QCD}} (\alpha)} \biggl] \equiv \overline m_q(\mu_0)\exp\!\big[\tilde{\omega}^{(N,K)}(\mu_0,\mu)\big],
\end{equation}
which defines the running kernel $\tilde\omega^{(N,K)}(\mu_0,\mu)$.
The \REvolver{} algorithm determines $\tilde\omega^{(N,K)}(\mu_0,\mu)$ from the Taylor series of the integrand
in $\alpha$ where each term is integrated individually. The result can be written in the form [\,$a_i=\alpha_s(\mu_i)/(4\pi)$\,]
\begin{equation}
\label{eq:wTilde}\!\!\!
\tilde{\omega}^{(N,K)}(\mu_1, \mu_2)\! = \!- \frac{1}{\beta_0} \! \biggl[ \gamma_0 \log \biggl(
\frac{a_2}{a_1} \biggr) \!+ (1-\delta_{N,1}\delta_{K,1})\!\sum_{n = 1} (a_2^n - a_1^n)\,d^{(N,K)}_n \biggr],
\end{equation}
where the coefficients $d^{(N,K)}_n$, defined for $n>0$, are non-zero if either $N$ or $K$ are larger than $1$. We also have $d^{(1,K)}_n = \gamma_n/\beta_0$ if $n<K$ and zero otherwise. They are given by the following expression
\begin{equation}
\label{eq:dtilde}
d^{(N,K)}_n \equiv \frac{1}{n}\! \sum^{\min(n,K-1)}_{i = 0} c_{n-i}^N \,\gamma_i \,,
\end{equation}
where the terms $c^N_i$ arise in the expansion of the inverse $\beta$-function~(\ref{eq:betaInv}).
\REvolver{} carries out the sum in Eq.~\eqref{eq:wTilde} until the relative size of the last computed term with respect to the associated partial sum is below the user specified precision.
The convergence radius of the infinite sum is the same as for the inverse $\beta$-function in Eq.~\eqref{eq:betaInv}. In practice, if convergence is found when computing $\alpha_s$, the sum for the computation of $\tilde{\omega}^{(N,K)}$ is going to
be convergent as well. If no convergence is found, \texttt{NaN} is returned.

The RGI-mass $\hat m_q$ is obtained from the $\MSbar$ mass $\overline m_q(\mu)$ by removing its
scale-dependence writing $\tilde{\omega}^{(N,K)}(\mu_0, \mu)=\tilde{\omega}^{(N,K)} (\mu)-\tilde{\omega}^{(N,K)} (\mu_0)$ with
\begin{equation}
\label{eq:wTilde2}
\tilde{\omega}^{(N,K)}(\mu) = - \frac{1}{\beta_0} \biggl[ \gamma_0 \log\! \biggl(
\frac{a_\mu}{\pi} \biggr) + (1-\delta_{N,1}\delta_{K,1})\sum_{n = 1} d^{(N,K)}_n a_\mu^n \biggr].
\end{equation}
This leads to the definition $\hat m_q \equiv \overline m_q(\mu)\exp[- \tilde{\omega} (\mu)]$. \REvolver{} uses the \mbox{$N$-loop} exact solution for $a_\mu$ and the $K$-loop solution for $m_q(\mu)$, which ensures that the value obtained for the RGI-mass is $\mu$-independent. In \REvolver{} the computation of $\tilde{\omega}(\mu)$ is carried out in the same way as for the $\MSbar$ mass explained above. For the numerical calculation $\mu$ is set to the standard running mass, $\mu=\overline m_q$.

\subsection{MSR mass}
\label{sec:MSRmass}
The $M$-loop renormalization group equation for the MSR mass $m_q^{\rm MSR}(R)$ (where $M$ is the integer value specified in the variable \texttt{runMSR}) has the form
\begin{equation}
\label{eq:Revolution}
\frac{\text{d} m_q^{\rm{MSR}} (R)}{\text{d} \ln(R)} = - R\,\gamma_R^M[\alpha_s(R)] = - R\sum_{n = 0}^{M-1} \gamma^R_n \biggl[\frac{\alpha_s(R)}{4 \pi} \biggr]^{n+1}\,,
\end{equation}
and implies logarithmic and linear scale dependence.
The coefficients $\gamma^R_n$ are obtained from the series relating the pole and MSR masses using Eq.~(3.2) of Ref.~\cite{Hoang:2017suc}. The \REvolver{} algorithm provides (within the user specified precision) the exact solution of Eq.~\eqref{eq:Revolution} employing the $N$-loop strong coupling described in Sec.~\ref{sec:strongcoupling}. In the following we provide details of the algorithm's mathematical derivation.

Switching to the $t$-variable formalism~\cite{Hoang:2008yj,Hoang:2017suc} one obtains re-scaled $\gamma_R$ coefficients:
\begin{equation}
\gamma_R^M\!\biggl(- \frac{2 \pi}{\beta_0} \frac{1}{t}\biggr)\equiv \sum_{n = 0}^{M-1} \tilde{\gamma}^R_n (- t)^{- n - 1}, \qquad
\tilde{\gamma}^R_n =\frac{\gamma^R_n}{(2 \beta_0)^{n + 1}} \,.
\end{equation}
These can be used to define the series coefficients $S_j^{(N,M)}$ in the expansion of the product function
$\gamma_R^M\!(- \frac{2 \pi}{\beta_0} \frac{1}{t}) \hat{b}^N\!(t) e^{- \tilde{G}^N\!(t)} \equiv \sum_{j = 0} S^{(N,M)}_j (- t)^{- j-1} $
in powers of $1/t$, see Sec.~\ref{sec:strongcoupling}. The coefficients have the form
\begin{equation}\label{eq:S-recursive}
S^{(N,M)}_j = \! \! \sum_{k = 0}^{\min (M-1, j)} \! \! \tilde{\gamma}^R_k
\sum_{i = 0}^{j - k} (- 1)^i \,\tilde{b}^N_i \, \tilde{g}^N_{j - i - k}\,,
\end{equation}
which implies that $S^{(1,M)}_k = \tilde{\gamma}^R_k$ for $k<M$ and $S^{(1,M)}_k = 0$ for $k\ge M$.
The difference of MSR masses at the renormalization scales $R_1$ and $R_2$, \mbox{$\Delta_{\rm{MSR}} (R_2, R_1)\equiv m_q^{\rm{MSR}} (R_2) - m_q^{\rm{MSR}} (R_1)$}, can be computed from Eq.~\eqref{eq:Revolution} by integrating over $\alpha_s$ (or $t$) using the relation $R = \Lambda^N_{\rm{QCD}} e^{-G^N\!\bigl[- \frac{2 \pi}{\beta_0} \frac{1}{\alpha_s^N\!(R)}\bigr]}$
\begin{align}
\label{eq:DeltaR1R2}
&\Delta^{(M,N)}_{\rm{MSR}} (R_2, R_1) =
- \Lambda^N_{\rm{QCD}}\!\int_{\alpha_1}^{\alpha_2} \!\! \frac{\text{d} \alpha}{\beta^N (\alpha)}
e^{-G^N\!\bigl[- \frac{2 \pi}{\beta_0} \frac{1}{\alpha}\bigr]} \gamma^M_R (\alpha)\\
&=\Lambda^N_{\mathrm{QCD}}\int_{t_1}^{t_2} \text{d} t \,\hat{b}^N (t) e^{- \tilde{G}^N\!(t)}
\gamma_R\biggl(- \frac{2 \pi}{\beta_0} \frac{1}{t}\biggr) e^{- t} (- t)^{- \hat{b}_1^N}\nonumber\\
&=\Lambda^N_{\mathrm{QCD}}\sum_{j = 0} S^{(N,M)}_j \! \!\int_{t_1}^{t_2} \!\text{d} t\, e^{- t} (- t)^{- 1 - \hat{b}_1^N -j} \nonumber \\
& \equiv\Lambda^N_{\mathrm{QCD}}\sum_{j = 0} S^{(N,M)}_j[I (\hat{b}^N_1, j, t_2) - I (\hat{b}^N_1, j, t_1)]\,,\nonumber
\end{align}
where in the first line we have used
$\beta^N(\alpha)\text{d} R = \Lambda^N_{\rm{QCD}}e^{-G^N\!\bigl[- \frac{2 \pi}{\beta_0} \frac{1}{\alpha_s^N\!(R)}\bigr]} \text{d} \alpha$, in the second
we have switched variables to $t$ and in the third we used the definition of the coefficients in Eq.~\eqref{eq:S-recursive}.
Using integration by parts repeatedly one can write $I (\hat{b}^N_1, j, t)$
in terms of $I (\hat{b}^N_1, 0, t)$
\begin{equation}\label{eq:IBP}
I (\hat{b}^N_1, j, t)=\frac{1}{(1 + \hat{b}^N_1)_j} \!\biggl[I (\hat{b}^N_1, 0, t) + e^{- t} \sum_{i = 1}^j
(1 + \hat{b}^N_1)_{i-1} (- t)^{- i - \hat{b}^N_1}\biggr],
\end{equation}
with $(a)_n=\Gamma(a+n)/\Gamma(a)$ being the Pochhammer symbol. The exponential contained in $I (\hat{b}^N_1, 0, t)$
can be expanded in powers of $t$ (with infinite convergence radius) and each term can be integrated separately.
In practice, the algorithm computes the difference
\begin{equation}
\label{eq:Idiffdef}
I (\hat{b}^N_1, 0, t_2) - I (\hat{b}^N_1,0, t_1) = \sum_{i = 0} \frac{1}{i !} \frac{(- t_2)^{i -
\hat{b}^N_1} - (- t_1)^{i - \hat{b}^N_1}}{\hat{b}^N_1- i}\,,
\end{equation}
terminating the infinite sum once the specified precision is achieved, and the result for this difference is saved, such that it needs to be computed only once for each computation of $\Delta^{(M,N)}_{\rm{MSR}} (R_2, R_1)$. In case $\hat b_1^N=0$ the $i=0$ term in Eq.~\eqref{eq:Idiffdef} is replaced by $\ln(t_1/t_2)$.
Furthermore, the values of the Pochhammer symbols in the formula Eq.~\eqref{eq:IBP} are computed only once in one \Core{} object and stored for later use.
Likewise, since the sum in Eq.~(\ref{eq:IBP}) depends on $j$ only through its upper limit, its value is stored and updated as $j$ grows by one unit.

\REvolver{} computes the (infinite) sum over $j$ in Eq.~(\ref{eq:DeltaR1R2}) by truncating
it once the desired precision is achieved. The convergence radius of the $j$ sum coincides with that of the inverse $\beta$-function explained in Sec.~\ref{sec:strongcoupling}.
So, if convergence is found for the evolution of the strong coupling $\alpha_s$, the sum in Eq.~(\ref{eq:DeltaR1R2}) converges as well. If no convergence is found for the sum or for the strong coupling, \texttt{NaN} is returned. \REvolver{} stores in a member vector the values of $S_j^{(M,N)}$ after they are computed the first
time, such that their values can be recycled in later evaluations of R-evolution.

\subsection{Asymptotic Pole Mass Coefficients}
\label{sec:asymptotic}

For the asymptotic formulae for the ${\cal O}(\Lambda_{\rm QCD})$ renormalon-dominated perturbative coefficients of the relation between the MSR and pole masses (see Eqs.~(2.1) and (2.3) in Ref.~\cite{Hoang:2017suc}), always defined in the flavor number scheme where all massive quarks are integrated out,
we use the expression
\begin{equation}
a_n
= (2\beta_0)^n\! \sum_{k=0}^{k_{\rm max}[n-1]} \! S^{(N,M)}_k\!\!
\sum_{\ell=0}^{\ell_{\rm max}[n - 1-k]} \!\!g^N_\ell
(1+\hat b_1^N + k)_{n-1-\ell -k}\,,
\label{eq:anasymptotic}
\end{equation}
where
\begin{align}
k_{\rm max}[m] \, = \,
& {\rm min}[m,M-1]\,, \\
\ell_{\rm max}[m] \, = \,
& {\rm max}[ {\rm min}(m,N-2) , 0]\,.\nonumber
\end{align}
The integers $N$ and $M$ are explained in App.~\ref{sec:strongcoupling} and \ref{sec:MSRmass}.
The asymptotic formula~\eqref{eq:anasymptotic} depends on the number of massless quarks (entering the expressions for the QCD $\beta$-function coefficients $\beta_i$ and the coefficients $S^{(N,M)}_k$ and $g^N_\ell$) and was derived in Ref.~\cite{Hoang:2017suc} [\,see Eq.~(4.19) in that reference\,]. It has the property that it reproduces the exact coefficients $a_n$ up to ${\cal O}(\alpha_s^{{\rm min}[M,N-1]})$. In contrast to the expression given in Ref.~\cite{Hoang:2017suc}, Eq.~\eqref{eq:anasymptotic} displays concrete truncation prescriptions for the sums as a function of the loop order variables.

\subsection{Renormalon Sum Rule}
\label{sec:sumrule}
To extract the pole mass renormalon normalization $N_{1/2}$ from a \Core{}, the following formula is used:
\begin{equation}
\label{eq:N12def}
N_{1/2} =
\frac{\beta_0}{2 \pi}\sum_{k=0}^{M}\frac{S_k^{(N,M)}}{(1+\hat b_1^N)_k} \;,
\end{equation}
The formula was derived in Ref.~\cite{Hoang:2017suc}, and Eq.~(\ref{eq:N12def}) is adapted to account for the optional parameters
used in \REvolver{}.
In analogy to Eq.~\eqref{eq:anasymptotic}, $N_{1/2}$ is always evaluated in the flavor number scheme where all massive quarks are integrated out.
The expression for the coefficients $S^{(N,M)}_k$ is given in Eq.~\eqref{eq:S-recursive} while $b_1^N$ has been defined in Eq.~\eqref{eq:hatbetaN}.

\bibliographystyle{elsarticle-num}
\bibliography{biblio}

\end{document}